\documentclass[epj]{svjour}
\usepackage{graphicx,psfrag}
\usepackage{mathrsfs}
\usepackage{amsmath,amsfonts,amssymb}
\usepackage{multirow}
\usepackage{comment}
\usepackage{ulem}
\usepackage{hyperref}
\usepackage{graphicx}
\usepackage{acronym}
\usepackage{pifont}

\newcommand{\be}{\begin{equation}}
\newcommand{\ee}{\end{equation}}
\newcommand{\bea}{\begin{eqnarray}}
\newcommand{\eea}{\end{eqnarray}}
\newcommand{\bel}{\begin{align}}
\newcommand{\eel}{\end{align}}

\def\gccm{{\rm g\,cm^{-3}}}

\def\GMc2{{\rm G M_{\odot} c^{-2}}}

\usepackage{color}
\definecolor{cyan}{rgb}{0,0.9,0.9}
\definecolor{orange}{rgb}{0.9,0.5,0}
\definecolor{magenta}{rgb}{1,0,1}
\definecolor{purple}{rgb}{0.8,0.4,0.8}
\definecolor{gray}{rgb}{0.8242,0.8242,0.8242}

\newacro{ADM}{Arnowitt-Deser-Misner}
\newacro{AMR}{adaptive mesh-refinement}
\newacro{BH}{black hole}
\newacro{BBH}{binary black-hole}
\newacro{BHNS}{black-hole neutron-star}
\newacro{BNS}{binary neutron star}
\newacro{CCSN}{core-collapse supernova}
\newacroplural{CCSN}[CCSNe]{core-collapse supernovae}
\newacro{CMA}{consistent multi-fluid advection}
\newacro{DG}{discontinuous Galerkin}
\newacro{HMNS}{hypermassive neutron star}
\newacro{EM}{electromagnetic}
\newacro{ET}{Einstein Telescope}
\newacro{EOB}{effective-one-body}
\newacro{EOS}{equation of state}
\newacroplural{EOS}[EOSs]{equations of state}
\newacro{FF}{fitting factor}
\newacro{GR}{general relativity}
\newacro{GRLES}{general-relativistic large-eddy simulation}
\newacro{GRHD}{general-relativistic hydrodynamics}
\newacro{GRMHD}{general-relativistic magnetohydrodynamics}
\newacro{GW}{gravitational wave}
\newacroplural{GW}[GWs]{gravitational waves}
\newacro{ILES}{implicit large-eddy simulations}
\newacro{LIA}{linear interaction analysis}
\newacro{LES}{large-eddy simulation}
\newacroplural{LES}[LES]{large-eddy simulations}
\newacro{MRI}{magnetorotational instability}
\newacro{NR}{numerical relativity}
\newacro{NS}{neutron star}
\newacroplural{NS}[NSs]{neutron stars}
\newacro{PNS}{protoneutron star}
\newacro{SASI}{standing accretion shock instability}
\newacro{SGRB}{short $\gamma$-ray burst}
\newacro{SN}{supernova}
\newacroplural{SN}[SNe]{supernovae}
\newacro{SNR}{signal-to-noise ratio}

\begin{document}
\title{Thermodynamics conditions of matter in neutron star mergers}
\author{Albino Perego\inst{1,2} \and Sebastiano Bernuzzi \inst{3} \and David Radice \inst{4,5}} 
%
\offprints{albino.perego@unitn.it}          
\institute{Department of Physics, Trento University, Via Sommarive 14, 38123 Trento, Italy \and 
           Istituto Nazionale di Fisica Nucleare, Sezione di Milano-Bicocca, Piazza della Scienza 20100, Milano, Italy \and 
           Theoretisch-Physikalisches Institut, Friedrich-Schiller-Universit\"at Jena, 07743, Jena, German \and 
           Institute for Advanced Study, 1 Einstein Drive, Princeton, NJ 08540, USA \and 
           Department of Astrophysical Sciences, Princeton University, 4 Ivy Lane, Princeton, NJ 08544, USA}
\date{Received: date / Revised version: date}
%
\abstract{
  Matter in neutron star collisions can reach densities up to 
  few times the nuclear saturation threshold ($\rho_0\simeq2.7\times
  10^{14}$~$\gccm$) and temperatures up to one hundred MeV. 
  Understanding the structure and composition of such matter
  requires many-body nonperturbative calculations that are currently highly
  uncertain.
  Unique constraints on the neutron star matter are provided by 
  gravitational-wave observations aided by numerical relativity simulations.
  In this work, we explore the thermodynamical conditions of
  matter and radiation along  
  the merger dynamics. We consider 3 microphysical equation of state
  models and numerical relativity simulations including an approximate 
  neutrino transport scheme.
  The neutron star cores collision and their multiple
  centrifugal bounces heat the initially cold matter to several tens of
  MeV. Streams of hot matter with initial densities $\sim1-2\rho_0$
  move outwards and cool due to decompression and neutrino emission. 
  The merger can result in a neutron star remnant with 
  densities up to $3-5\rho_0$ and temperatures 
  $\sim 50$~MeV. The highest temperatures are confined in an approximately
  spherical annulus at 
  densities $\sim\rho_0$. Such temperatures favour positron-neutron
  capture at densities $\sim\rho_0$, thus leading to a neutrino emission
  dominated by electron antineutrinos. 
  We study the impact of trapped neutrinos on the remnant matter's pressure,
  electron fraction and temperature and find that it has a negligible
  effect. 
  Disks around neutron star or black hole remnant are 
  neutron rich and not isentropic, but they differ
  in size, entropy and lepton fraction depending on the nature of the central object.
  In presence of a black hole, disks are smaller and mostly
  transparent to neutrinos; in presence of a massive neutron star, 
  they are more massive, geometrically and optically thick.  
} 
\maketitle

\section{Introduction}
\label{intro}

Simulation in \ac{NR} predict that the outcome of
\ac{BNS} mergers on dynamical timescales is the prompt formation of a
black hole via gravitational collapse or of a remnant \ac{NS}
\cite{Shibata:1999wm}. In the
latter case, the cold, $\beta$-equilibrated matter of the individual
\acp{NS} at typical densities
\footnote{We consider here \acp{NS} of
``fiducial'' mass of $M_\text{NS}\sim1.35-1.4~M_{\odot}$.}
of $\rho_\text{NS}\sim2-3\rho_0$
is pushed to $\rho\sim5-6\rho_0$ and
temperatures up to $T\sim50-100$~MeV in the remnant
(see e.g. \cite{Sekiguchi:2011mc,Bernuzzi:2015opx,Radice:2018pdn}).
Those extreme conditions are reached on a timescale of milliseconds,
and potentially differ from those of proto NS in supernova remnants (e.g. \cite{Roberts.etal:2012}).

Our knowledge of the properties of cold matter at supranuclear densities 
are still affected by large uncertainties, see e.g. \cite{Lattimer:2012nd,Ozel:2016oaf}.
Nevertheless they are key in determining the internal structure and 
mass-radius relation of \acp{NS}. Despite many theoretical effords, also the behavior
of nuclear matter at finite temperature is poorly known, see e.g. \cite{Oertel:2016bki}
and references therein.
The first composition- and temperature-dependent versions of the equation of state
(EOS) widely used in simulations of binary NS mergers and core-collapse supernovae
were the Lattimer-Swesty (LS) EOS \cite{Lattimer:1991nc} and the Shen EOS \cite{Shen:1998gq}.
Most of the EOS currently used and compatible with present nuclear constraints are based 
on the relativistic mean field (RMF) approach \cite{Oertel:2016bki}.
However, more realistic EOSs based on detailed microscopic interactions and many-body 
techniques are becoming available, e.g. \cite{Bombaci:2018ksa}.

A detailed knowledge of the matter interactions in the merger
remnant is pivotal for characterizing the dynamics and the related
gravitational and electromagnetic radiation signatures. 
For example, %
tidal interactions during the inspiral are
parametrized by NS compactness and EOS-dependent tidal polarizability coefficients
\cite{Damour:1983a,Flanagan:2007ix,Hinderer:2009ca,Damour:2009wj,Damour:2012yf,Bernuzzi:2012ci}; the properties of the merger
dynamics dependent primarily on those coefficients
\cite{Bernuzzi:2014kca,Zappa:2017xba}.
Moreover, the maximum mass of nonrotating \ac{NS} supported by the EOS
determines the threshold for the prompt gravitational collapse of the remnant
\cite{Hotokezaka:2011dh,Bauswein:2013jpa}.
Similarly, the EOS characterizes the remnant \ac{NS}, the disk, the ejecta, and the neutrino emission properties
(e.g. \cite{Radice:2017lry,Radice:2018pdn}), thus influencing the nucleosynthesis and ultimately 
the properties of the kilonova emission \cite{Radice:2018pdn,Radice:2018xqa}.
The timescale for the gravitational collapse to a black hole, as well as the mass and the lifetime of the disk,
can also play a pivotal role in launching a relativistic jet and producing a (short) gamma-ray burst, see e.g.
\cite{Eichler:1989ve,Nakar:2007yr,Rosswog:2015nja,Just:2016ApJ...816L..30J}.
The appearance of phase transitions and extra degrees of
freedom, including hyperons and quarks, can soften the EOS in the merger
remnant leading to collapse and/or signatures in the \acp{GW} signal
\cite{Sekiguchi:2011mc,Radice:2016rys,Most:2018eaw,Bauswein:2018bma}.   

\ac{NR} simulations are the only means to build models and to
interpret the observations that 
set constraints on the \ac{NS} properties and thus on the EOS. 
The measurement of tidal polarizability parameters 
in the GW170817 signal constrained the EOS at $\rho\lesssim\rho_\text{NS}$, e.g.
\cite{TheLIGOScientific:2017qsa,Abbott:2018wiz,Abbott:2018exr,De:2018uhw}.
\ac{NR} simulations support the development of tidal waveform models 
for such measurement,
e.g. \cite{Baiotti:2011am,Bernuzzi:2012ci,Radice:2013hxh,Hotokezaka:2015xka,Nagar:2018zoe}. 
Similarly, the post-merger signal from the remnant, although not detected in GW170817
\cite{Abbott:2017dke}, could also place strong constraints on
the physics at extreme densities $\rho\lesssim5-6\rho_0$
\cite{Bauswein:2011tp,Takami:2014zpa,Bernuzzi:2015rla,Radice:2016rys,Yang:2017xlf,Chatziioannou:2017ixj}. 

Simulations results combined with the ejecta properties inferred from AT2017gfo (the kilonova 
associated with GW170817) led to upper
bounds on the maximum mass of \ac{NS} by noticing that the energy of the ejecta 
inferred from optical and infrared data are
likely incompatible with both a prompt collapse and  a long lived
remnant \cite{Margalit:2017dij} (see also \cite{Shibata:2017xdx,Rezzolla:2017aly,Ruiz:2017due}). 
Similar arguments, combined with empirical relations between \ac{NS}
radii and the threshold mass for prompt collapse mentioned above, tentatively rule out EOSs
predicting very small \ac{NS} radii \cite{Bauswein:2017vtn}.
The correlation between the disk mass and the tidal
polarizability parameter observed in \ac{NR} simulations can complement
the constraints on the tidal coefficients  in
joint observation of GW and \ac{EM} radiation \cite{Radice:2017lry,Radice:2018ozg}.

In this work we study in details the thermodynamical conditions of
matter and radiation during binary NS (BNS) mergers. We explore three microphysical EOS models
by simulating the matter flow along the general relativistic merger
dynamics and including weak interactions.
Our work aims at quantitatively investigating the typical densities,
temperatures, and electron fractions that can be reached in real astrophysical 
mergers and their intrinsic variability. We also explore the properties and the 
potential role of a gas of dynamically trapped neutrinos deep inside the 
merger remnant.
We finally point out the robust features and some of the main
uncertainties of current state-of-art simulations. 

The paper is structured as follows: in Section~\ref{sec:Method} we present the EOSs employed
in this work, the NR simulations setups, and the method used for the analysis of the
neutrino trapped components. Section~\ref{sec:histograms} is devoted to the presentation
of the thermodynamical conditions experienced by the astrophysical plasma as obtained by several 
BNS merger models, characterized by different EOSs, NS masses and physical ingredients (e.g. turbulent viscosity).
In Section~\ref{sec:neutrinos} we report the results of the analysis of the role of trapped
neutrinos, while in Section~\ref{sec:disks} we discuss the properties of
disks in BNS merger remnants. Finally, we conclude in Section~\ref{sec:conclusions}.

\section{Method}
\label{sec:Method}
\begin{table*}[t]
  \centering    
  \caption{Summary of the hadronic equation of state models considered in this work, with their nuclear matter properties at saturation density ($n_0$)
  and zero temperature: binding energy ($E_0$), incompressibility ($K$), symmetry energy ($S$), slope of the symmetry energy ($L$), radius of a 
  1.4 $M_{\odot}$ NS ($R_{1.4}$), maximum cold, irrotational NS mass ($M_{\rm max}$). Experimental and observational results were taken from the comprehensive
  review of \cite{Oertel:2016bki} and from \cite{Demorest:2010bx,Antoniadis:2013pzd,Radice:2018ozg}. }
  \begin{tabular}{cccccccccc}        
    \hline\hline
     EOS             & Brief description                 & $n_0$             & $E_0$ & $K$   & $S$   & $L$   & $R_{1.4}$ & $M_{\rm max}$ & Ref \\
                     &                                   & ${\rm [fm]^{-3}}$ & [MeV] & [MeV] & [MeV] & [MeV] & [km]      & $[M_{\odot}]$ &     \\ 
    \hline
    DD2              & NSE + RMF                         & 0.1491            & 16.02 & 243   & 31.67 & 55.04 & 13.2     & 2.42           & \cite{Typel:2009sy,Hempel:2009mc}\\
    LS220            & SNA + Liquid droplet Skyrme       & 0.1550            & 16.00 & 220   & 28.61 & 73.82 & 12.7     & 2.06           & \cite{Lattimer:1991nc}\\
    SFHo             &  NSE + RMF                        & 0.1583            & 16.19 & 245   & 31.57 & 47.10 & 11.9     & 2.06           & \cite{Steiner:2012rk} \\
    \hline   
                     & experiments and observations      & 0.15-0.16         & $\sim$16 & 220-315  & 28.5-34.9  & 30.6 - 86.8 & 11.2-13.4 & $>$ 1.97  &   \\
    \hline\hline
  \end{tabular}
 \label{tab:EOS}
\end{table*}

We consider three different nuclear EOS models:
HS(DD2) \cite{Typel:2009sy,Hempel:2009mc},
LS220 \cite{Lattimer:1991nc},
and HS(SFHo) \cite{Steiner:2012rk}, see Tab.~\ref{tab:EOS}
(in the following, we will refer to the first and third ones simply as 
DD2 and SFHo, respectively).
The LS220 EOS is based on a liquid droplet Skyrme
model with a value of 220~MeV for the modulus of the nuclear incompressibility.
It includes surface effects and models $\alpha$-particles as
an ideal, classical, non-relativistic gas. 
Heavy nuclei are treated using the so-called single nucleus approximation (SNA).
The DD2 and SFHo EOSs combine a statistical ensemble of several thousands of nuclei,
under the assumption of nuclear statistical equilibrium (NSE), with a 
RMF approach for the unbound nucleons to treat high-density nuclear matter
\cite{Hempel:2009mc}. 
The phase transition from nuclei to homogeneous nuclear matter at densities close to nuclear saturation density 
is achieved by an excluded volume mechanism. 
DD2 and SFHo
use different parameterizations and values for modeling the mean-field
nuclear interactions.
In particular, DD2 uses a linear, but density dependent coupling \cite{Typel:2009sy},
while the RMF parametrization of SFHo is motivated by neutron star
radius measurements from low-mass X-ray binaries (\cite{Steiner:2012rk} and references therein).

All three models have symmetry energies at saturation density within experimental
bounds. LS220 has a significantly steeper density dependence of its
symmetry energy than the other models. In all models, the finite
temperature behavior of the EOS is mainly determined by the nucleon
effective mass, $m_N^*$, with smaller effective masses leading to higher
temperatures for constant entropy. The LS220 EOS assumes that the
nucleon mass is the bare nucleon mass at all densities, while SFHo has
$m_N^*/m_N = 0.76$ at saturation density, and  DD2 
has $m_N^*/m_N = 0.56$, where $m_N$ is the bare nucleon mass.

These EOSs predict NS maximum masses and radii within the
range allowed by current astrophysical constraints, including the recent
LIGO-Virgo constraint on tidal deformability
\cite{TheLIGOScientific:2017qsa,Abbott:2018wiz,De:2018uhw,Abbott:2018exr}. 
SFHo, LS220, and DD2
support 2.06, 2.06, and 2.42 $M_\odot$ cold,
non-rotating maximum NS masses. The cold, non-rotating NS radius corresponding to
$1.4$~$M_{\odot}$ is $R_{1.4}=11.9$, 12.7, 13.2 km 
respectively. Since NS radii correlate with the
pressure at roughly twice saturation density \cite{Lattimer:2012nd}, we
refer to EOS having smaller $R_{1.4}$ as being ``softer'' and to
EOS having larger $R_{1.4}$ as being stiffer. In Tab.~\ref{tab:EOS} we
summarize the main nuclear properties of the cold EOS models used in this work.

\begin{table}[t]
  \caption{Fiducial binary neutron star configurations.
    We report EOS, individual gravitational masses and maximum rest-mass density of the initial data in units of $\rho_0\simeq2.7\times 10^{14}$~$\gccm$.
    All simulations were performed with a maximal grid resolution of 185~m, using the \texttt{M0} neutrino-transport approximation. Two simulations include
    the effect of physical viscosity and are indicated with ``V'' (see text for more details.)}
\label{tab:bns}
\begin{center}
\scalebox{1}{
\begin{tabular}{llccc}
\hline\hline
Simulation & EOS & $(M_1, M_2)$  &
$(\rho_\text{NS1},\rho_\text{NS2})$ \\
\hline
DD2\_M136136    & DD2   & (1.36, 1.36) & (2.1,2.1)  \\
LS220\_M135135  & LS220 & (1.35, 1.35) & (2.6,2.1)  \\
LS220\_M135135V & LS220 & (1.35, 1.35) & (2.6,2.6)  \\
LS220\_M140120V & LS220 & (1.40, 1.20) & (2.7,2.3) \\
SFHo\_M135135   & SFHo  & (1.35, 1.35) & (3.0,3.0) \\
\hline\hline
\end{tabular}
}
\end{center}
\end{table}

For each EOS model, irrotational BNS configurations in
quasi-circular orbit are computed by solving the general
relativistic initial data problem. We use the
\texttt{Lorene} pseudo-spectral code \cite{LoreneCode} and specify the initial
separation between the NS to at least $40\, {\rm km}$, corresponding
to $\sim2{-}3$ orbits before merger. 
The EOS used for the initial data are constructed from the minimum
temperature slice ($T\sim0.5-0.1$~MeV) of the
EOS table used for the evolution assuming neutrino-less
beta-equilibrium. To create the initial data tables we also subtract
from the pressure the contribution of photon radiation, which dominates
at the lowest densities due to the assumption of constant temperature.
Initial data properties are summarized in Tab.~\ref{tab:bns}.

The initial data are evolved according to Einstein's general relativistic
equations for the spacetime in the (3+1)D form described in \cite{Bernuzzi:2009ex,Hilditch:2012fp}
and coupled to general relativistic hydrodynamics. The (3+1)D simulations span the
merger and the remnant evolution for a timescale of at least 20-30~ms.
We use the \texttt{WhiskyTHC} code \cite{Radice:2012cu,Radice:2013hxh,Radice:2013xpa,Radice:2015nva}; all the technical
details are given in \cite{Radice:2018xqa}. 
All simulations involving the LS220 EOS and simulation DD2\_M136136
are reported here for the first time. 
All our simulations domain covers a cube of 3,024~km in diameter whose
center is at the center of mass of the binary. Our code uses
Berger-Oliger conservative adaptive mesh refinement (AMR;
\cite{Berger:1984zza}) with sub-cycling in time and
refluxing \cite{Berger:1989a,Reisswig:2012nc} as provided by the
\texttt{Carpet} module of the \texttt{Einstein Toolkit}
\cite{Schnetter:2003rb}. We setup an AMR grid structure with 7
refinement levels. The finest refinement level covers both NSs
during the inspiral and the remnant after the merger, and has a typical
resolution of $h \simeq 185\, {\rm m}$. 

For a subset of binaries,
we use the general-relativistic large eddy
simulations method (GRLES; \cite{Radice:2017zta}) to explore the
impact of subgrid-scale turbulent angular momentum transport.
The turbulent viscosity is parametrized as
$\sigma_T = \ell_{\rm mix} c_s$, where $c_s$
is the sound speed, and $\ell_{\rm mix}$ is a free parameter
that sets the intensity of the turbulence.  
In the context of accretion disk theory turbulent viscosity is usually expressed 
in terms of a dimensionless constant $\alpha$ related to
$\ell_{\rm mix}$ through the relation $\ell_{\rm mix} = \alpha\, c_s\,
\Omega^{-1}$, where $\Omega$ is the angular velocity of the fluid
\cite{Shakura:1972te}.
Recently, very high resolution general relativistic magnetohydrodynamics simulations of a NS merger 
were reported by \cite{Kiuchi:2017zzg}, who used sufficiently high seed magnetic fields $(10^{15}\, {\rm G})$
to resolve the MRI in the merger remnant. 
Averaged $\alpha$ values for different rest-mass density shells were also provided.
In our models, we combine values of $c_s$ and $\Omega$ directly obtained during the
simulations with their
estimate of $\alpha$ to compute $\ell_{\rm mix}$ as a function of density.

The high temperatures reached during the merger lead to the copious production of
neutrinos of all flavors. Electron (anti)neutrino production rates due 
to electron and positron captures on free protons and neutrons, respectively, 
are characterized by a $T^{5}$ dependence on matter temperature (e.g., 
\cite{Rosswog:2003rv}). 
Thermal processes producing all kinds of neutrinos have an even stronger
dependence,  
$\sim T^7$ (e.g., \cite{Itoh:1996ApJS..102..411I}).
For densities in excess of $\rho_{\rm lim} \sim 10^{12}{\rm g \,cm^{-3}}$ and temperatures
above a few MeV, the neutrino mean free path 
becomes smaller than the typical length scales of the system, and the 
diffusion timescale significantly exceeds the dynamical timescales.
Under these conditions, neutrino trapping occurs and neutrinos form 
a gas thermally and dynamically coupled with the baryonic fluid.
Diffusing neutrinos are eventually emitted at the last scattering 
surface (also called neutrino surface), located at a density close to $\rho_{\rm lim}$,
and propagate further in optically thin conditions.

The compositional and energy changes in the matter due to weak
reactions during the simulations are treated using a leakage scheme 
that tracks reactions involving electron $\nu_e$ and anti-electron
type $\bar{\nu}_e$ neutrinos separately
\cite{Galeazzi:2013mia,Radice:2016dwd}. Heavy-lepton neutrinos are
treated as a single effective species labeled as $\nu_x$.
The reactions included in the simulations are listed in
Tab.~\ref{tab:leakage}; the precise formulas implemented for neutrino
production and absorption rates and for scattering opacities can be found
in the references listed there.
The total neutrino opacities are used to compute
an estimate to the optical depth following the scheme
of \cite{Neilsen:2014hha}. The optical depth and local opacity are 
then employed to calculate the neutrino diffusion timescale, as 
described in \cite{Ruffert:1995fs,Radice:2018pdn}.
The neutrino emission rate from optically 
thick regions is then computed as the ratio between the local neutrino 
densities and the diffusion timescale. For the former quantity, we assume 
that the neutrinos follow Fermi-Dirac distributions with chemical potentials 
obtained assuming beta-equilibrium with thermalized neutrinos as in \cite{Rosswog:2003rv}.
Free-streaming neutrinos are
emitted at an average energy and then evolved according to the M0 scheme introduced in
\cite{Radice:2016dwd,Radice:2018pdn}. 
The M0 scheme is less sophisticated that the frequency-integrated M1
scheme adopted by others \cite{Sekiguchi:2015dma,Foucart:2015vpa}.
However, it is has the advantage of computational efficiency,
it incorporates an approximate treatment of gravitational and Doppler effects,
and is well adapted to the geometry found in NS mergers.
In particular, it is not affected by the well-known radiation shock artifact
that plagues the M1 scheme~\cite{Foucart:2018gis}.

\begin{table}
\caption{Weak reaction rates and references for their implementation.
We use the following notation: $\nu \in \{\nu_e, \bar{\nu}_e, \nu_{x}\}$
denotes a neutrino species, $\nu_{x}$ any heavy-lepton neutrinos, $N \in
\{n, p\}$ denotes a nucleon, and $A$ denotes a nucleus.}
\label{tab:leakage}
\begin{center}
  \begin{tabular}{ll}
\hline\hline
Reaction & Ref. \\ 
\hline
$\nu_e + n \leftrightarrow p + e^-$           & \cite{Bruenn:1985en} \\
$\bar{\nu}_{e} + p \leftrightarrow n + e^+$   & \cite{Bruenn:1985en} \\
$e^+ + e^- \rightarrow \nu + \bar{\nu}$       & \cite{Ruffert:1995fs} \\
$\gamma + \gamma \rightarrow \nu + \bar{\nu}$ & \cite{Ruffert:1995fs} \\
$N + N \rightarrow \nu + \bar{\nu} + N  + N$  & \cite{Burrows:2004vq} \\
$\nu + N \rightarrow \nu + N$                 & \cite{Ruffert:1995fs} \\
$\nu + A \rightarrow \nu + A$                 & \cite{Shapiro:1983du} \\
\hline\hline
\end{tabular}
\end{center}
\end{table}

\subsection{Modelling of trapped neutrinos}
\label{sec:Method_neutrinos}

Despite the fact that trapped neutrino density is employed to compute
emission rates from optically thick regions, the neutrino treatment adopted in
our simulations does not explicitly 
model the presence of a trapped component.
The temperature- and composition-dependent weak equilibrium that
establishes between the baryonic and leptonic fluids for $\rho \gtrsim \rho_{\rm lim}$ 
could potentially change the matter properties at those densities.
For example, we presently neglects the pressure and 
internal energy contributions, as well as possible changes in the electron fraction,
due to the presence of these neutrino gases.

In order to explore this effect and to assess its relevance, we postprocess
the results of our simulations by computing the thermodynamical conditions of the fluid in presence of a
neutrino gas in weak and thermal equilibrium. For consistency with the underlying simulations, we consider 
(anti)neutrino of all flavors modelled as three independent species: $\nu_e$, $\bar{\nu}_{e}$, and $\nu_{x}$.
Since neutral current reactions involving $\nu_x$-$\bar{\nu}_x$ pairs decouple at significantly
larger densities than charged current reactions involving $\nu_e$ and $\bar{\nu}_e$, we distinguish between
$\rho_{\rm lim,e} = 5 \times 10^{11} {\rm g\, cm^{-3}}$ for both $\nu_e$ and $\bar{\nu}_e$, 
and $\rho_{\rm lim,x} = 5 \times 10^{12} {\rm g\, cm^{-3}}$ for $\nu_x$.
\footnote{Note that we do not include the presence of a scattering atmosphere, which could be relevant for 
$\nu_x$ neutrinos.}
We first promote the electron fraction and the internal energy density obtained by the simulations 
and measured in the local comoving frame to the lepton and total (i.e., fluid+neutrinos) internal 
energy density:
\begin{eqnarray}
Y_{\rm e,sim} \rightarrow Y_{l} \nonumber \\
e_{\rm sim} \rightarrow u \nonumber \, .
\end{eqnarray}
Indeed, the leakage scheme prescriptions evolve the lepton fraction and total internal energy, 
rather than the electron fraction and the fluid internal energy (see, e.g., \cite{Perego:2015agy}). 
For a given matter density $\rho$, the electron fraction and the fluid energy obtained in the simulations 
univocally identify a temperature through the EOS, $T_{\rm sim}$. 
We then compute the electron fraction and the temperature
at equilibrium, $(Y_{e,{\rm eq}}, T_{\rm eq})$ , solving the following set of equations:
\begin{eqnarray}
  Y_{l} =  & Y_{e,{ \rm eq}} + Y_{\nu_e}(Y_{e,{\rm eq}}, T_{\rm eq}) - Y_{\bar{\nu}_e}(Y_{\rm e,eq}, T_{\rm eq}) \,  \\
  u         =  & e(Y_{e,{\rm eq}}, T_{\rm eq}) + \frac{\rho}{m_{\rm b}} \left[ Z_{\nu_e}(Y_{e,{\rm eq}}, T_{\rm eq})+ \right. \nonumber \\
               & + \left. Z_{\bar{\nu}_e}(Y_{e,{\rm eq}}, T_{\rm eq}) + 4 Z_{{\nu}_x}(T_{\rm eq}) \right] \, \\
  0  = & \eta_{\nu_e}(Y_{e,{\rm eq}}, T_{\rm eq}) - \eta_{e}(Y_{e,{\rm eq}}, T_{\rm eq}) +  \nonumber \\
       &   - \eta_{\rm p}(Y_{e,{\rm eq}}, T_{\rm eq}) + \eta_{\rm n}(Y_{e,{\rm eq}}, T_{\rm eq}) \, .
\end{eqnarray}

In the previous equations, $Y_{\nu_i}$ and $Z_{\nu_i}$ represent the neutrino particle and energy fractions, respectively, while
$\eta_i=\mu_i/T$ is the degeneracy parameter for the $i$ species. They are computed assuming weak and thermal equilibrium between all
the modelled species. The former is ensured 
by the equilibrium relationship between the relativistic chemical potentials $\mu$, 
i.e. $\mu_{\nu_e} = \mu_{e} + \mu_{p} - \mu_{n}$. The
latter is given by the assumption that the neutrino and matter temperature
coincide. In particular, we compute
\begin{eqnarray}
\label{eq: neutrino fractions}
 Y_{\nu_i}(\rho,Y_e,T) & = & \frac{4 \pi m_{\rm b}}{\rho (hc)^3} \left( {k_{\rm B}T} \right)^3 F_2(\eta_{\nu_i}) \exp \left(-\frac{\rho_{\rm lim,i}}{\rho} \right) \\
\label{eq: neutrino energy fractions} 
 Z_{\nu_i}(\rho,Y_e,T) & = & \frac{4 \pi m_{\rm b}}{\rho (hc)^3} \left( {k_{\rm B}T} \right)^4 F_3(\eta_{\nu_i}) \exp \left(-\frac{\rho_{\rm lim,i}}{\rho} \right)
\end{eqnarray}
where $F_k(x)$ is the Fermi function of order $k$, and we further assume that $\mu_{\bar{\nu}_e} = - \mu_{\nu_e}$ and $\mu_{\nu_x} = 0$.
The density dependent exponential term ensures that trapped neutrinos are only present for $\rho \gtrsim \rho_{\rm lim}$ (see e.g. \cite{Kaplan:2013wra}).
Trapped neutrinos do provide a pressure as an ultra-relativistic gas, i.e. $P_{\nu} = (\rho Z_{\nu})/(3 m_{\rm b})$. 
We thus compare the pressure obtained in the simulation, 
\begin{equation}
P_{\rm sim} \equiv P_{\rm fluid}(\rho, Y_{e,{\rm sim}},T_{\rm sim}) \, , 
\end{equation}
with the total pressure at equilibrium,
\begin{eqnarray}
P_{\rm eq} & \equiv & P_{\rm fluid}(\rho, Y_{e,{\rm eq}},T_{\rm eq}) + \nonumber \\
             &   & + \frac{\rho}{3} \left( Z_{\nu_e }+Z_{\bar{\nu}_e}+4Z_{\nu_x} \right) (Y_{e,{\rm eq}},T_{\rm eq}) \, .
\end{eqnarray}

\section{Thermodynamical evolution}
\label{sec:histograms}

\begin{figure*}[t]
  \centering 
    \includegraphics[width=0.9\textwidth]{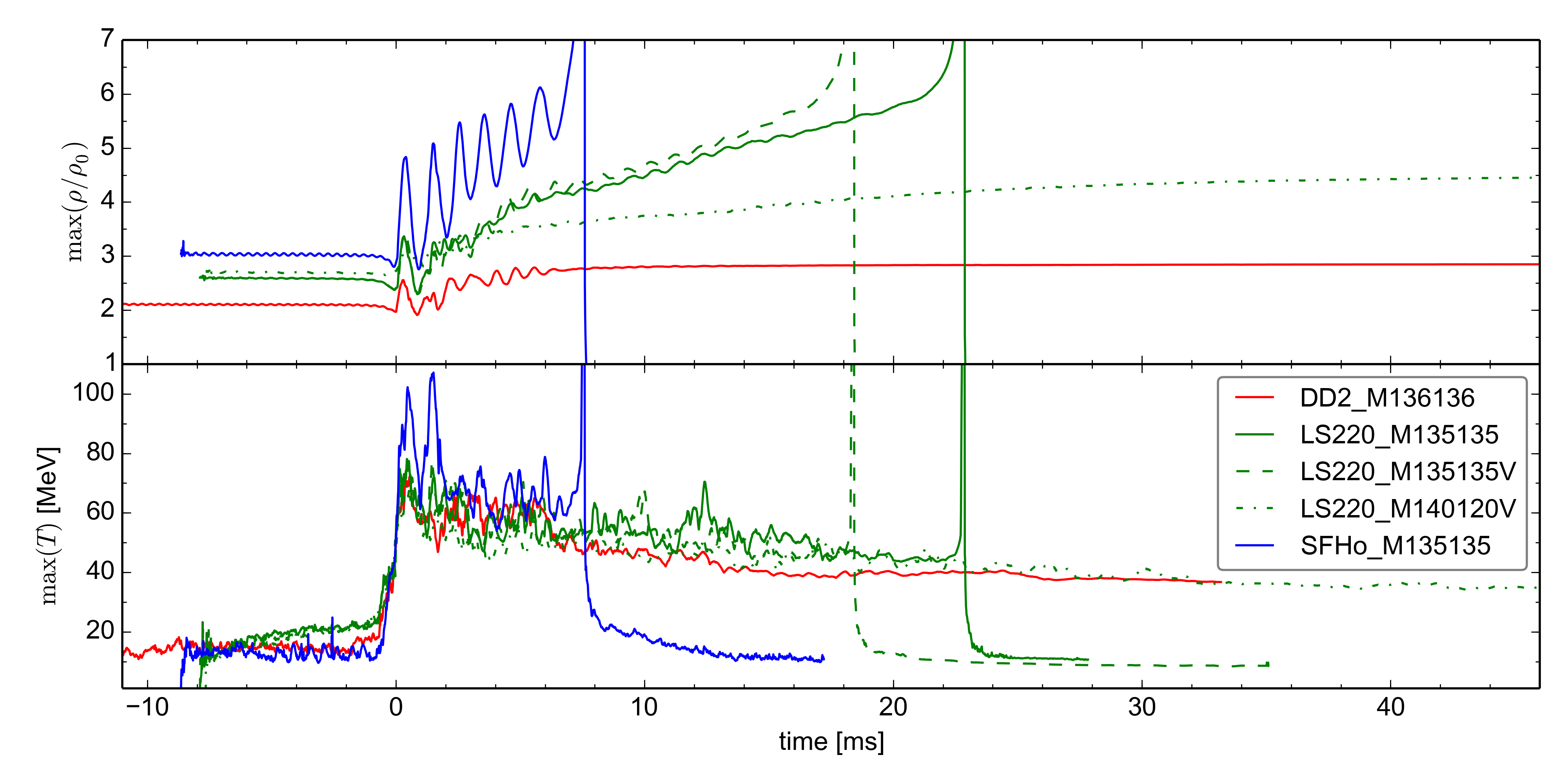}
    \caption{Evolution of the maximum rest-mass density (top) and of
      the maximum temperature (bottom) for all the binaries
      considered in this work. Peak temperatures correspond the first
      bounce of the NS cores, and are reached at typical densities
      $\rho\sim\rho_0$. Collapse to black hole happens for
      LS220\_M135135 and SFHo\_M135135; during collapse the maximum
      density reaches $(\max\rho) \gtrsim5\rho_0$.}
    \label{fig:rhoT}
\end{figure*}

We now discuss results for the four fiducial binary neutron star mergers 
reported in Tab.~\ref{tab:bns}.
The binaries have masses and EOS
compatible with GW170817 \cite{TheLIGOScientific:2017qsa} and, correspondingly, the maximal densities in
each NS are $2-3$ times the nuclear saturation point,
$\rho_{\rm NS_{1,2}}\sim2-3\rho_0$, with softer EOS having larger maximum densities.

The maximum density and temperature evolutions for each model are 
reported in Fig.~\ref{fig:rhoT}.
The moment $t=0$ corresponds to the peak of the gravitational-wave
amplitude (i.e. the end of the chirp signal) and it is conventionally
refereed as {\it moment of merger} (or shortly {\it merger}, where it
cannot be confused with the merger-phase of the coalescence) \cite{Bernuzzi:2012ci}.
During the last orbits ($t<0$) the maximum density and temperature are
approximately constant, the latter being 
$T\sim10$~MeV. These high temperature are reached close to the surface of the stars 
and are a numerical artifact.

At $t\sim0$ the two NS cores merge, the maximum density increases
rapidly up to $\sim1.5\rho_\text{NS}$ and temperature increases due to
(physical) hydrodynamical shocks and matter compression. 
In the considered binaries, no prompt collapse happens and the two NS cores 
bounce multiple times on a timescale of few milliseconds. During the 
first of these bounces, the temperature peaks at $70-100$~MeV and 
most of the dynamical ejecta is expelled \cite{Hotokezaka:2012ze,Bauswein:2013yna,Radice:2018pdn}.
The core bounces are more violent in binaries with softer EOS, and produce larger
density and temperature oscillations. Note that binary remnants are
closer to the gravitational collapse instability point for softer EOSs
\cite{Shibata:1999wm,Stergioulas:2011gd,Bernuzzi:2013rza}.
Hot material from the collisional interface between
the NSs is squeezed out of the remnant and settles into an accretion disk \cite{Radice:2018pdn}.
The remnant is still deformed into a bar ($m=2$ mode) and spiral arms are launched inside the disk.
On a timescale of
$10-20$~ms, the characteristic timescale of the GW transient, 
binaries with softer EOS collapse to a rotating black hole
with typical dimensionless Kerr parameter of $\sim0.7-0.8$ \cite{Bernuzzi:2015opx,Zappa:2017xba}.
Binaries with stiffer EOS generate long-lived NS remnants whose
dynamics is determined on timescales much longer than those presently
simulated in numerical relativity \cite{Radice:2018xqa}.

\begin{figure*}[ht]
  \noindent\makebox[\textwidth]{%
    \includegraphics[width=.33\textwidth]{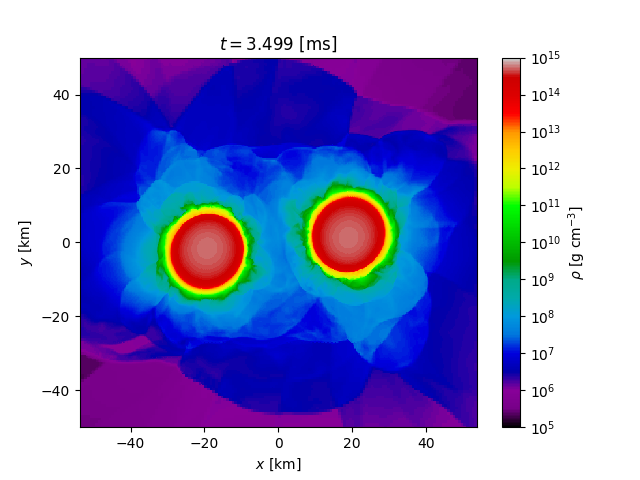}
    \includegraphics[width=.33\textwidth]{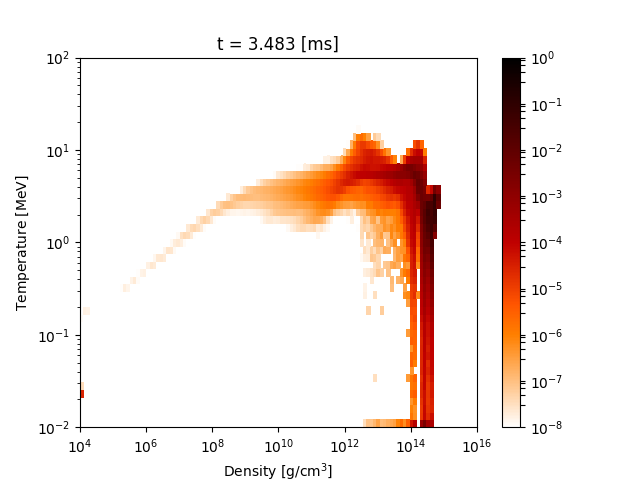}
    \includegraphics[width=.33\textwidth]{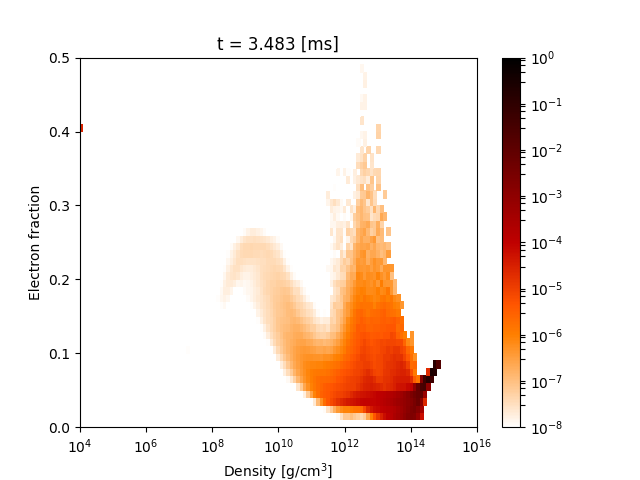}
  }\\
  \noindent\makebox[\textwidth]{%
  \includegraphics[width=.33\textwidth]{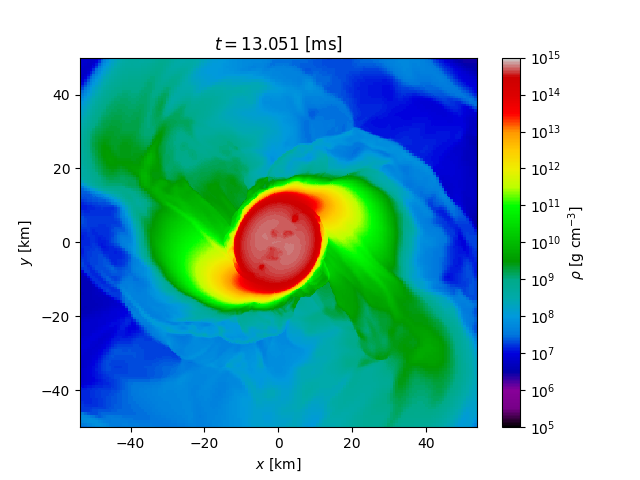}
  \includegraphics[width=.33\textwidth]{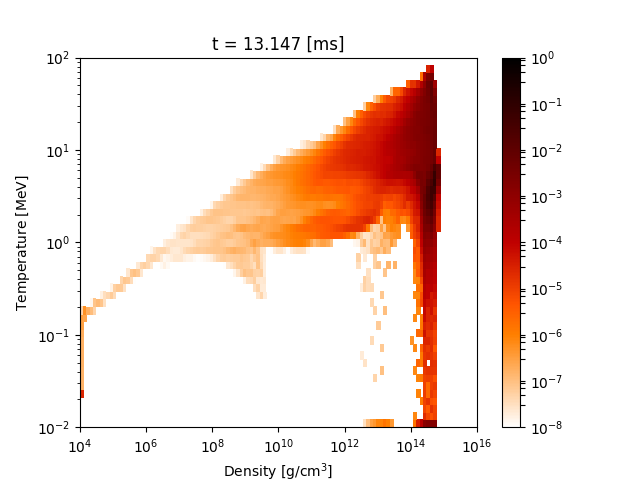}
  \includegraphics[width=.33\textwidth]{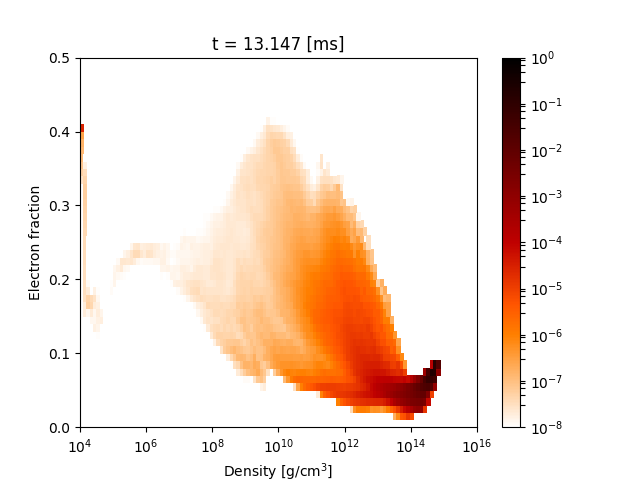}
  }\\
  \noindent\makebox[\textwidth]{%
  \includegraphics[width=.33\textwidth]{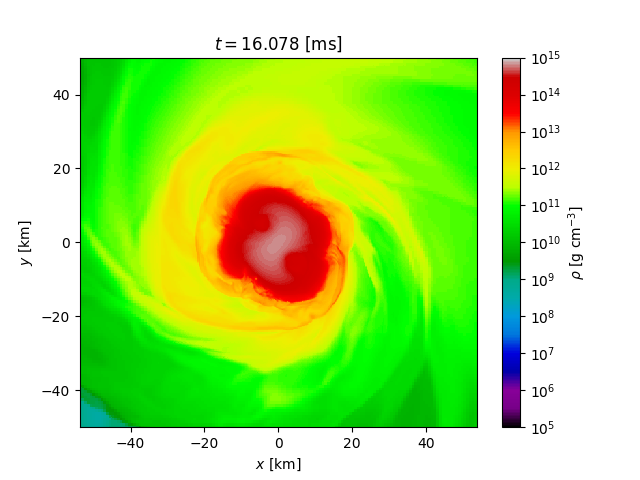}
  \includegraphics[width=.33\textwidth]{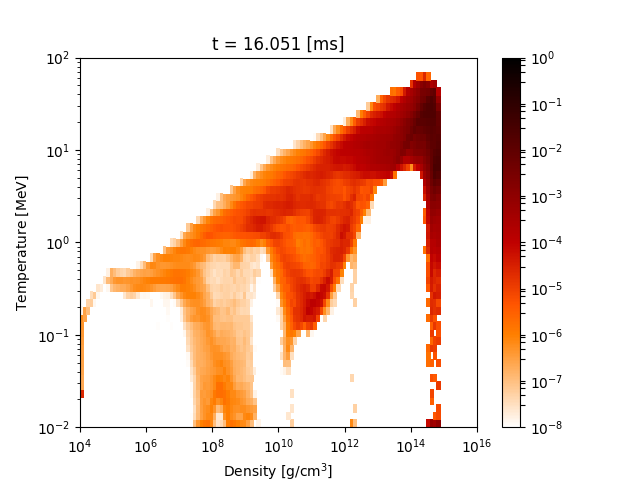}
  \includegraphics[width=.33\textwidth]{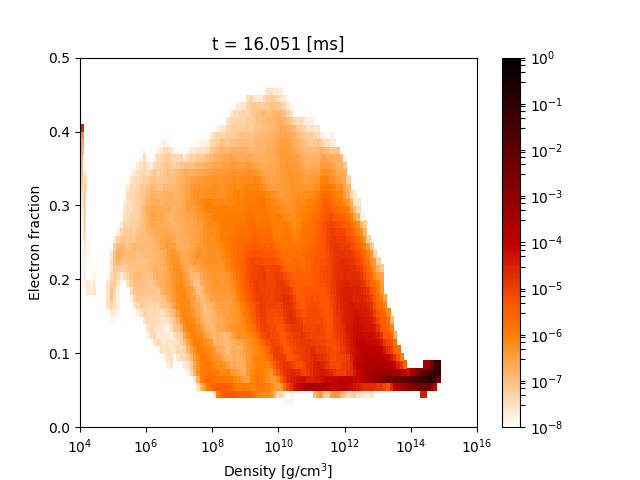}
  }\\
  \noindent\makebox[\textwidth]{%
  \includegraphics[width=.33\textwidth]{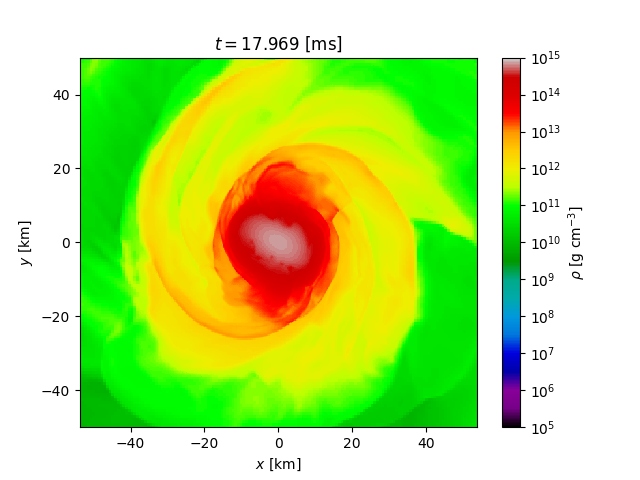}
  \includegraphics[width=.33\textwidth]{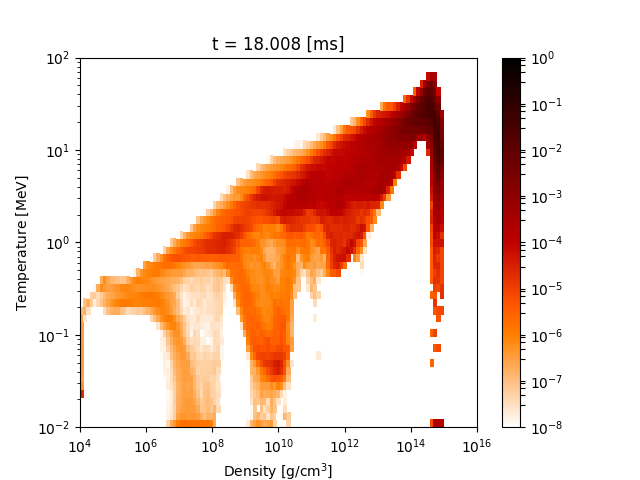}
  \includegraphics[width=.33\textwidth]{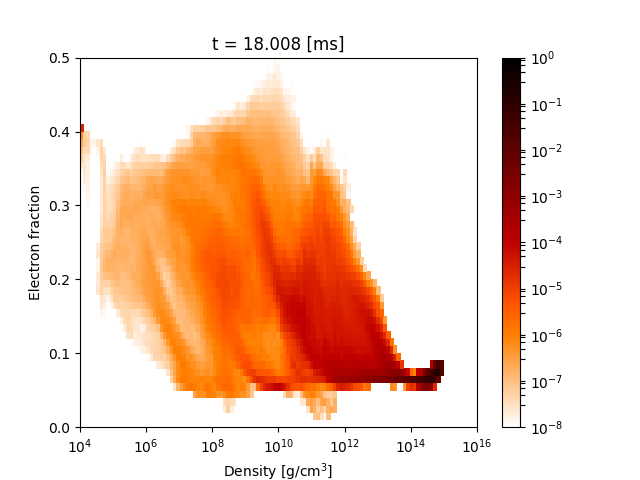}
  }\\
  \noindent\makebox[\textwidth]{%
    \includegraphics[width=.33\textwidth]{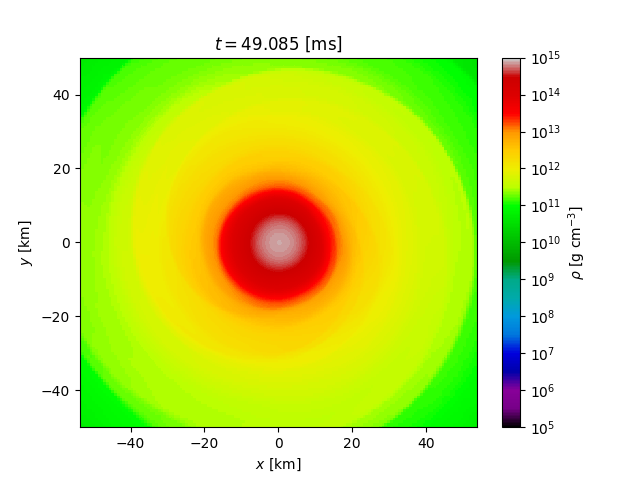}
  \includegraphics[width=.33\textwidth]{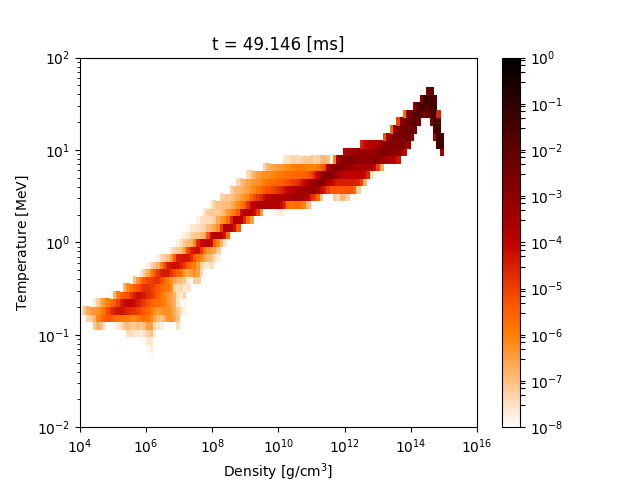}
  \includegraphics[width=.33\textwidth]{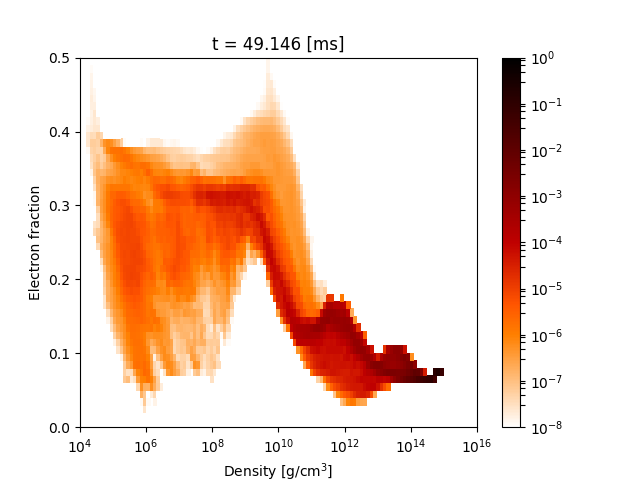}
  }\\
  \caption{Rest-mass density on the orbital plane (left), and corresponding histograms on the density-temperature (middle)
    and density-electron fraction (right) planes for the simulation DD2\_M136136. The five rows correspond to the different phases of the merger; 
    moving from the top to the bottom: inspiral phase, time of the temperature peak, 3-4 ms after the temperature peak,
    time close to the massive NS collapse (if present), end of the simulation. 
    The time reported above each panel refers to the time from the beginning of the simulation in ms. 
    Note 2D snapshot do not exactly correspond to same times.}
  \label{fig:histo_DD2}
\end{figure*}

\begin{figure*}[t]
  %
  \noindent\makebox[\textwidth]{%
    \includegraphics[width=.33\textwidth]{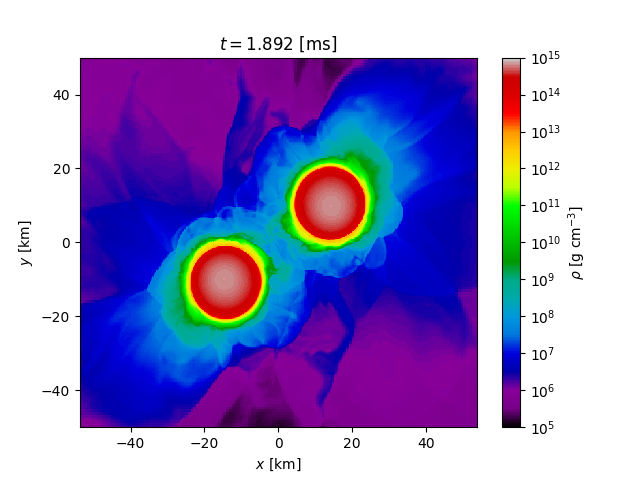}
    \includegraphics[width=.33\textwidth]{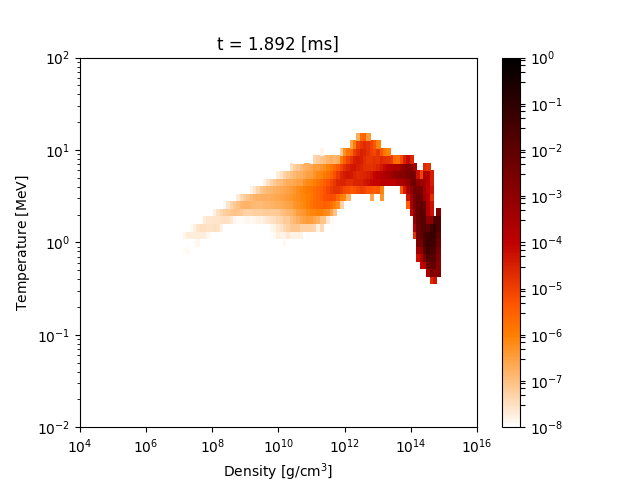}
    \includegraphics[width=.33\textwidth]{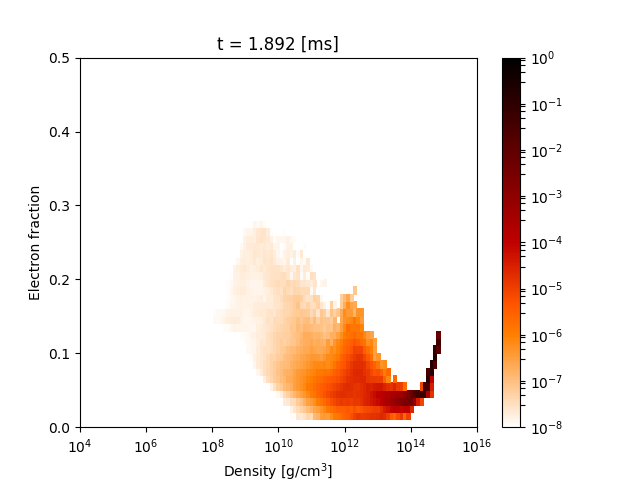}
  }\\
  \noindent\makebox[\textwidth]{%
  \includegraphics[width=.33\textwidth]{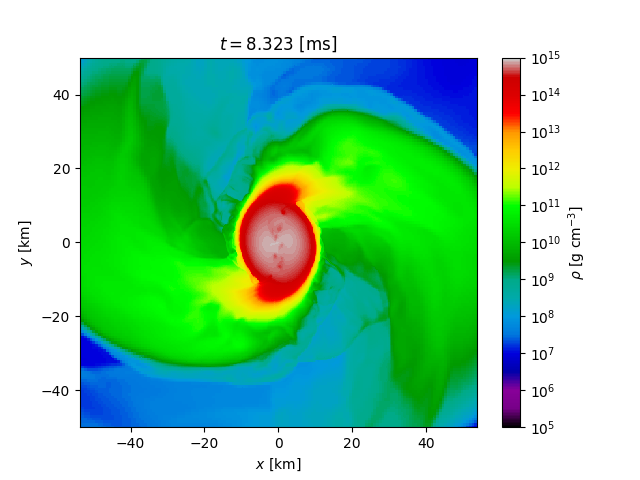}
  \includegraphics[width=.33\textwidth]{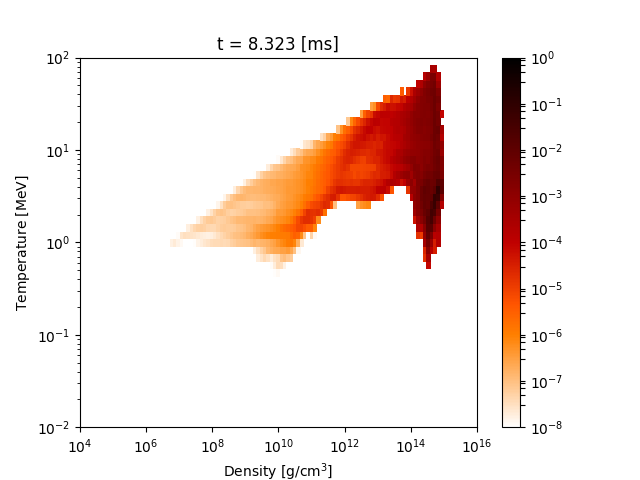}
  \includegraphics[width=.33\textwidth]{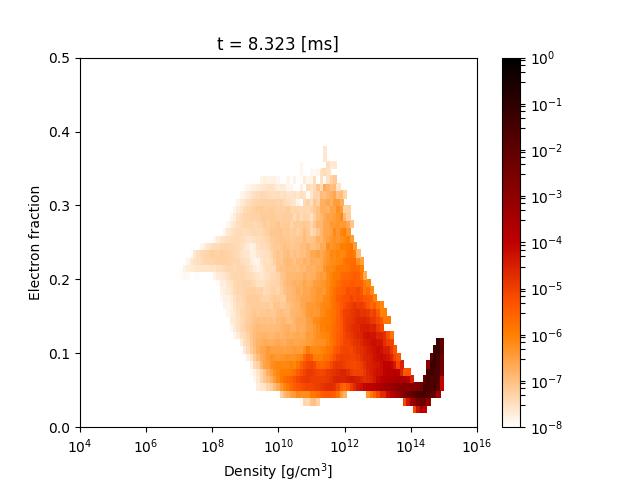}
  }\\
  \noindent\makebox[\textwidth]{%
  \includegraphics[width=.33\textwidth]{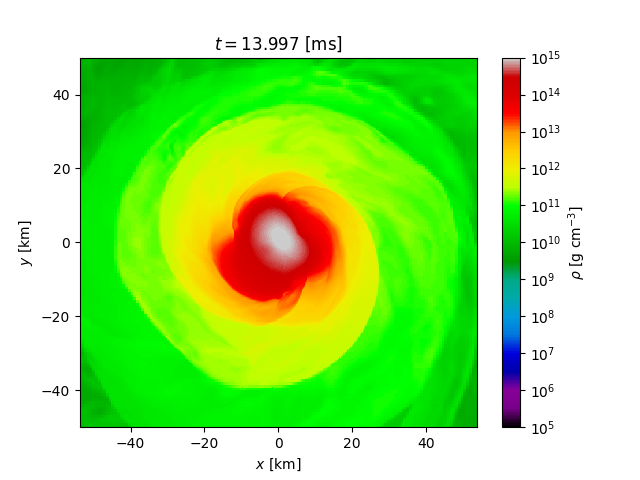}
  \includegraphics[width=.33\textwidth]{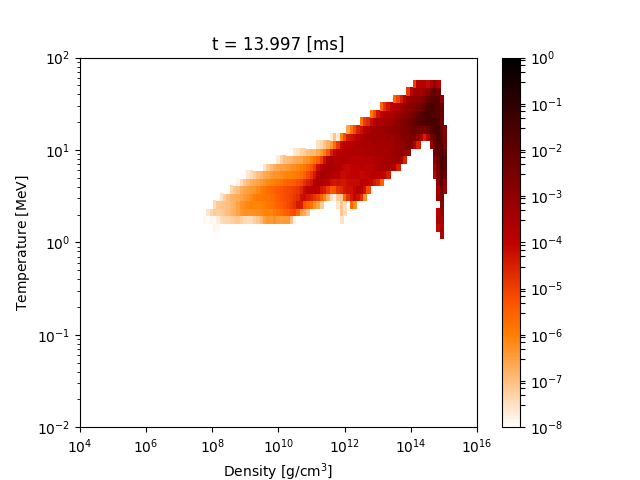}
  \includegraphics[width=.33\textwidth]{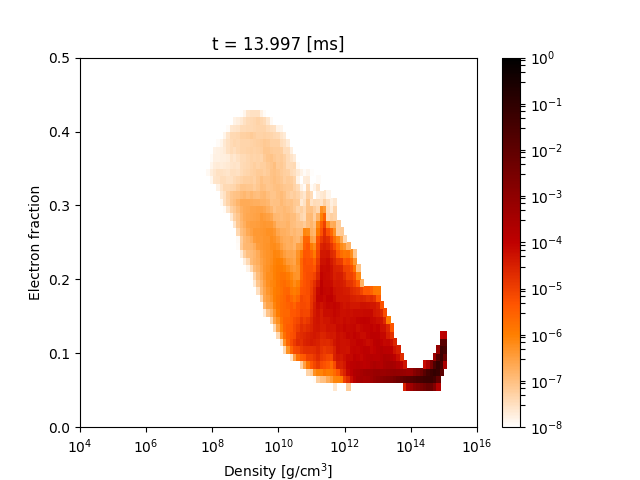}
  }\\
  \noindent\makebox[\textwidth]{%
  \includegraphics[width=.33\textwidth]{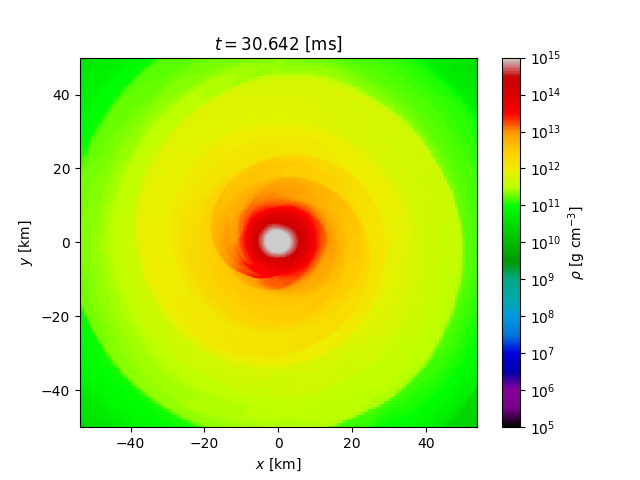}
  \includegraphics[width=.33\textwidth]{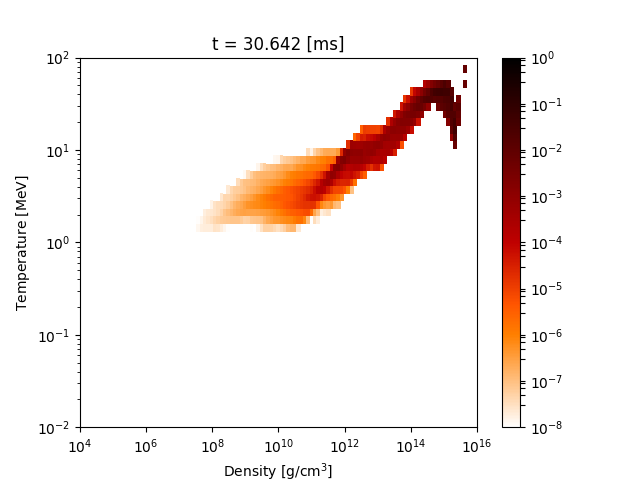}
  \includegraphics[width=.33\textwidth]{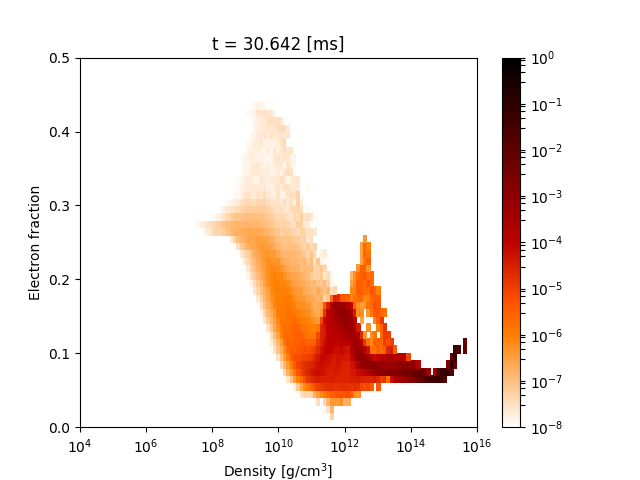}
  }\\
  \noindent\makebox[\textwidth]{%
  \includegraphics[width=.33\textwidth]{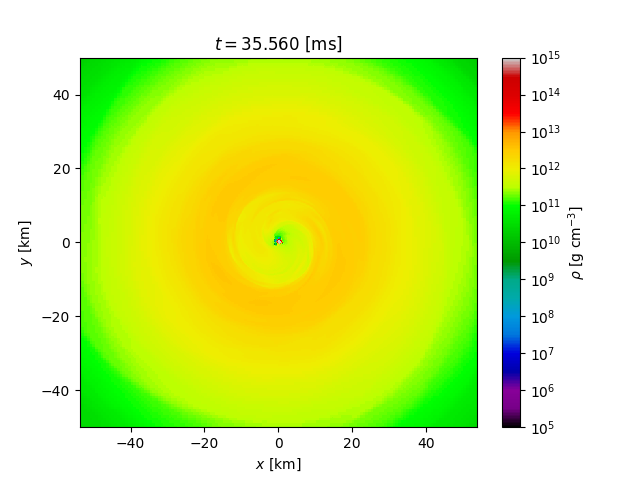}
  \includegraphics[width=.33\textwidth]{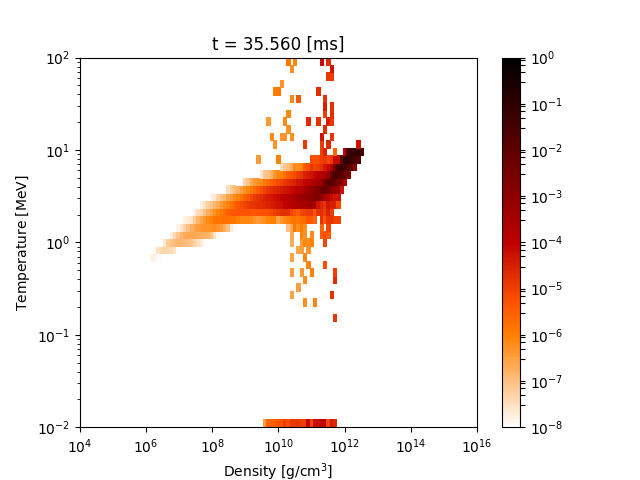}
  \includegraphics[width=.33\textwidth]{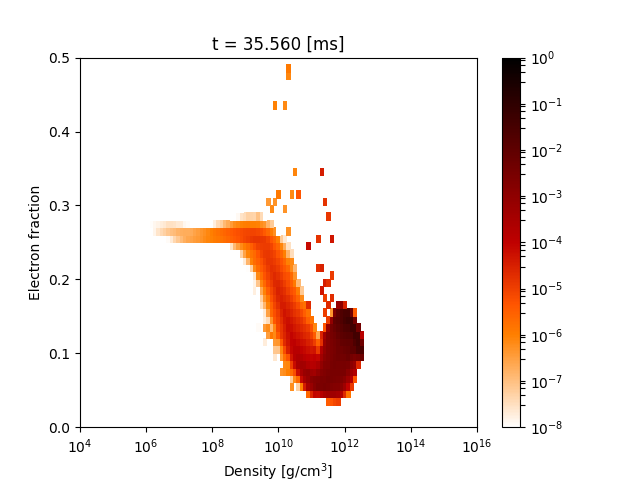}
  }\\
  \caption{Same as in Fig.~\ref{fig:histo_DD2}, but for simulation LS220\_M135135.}
  \label{fig:histo_LS135135}
\end{figure*}

\begin{figure*}[t]
  %
  \noindent\makebox[\textwidth]{%
    \includegraphics[width=.33\textwidth]{logrho_cgs_xy_r4_040960.png}
    \includegraphics[width=.33\textwidth]{hist_dens_temp_000040960.png}
    \includegraphics[width=.33\textwidth]{hist_dens_ye_000040960.png}
  }\\
  \noindent\makebox[\textwidth]{%
  \includegraphics[width=.33\textwidth]{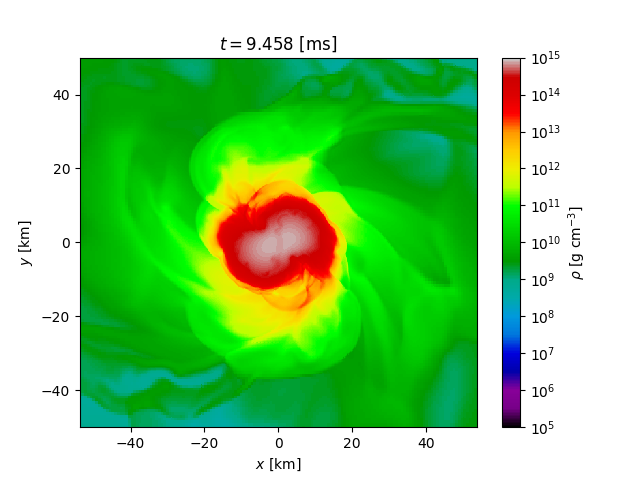}
  \includegraphics[width=.33\textwidth]{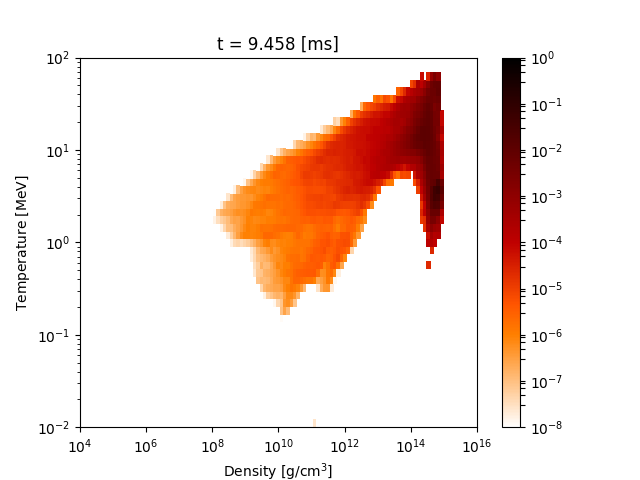}
  \includegraphics[width=.33\textwidth]{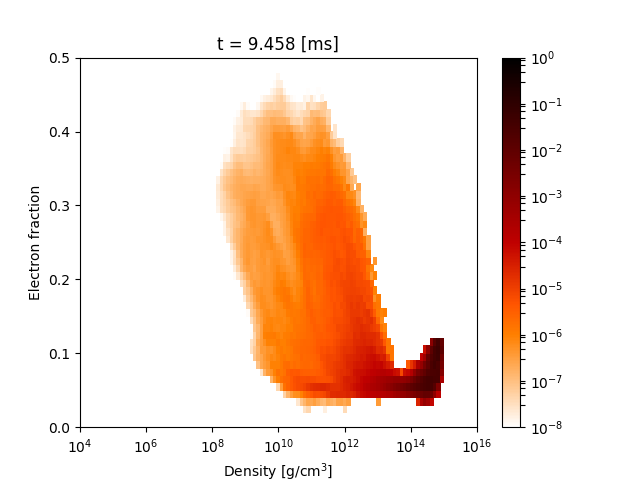}
  }\\
  \noindent\makebox[\textwidth]{%
  \includegraphics[width=.33\textwidth]{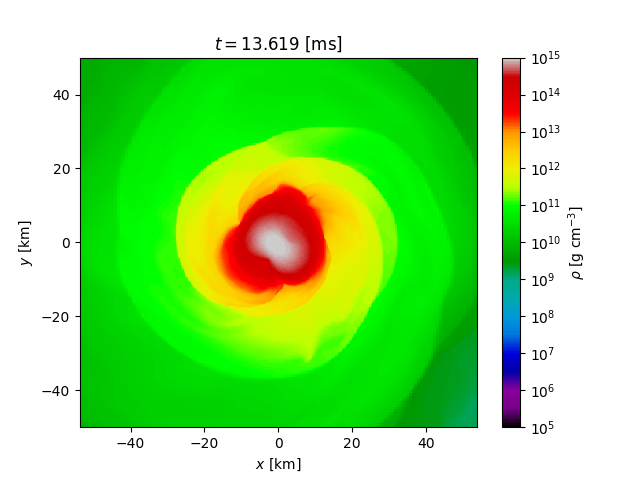}
  \includegraphics[width=.33\textwidth]{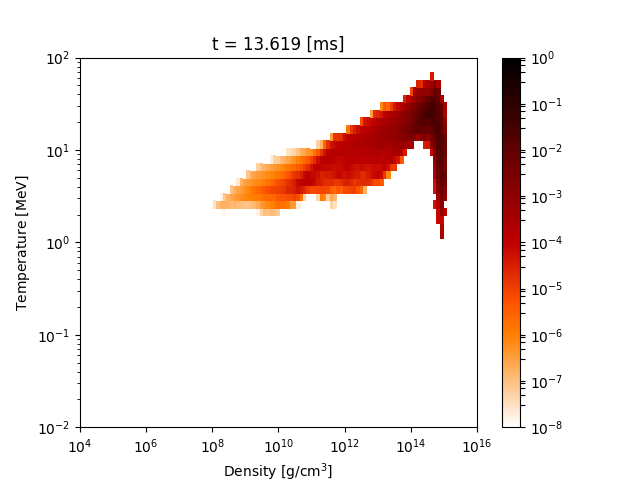}
  \includegraphics[width=.33\textwidth]{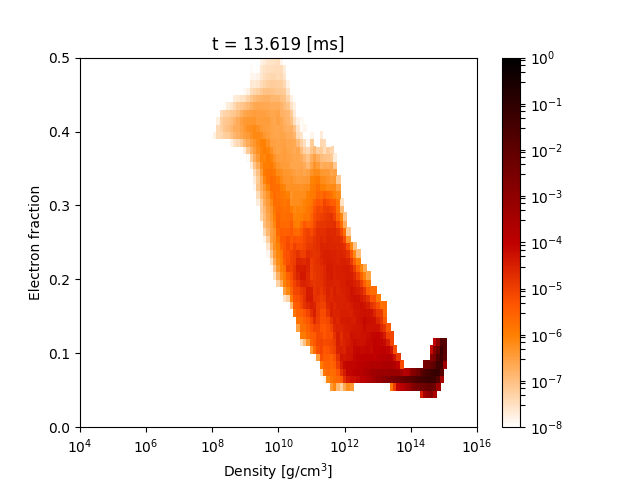}
  }\\
  \noindent\makebox[\textwidth]{%
  \includegraphics[width=.33\textwidth]{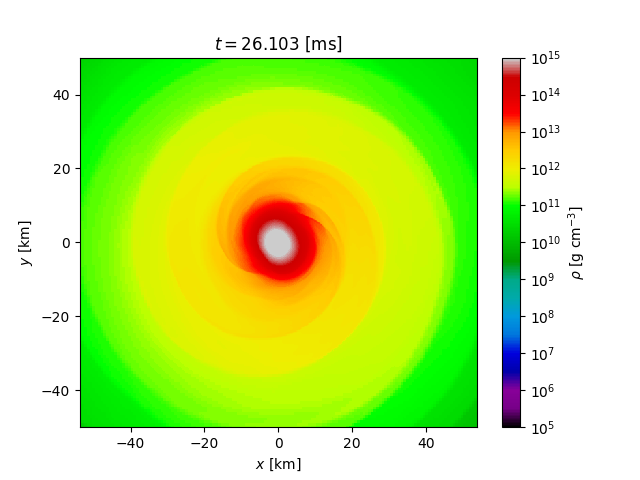}
  \includegraphics[width=.33\textwidth]{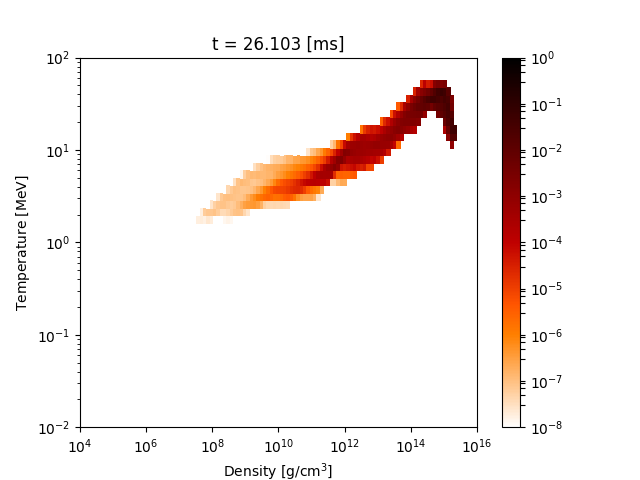}
  \includegraphics[width=.33\textwidth]{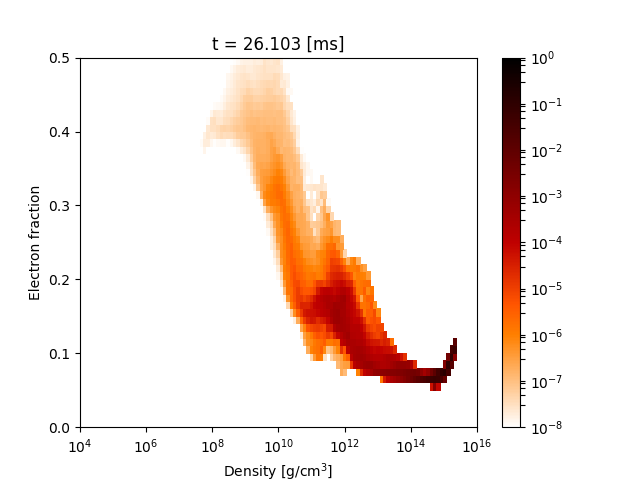}
  }\\
  \noindent\makebox[\textwidth]{%
  \includegraphics[width=.33\textwidth]{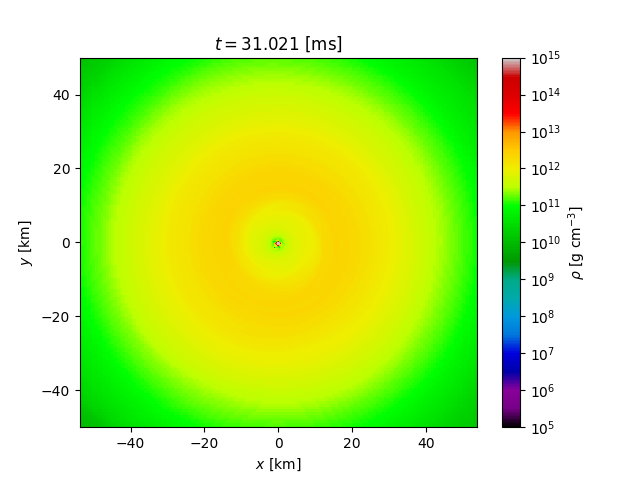}
  \includegraphics[width=.33\textwidth]{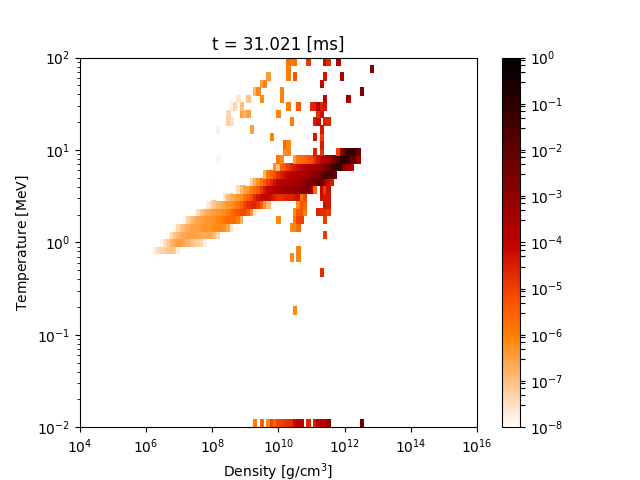}
  \includegraphics[width=.33\textwidth]{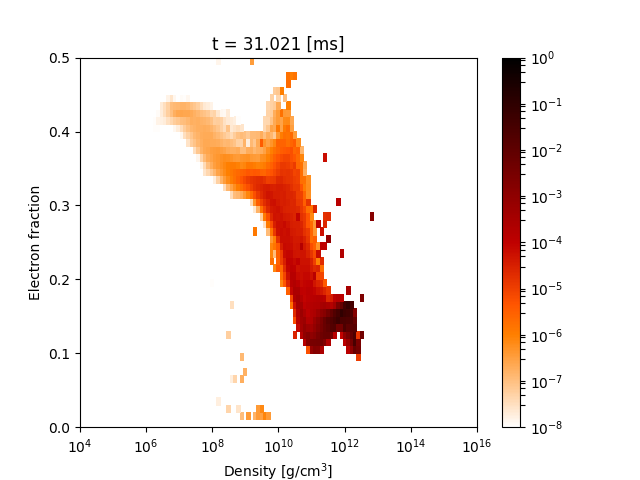}
  }\\
  \caption{Same as in Fig.~\ref{fig:histo_DD2}, but for simulation LS220\_M135135V.}  
  \label{fig:histo_LS135135_VIS}
\end{figure*}

\begin{figure*}[t]
  %
  \noindent\makebox[\textwidth]{%
    \includegraphics[width=.33\textwidth]{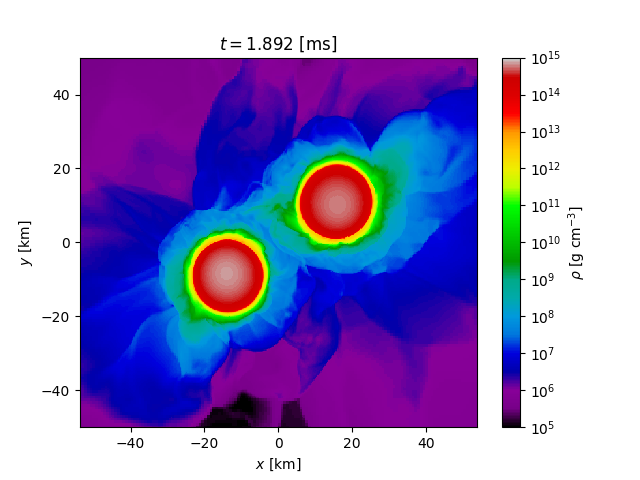}
    \includegraphics[width=.33\textwidth]{hist_dens_temp_000040960.png}
    \includegraphics[width=.33\textwidth]{hist_dens_ye_000040960.png}
  }\\
  \noindent\makebox[\textwidth]{%
  \includegraphics[width=.33\textwidth]{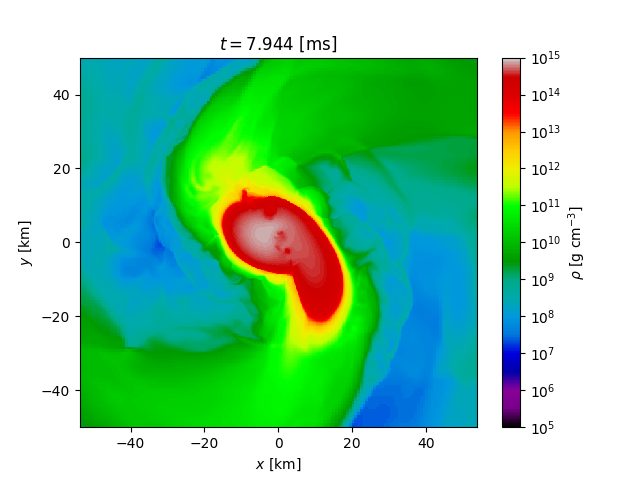}
  \includegraphics[width=.33\textwidth]{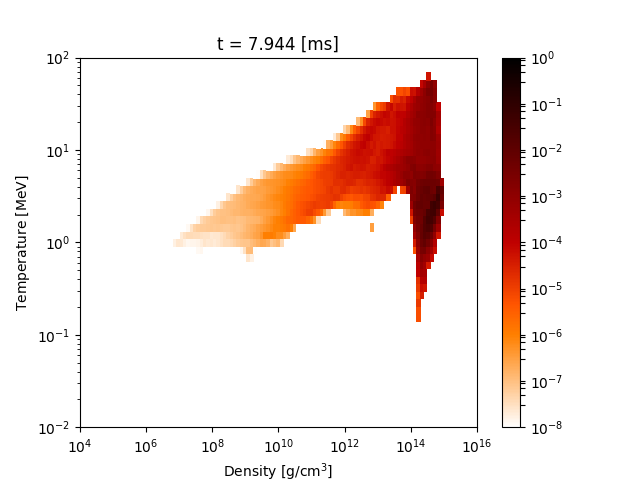}
  \includegraphics[width=.33\textwidth]{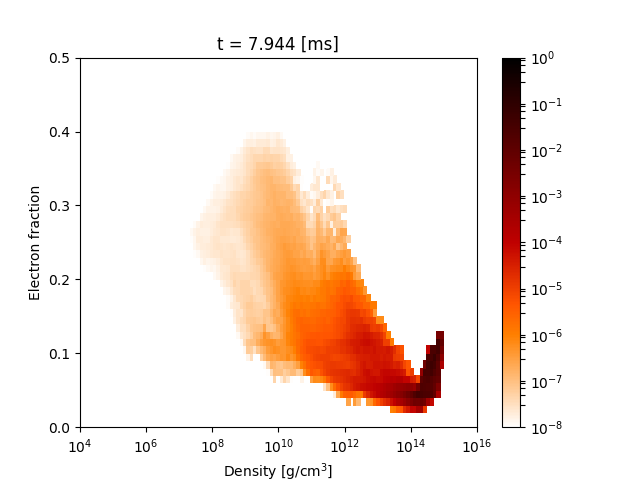}
  }\\
  \noindent\makebox[\textwidth]{%
  \includegraphics[width=.33\textwidth]{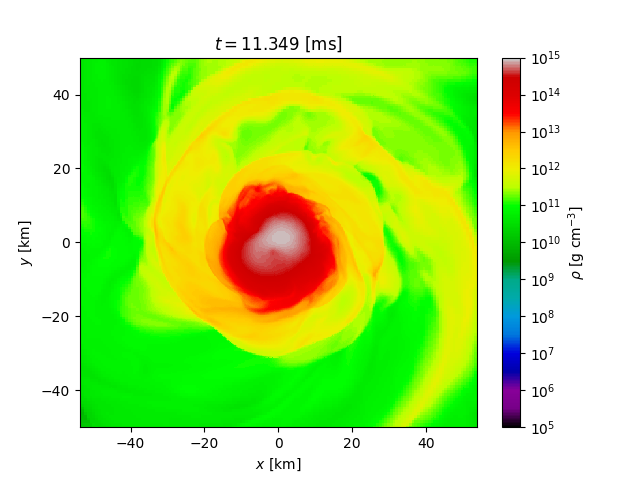}
  \includegraphics[width=.33\textwidth]{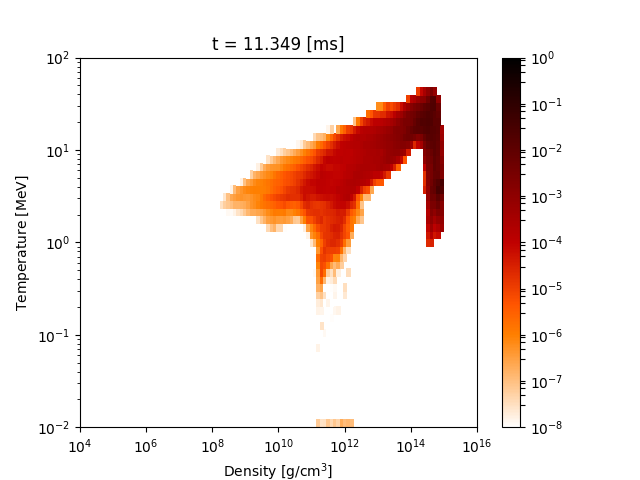}
  \includegraphics[width=.33\textwidth]{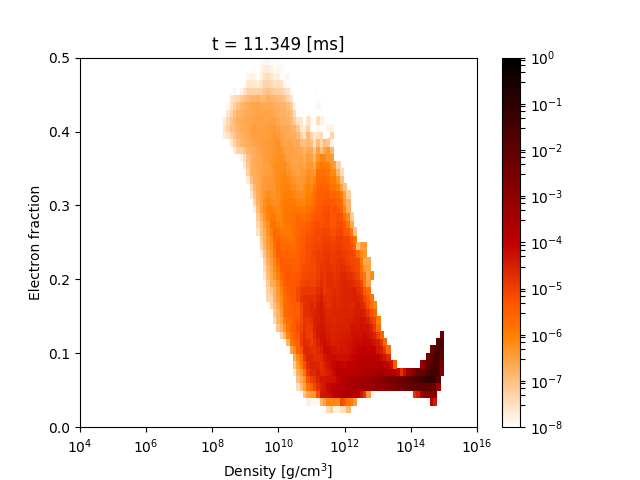}
  }\\
  \noindent\makebox[\textwidth]{%
  \includegraphics[width=.33\textwidth]{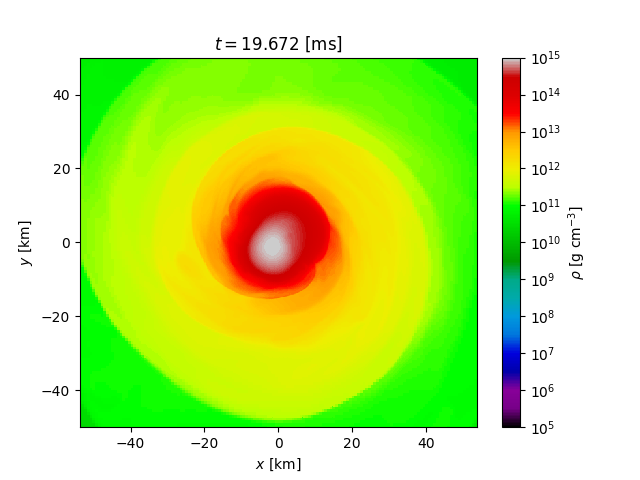}
  \includegraphics[width=.33\textwidth]{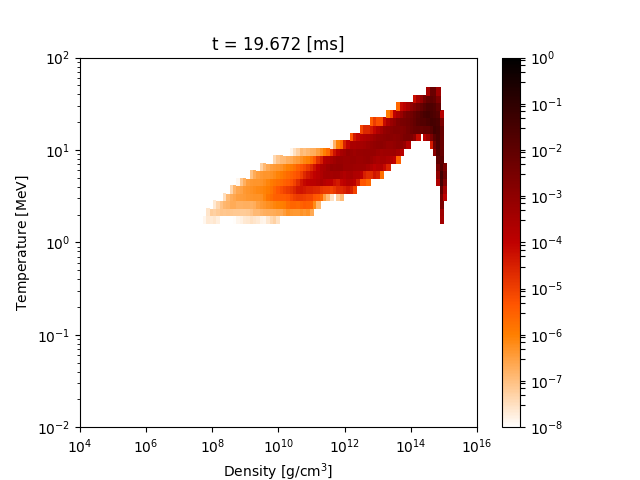}
  \includegraphics[width=.33\textwidth]{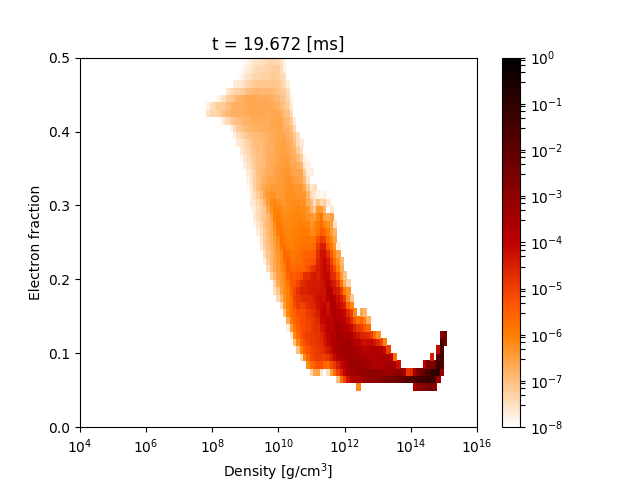}
  }\\
  \noindent\makebox[\textwidth]{%
  \includegraphics[width=.33\textwidth]{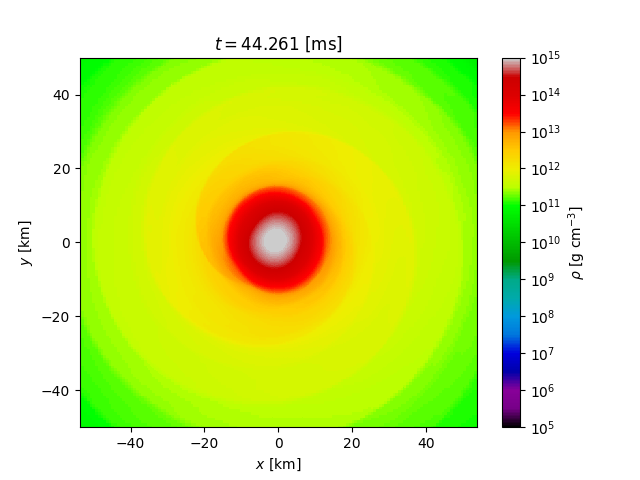}
  \includegraphics[width=.33\textwidth]{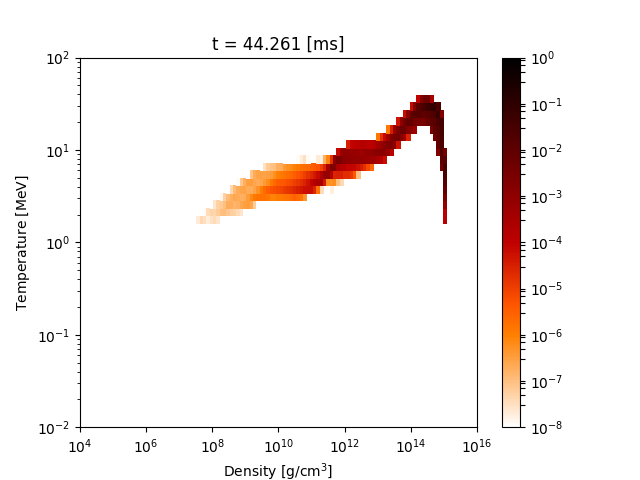}
  \includegraphics[width=.33\textwidth]{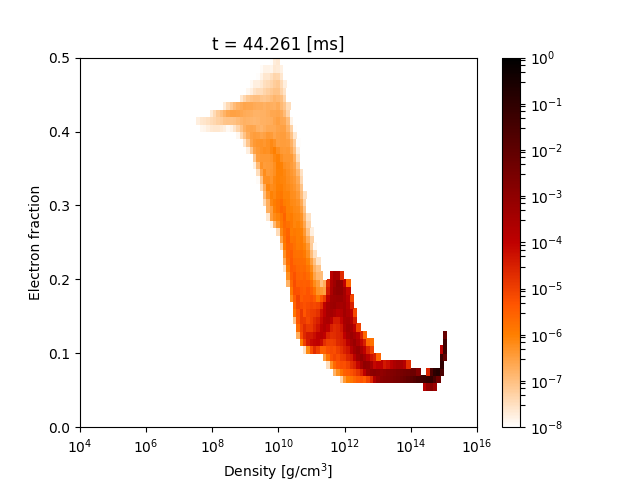}
  }\\
  \caption{Same as in Fig.~\ref{fig:histo_DD2}, but for simulation LS220\_M140120V.}    
  \label{fig:histo_LS140120_VIS}
\end{figure*}

\begin{figure*}[t]
  %
  \noindent\makebox[\textwidth]{%
    \includegraphics[width=.33\textwidth]{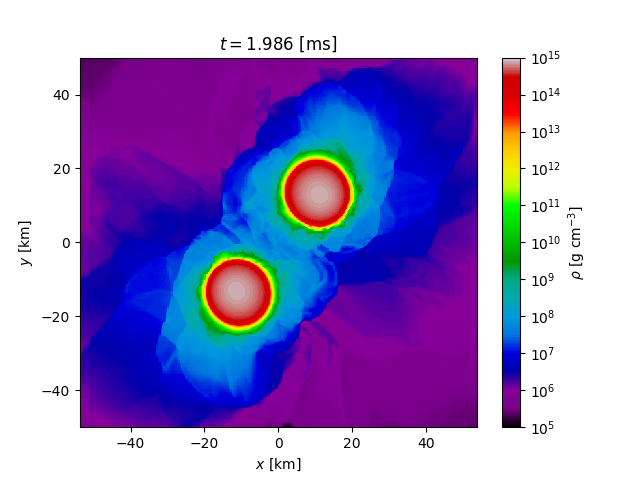}
    \includegraphics[width=.33\textwidth]{hist_dens_temp_000040960.png}
    \includegraphics[width=.33\textwidth]{hist_dens_ye_000040960.png}
  }\\
  \noindent\makebox[\textwidth]{%
  \includegraphics[width=.33\textwidth]{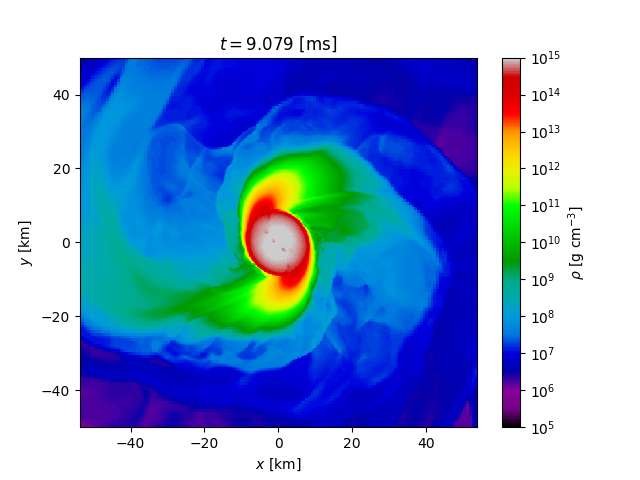}
  \includegraphics[width=.33\textwidth]{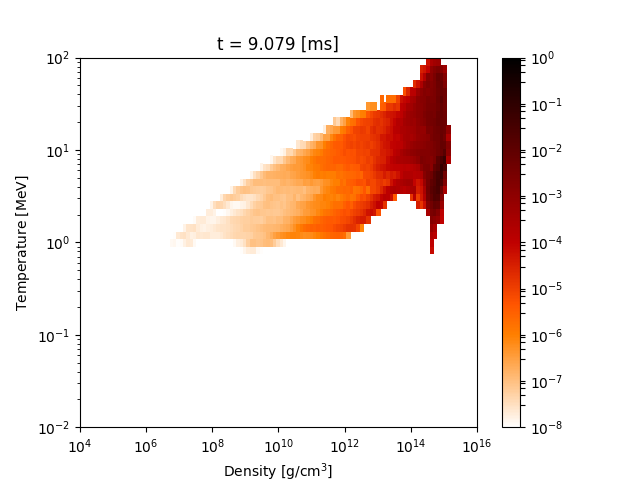}
  \includegraphics[width=.33\textwidth]{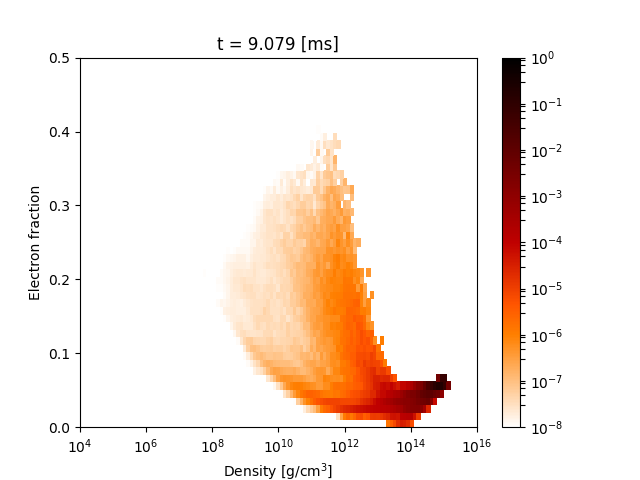}
  }\\
  \noindent\makebox[\textwidth]{%
  \includegraphics[width=.33\textwidth]{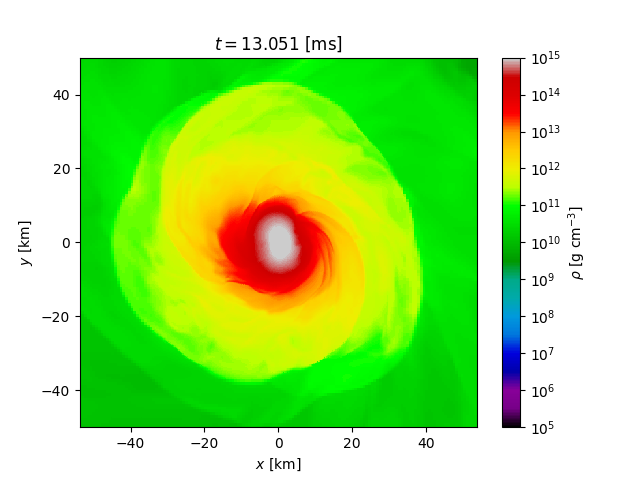}
  \includegraphics[width=.33\textwidth]{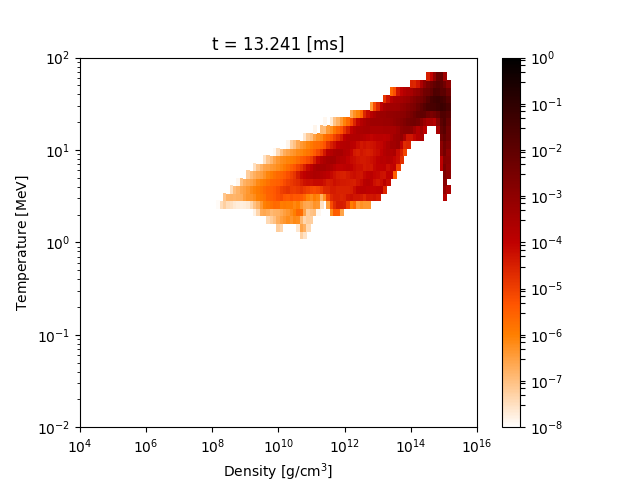}
  \includegraphics[width=.33\textwidth]{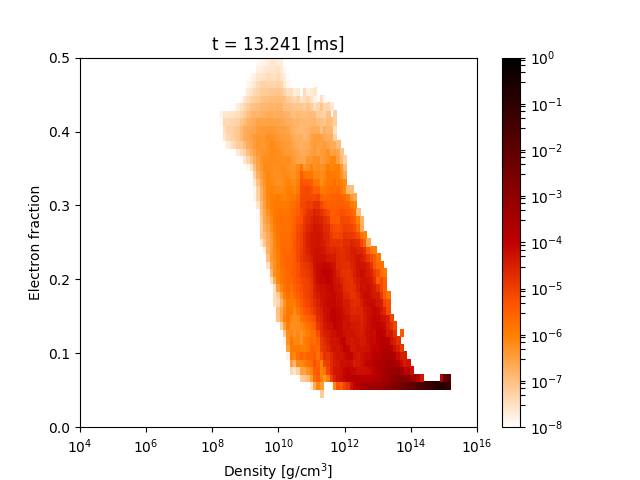}
  }\\
  \noindent\makebox[\textwidth]{%
  \includegraphics[width=.33\textwidth]{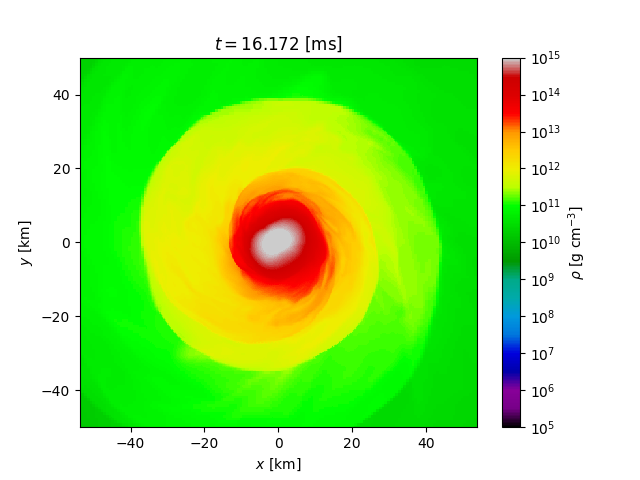}
  \includegraphics[width=.33\textwidth]{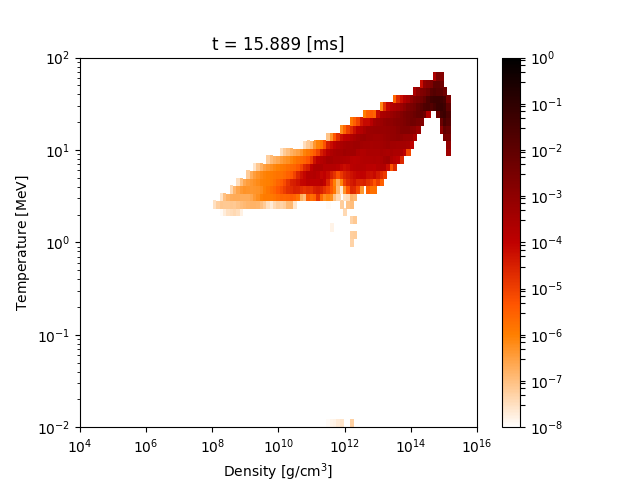}
  \includegraphics[width=.33\textwidth]{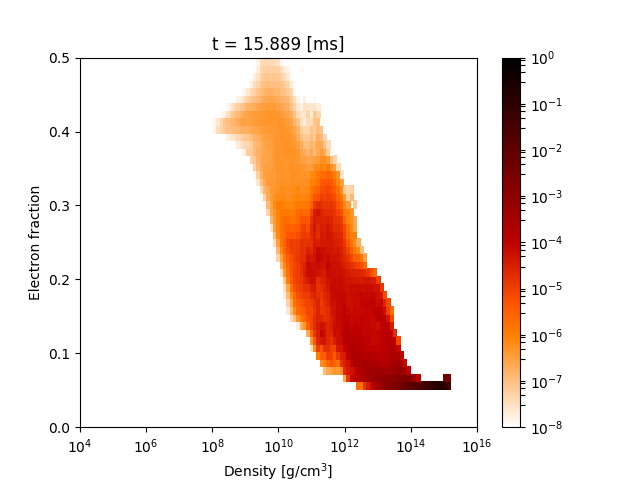}
  }\\
  \noindent\makebox[\textwidth]{%
  \includegraphics[width=.33\textwidth]{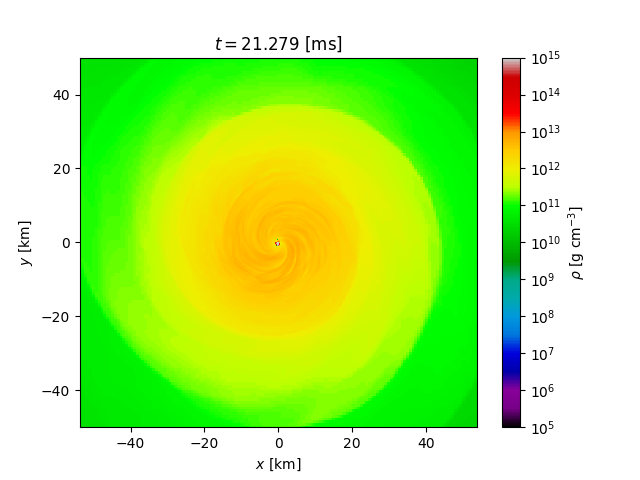}
  \includegraphics[width=.33\textwidth]{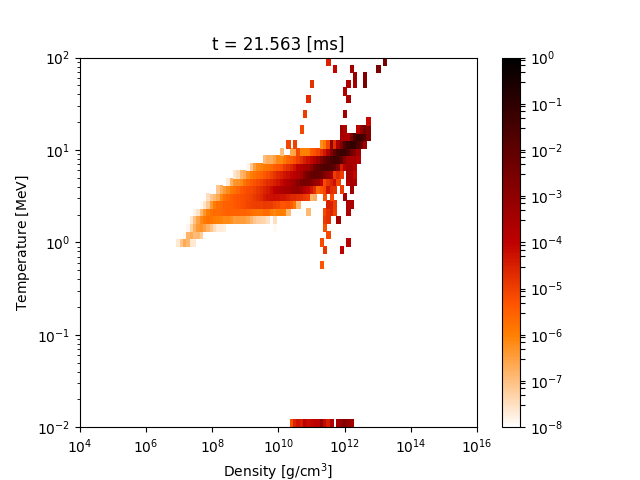}
  \includegraphics[width=.33\textwidth]{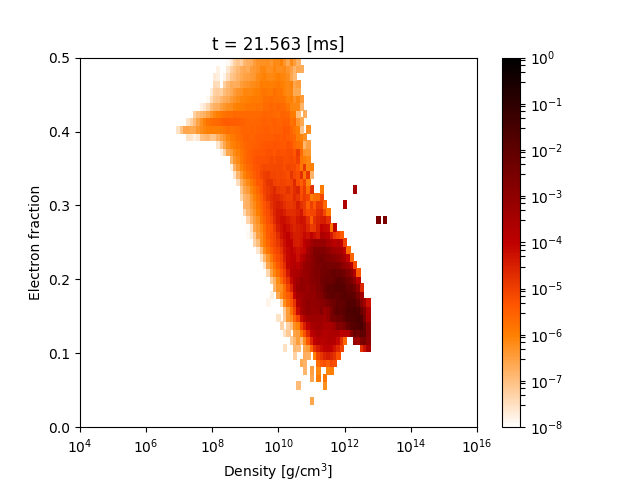}
  }\\
  \caption{Same as in Fig.~\ref{fig:histo_DD2}, but for simulation SFHo\_M135135.}
  \label{fig:histo_SFHo}
\end{figure*}

\begin{figure*}[t]
  %
  \noindent\makebox[\textwidth]{%
    \includegraphics[width=.35\textwidth]{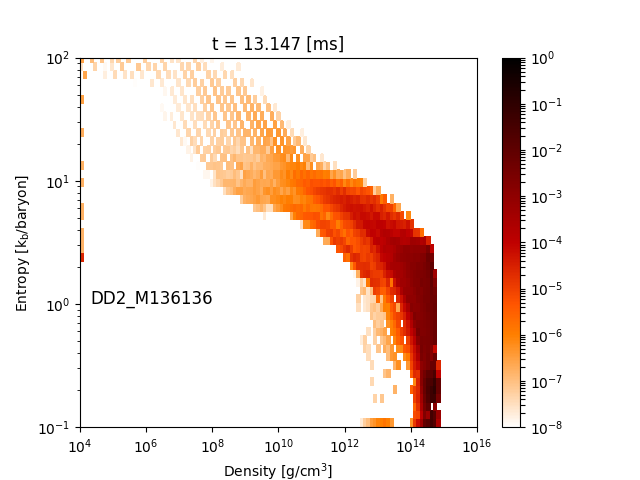}  
    \includegraphics[width=.35\textwidth]{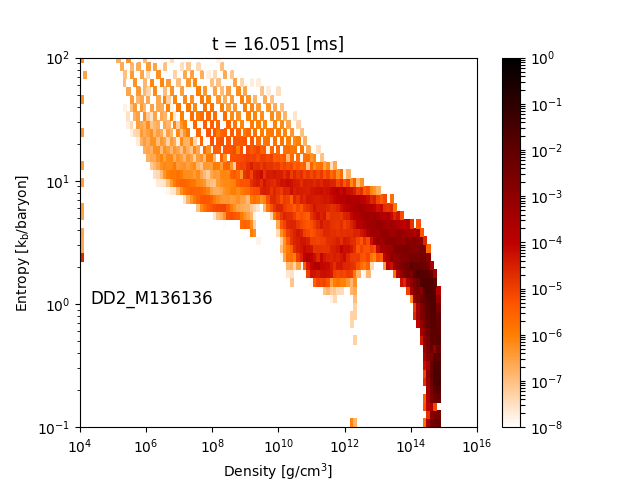}
    \includegraphics[width=.35\textwidth]{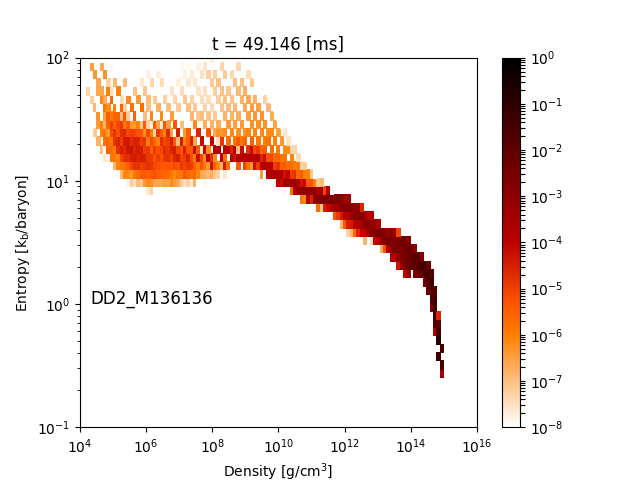}   
  }\\
  \noindent\makebox[\textwidth]{%
    \includegraphics[width=.35\textwidth]{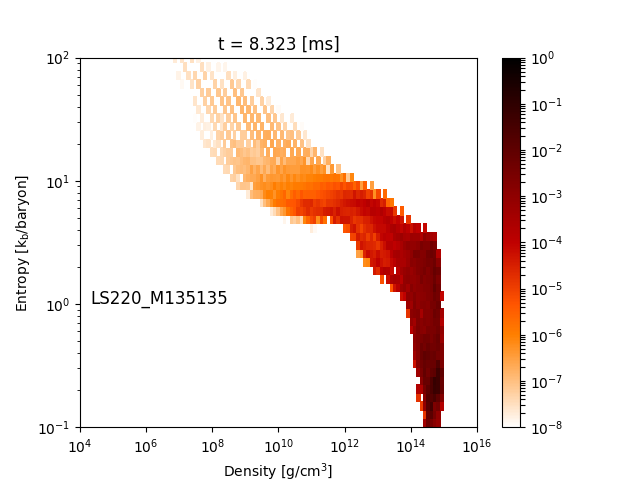}
    \includegraphics[width=.35\textwidth]{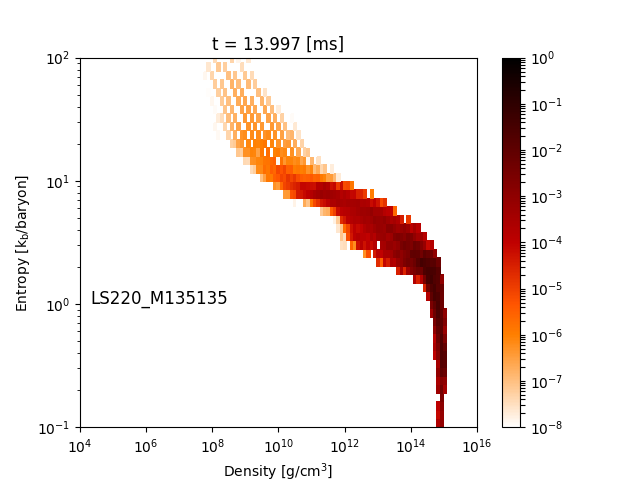}
    \includegraphics[width=.35\textwidth]{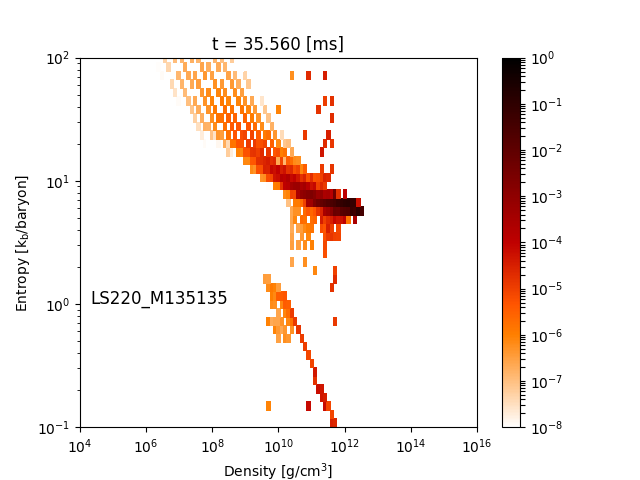}   
  }\\  
  \noindent\makebox[\textwidth]{%
    \includegraphics[width=.35\textwidth]{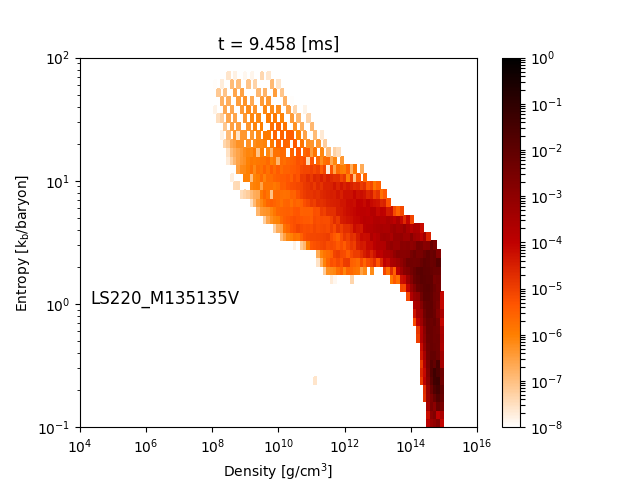}  
    \includegraphics[width=.35\textwidth]{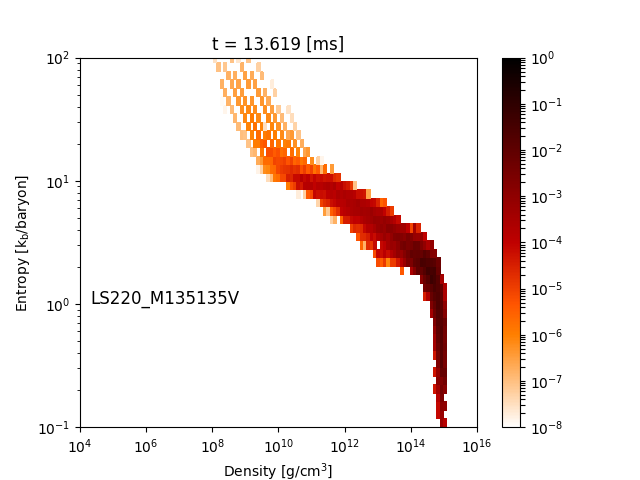}
    \includegraphics[width=.35\textwidth]{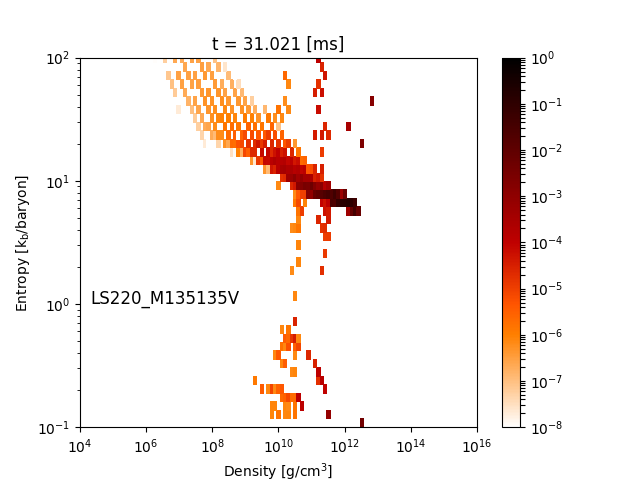}   
  }\\  
  \noindent\makebox[\textwidth]{%
    \includegraphics[width=.35\textwidth]{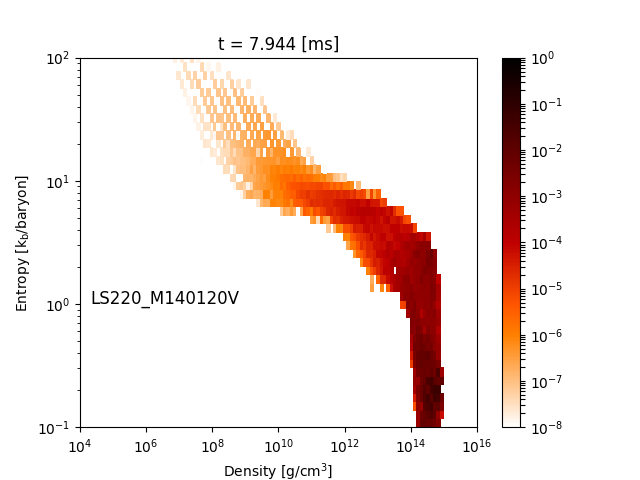}  
    \includegraphics[width=.35\textwidth]{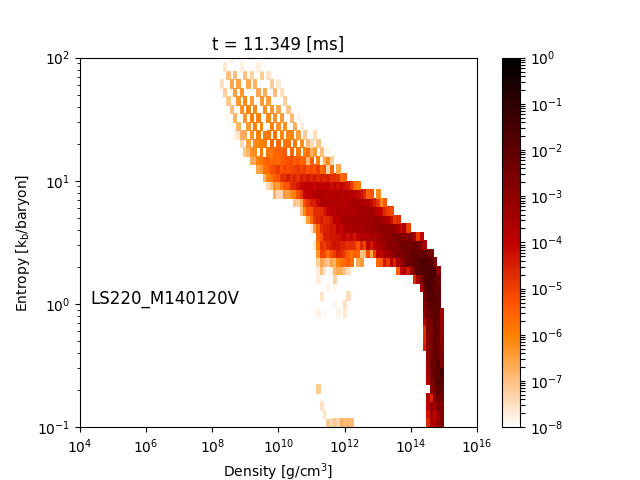}
    \includegraphics[width=.35\textwidth]{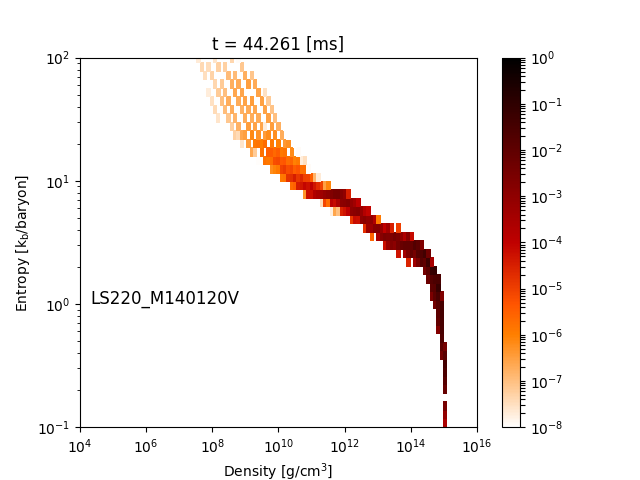}   
  }\\    
  \noindent\makebox[\textwidth]{%
    \includegraphics[width=.35\textwidth]{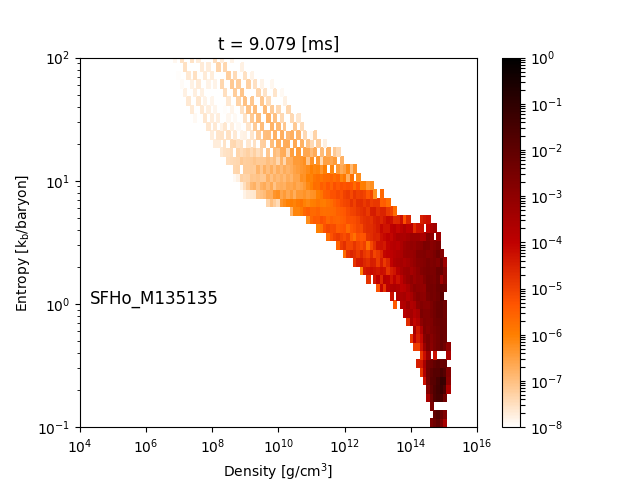}  
    \includegraphics[width=.35\textwidth]{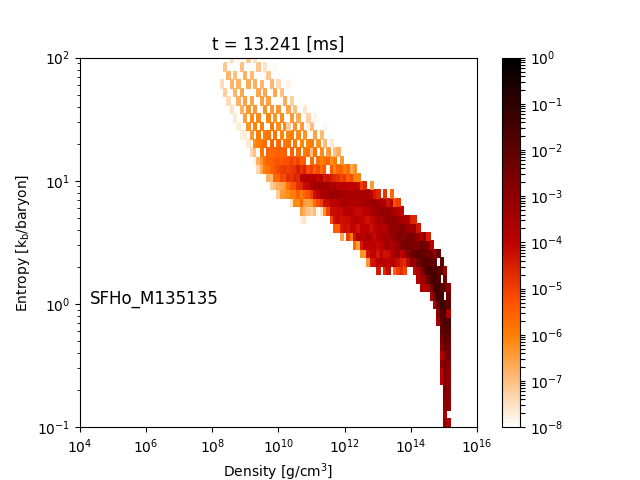}
    \includegraphics[width=.35\textwidth]{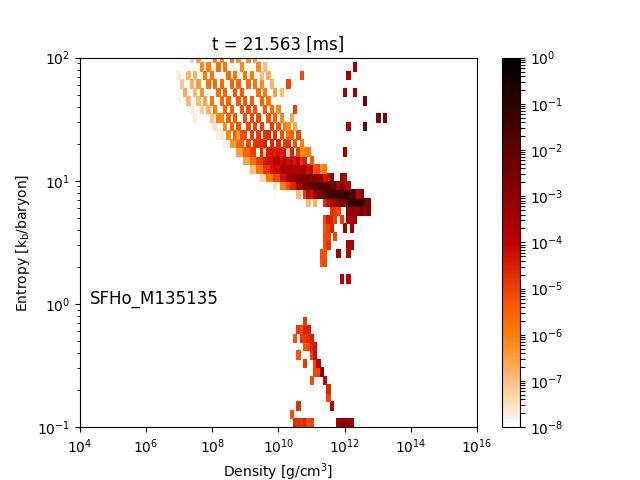}   
  }\\      
  \caption{Histograms of the baryonic mass in the density-entropy plane from the second (left), third
    (middle) and fifth (right) snapshots for all the five reference
    simulations reported in Table~\ref{tab:bns} and in Figs.~\ref{fig:histo_DD2}-\ref{fig:histo_SFHo}.
    The unshocked cores of the NSs retain their initial low entropy, while shocked decompressed matter 
    expanding towards lower densities experience a significant entropy increase.}
  \label{fig:histo_entropy}
\end{figure*}

\begin{figure}
  \includegraphics[width=0.49\textwidth]{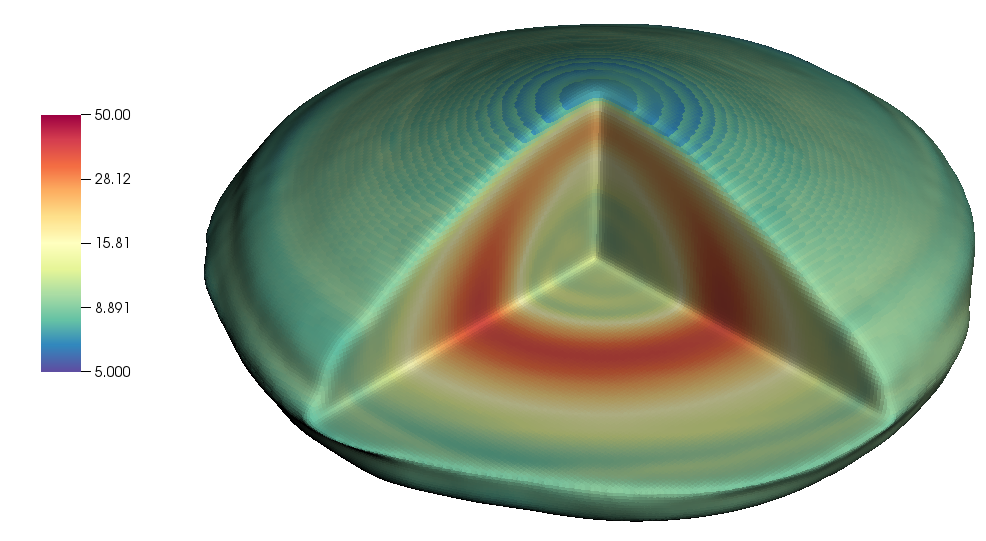}
  \caption{Temperature distribution (MeV) in the massive NS produced by the
    DD2\_M1361364 binary at $t = 34.2 {\rm ms}$ after the merger. The
    surface of the massive NS is defined as the point where the rest mass
    density $\rho$ exceeds $10^{13} {\rm g}\ {\rm cm}^{-3}$. Material
    at lower densities is made transparent in this visualization. Massive NSs
    produced in binary mergers are rotationally compressed along the
    rotation axis (vertically in the figure). High temperatures are
    confined to a spherical annulus at intermediate densities
    ${\sim}10^{14} {\rm g}\ {\rm cm}^{-3}$ formed of material
    originally at the collisional interface between the NSs.} 
  \label{fig:DD2.M13641364.temp}
\end{figure}

Let us discuss the thermodynamical conditions
experienced by the matter during the different phases of the merger. 
For the moment, we do not include the effect of trapped neutrinos
at $\rho > \rho_{\rm lim}$,
but we simply consider the outcome of our simulations.
Under the assumption that matter is always in Nuclear Statistical
Equilibrium (NSE), 
we focus on matter density, $\rho$, temperature, $T$, and 
electron fraction, $Y_{e}$. We consider 3D spatial hypersurfaces 
at specific coordinate times and produce histograms of the baryonic-mass 
as distributed at different $\rho$, $T$, and $Y_{e}$, 
i.e. $M_{\rm b}(\rho,T,Y_{e})$. We used the following ranges:
\begin{align}
\rho &\in [10^4,10^{16}]~\gccm \nonumber\\
T &\in[10^{-2},10^2]~{\rm MeV} \nonumber\\ 
Y_e &\in [0,0.5] \nonumber\ ,  
\end{align}
binning them in $120\times50\times50$ intervals.
For the density and temperature, the bins are uniformly displaced in the logarithm, 
while we use a linear scale for the electron fraction.
The artificial atmosphere of the simulations is set to density
level $\rho_\text{atm}\sim10^4$~$\gccm$. 
Similar plots have been presented in \cite{Bacca:2011qd,Fischer:2013eka} for core-collapse supernova
simulations, in \cite{Most:2018eaw,Lalit:2018dps} for BNS mergers, and in \cite{Perego:2014fma} for 
BNS merger aftermath~\footnote{Note the LS220 simulations presented
  here are a longer and higher resolution versions of those shown in~\cite{Lalit:2018dps}.}.
However, a detailed discussion is, to the best of our knowledge,
missing in the literature.

The five rows of Figs.~\ref{fig:histo_DD2}-\ref{fig:histo_SFHo}~\footnote{Time-dependent (animated) versions can be found at\\
  \url{http://www.computational-relativity.org/}.} refer to snapshots 
covering the entire available binary evolution. Moving from top to
bottom: 
(1) early time during the orbital phase, 
(2) the time corresponding to the temperature peak, 
(3) about 3-4 milliseconds after the temperature peak, 
(4) a later time close to collapse (if collapse happens),
(5) time close to simulation end. 
Each row shows a snapshot of rest-mass density on the orbital plane and 
two thermodynamical condition histograms in which two of three variables 
($(\rho,T)$ and $(\rho,Y_{e})$, respectively)
are shown and the third is integrated over its range.
  
During the last orbits (snapshots (1)) most of the baryon mass is clearly distributed around
the maximum NS densities at $T\sim1-3$~MeV, with an electron fraction
still close to the cold, $\nu$-less weak equilibrium,
$0.01 \lesssim Y_e \lesssim 0.12$.
In all models we recognize the typical increase of $Y_{e}$ moving 
from $\rho \sim \rho_{0}$ to a few times $\rho_0$ due to the increase 
in the symmetry energy.
The highest temperatures are reached at
densities $\rho\sim10^{12}{\rm g~cm^{-3}}$, but as mentioned above this is a
numerical artifact.

After merger ($t\gtrsim0$, snaphshots (2)), the innermost part of the
merging cores, $\rho \gtrsim \rho_\text{NS}$, 
does not experience violent shocks ($ s \lesssim 1~k_{\rm B}~{\rm
  baryon}^{-1}$, where $s$ 
is the specific matter entropy). Thus the increase of the maximum density 
in the center of the remnant is associated with a marginal increase in 
the temperature (up to a few tens of MeV), mainy due to matter compression 
of degenerate nuclear matter well above nuclear saturation density.
Indeed, most of the matter in the cores resides below the semi-degenerate
transition line represented by 
\begin{equation}
\label{eq: T Fermi}
T=T_{\rm F}(\rho) = \frac{\hbar^2}{2m_b} \left( \frac{3 \pi^2 \rho}{m_b} \right)^{2/3} , 
\end{equation}
where $T_{\rm F}$ is the Fermi temperature of an ideal baryon gas of mass $m_b$.
At merger ($t\sim0$) the electron fraction at the highest densities is
still frozen to the initial $\nu$-less,  
cold weak equilibrium value. 
Due to the high opacities of these regions during and after the merger, 
$Y_{e}$ does not change significantly on the simulated timescale 
due to the emission of neutrinos.
Only later, when the cores experience a
significant compression and mixing, the $Y_e$ profiles become 
more uniform for $\rho \gtrsim \rho_\text{NS}$.

On the contrary, the densities at which $T$ peaks correspond to the regions close to
the binary center of mass where shocks are generated and material is pushed outwards
(snaphshots (2)-(3)).
The highest temperatures are reached at densities
slightly below 
$\rho_\text{NS}$ and characterized by the lowest 
$Y_e$. In particular, the DD2 model peaks at $\sim70$~MeV, 
LS220 at $\sim80$~MeV, and SFHo at $\sim100$~MeV.
This matter, initially located in the outer layers of the two cores, 
expands into the spiral arms. 
The temperature lowers as the spiral-arms expand and decompress, forming
an envelope around the merging cores. As a consequence of the fast expansion,
the temperature drops to $\sim 10$~MeV as the density reaches a few times
$10^{13}~{\rm g~cm^{-3}}$ (snapshot (3)). The EOS is dominated by 
the non-degenerate nucleons ($T \gtrsim T_{\rm F}$) 
and the matter expanding adiabatically 
satisfies the relation $T^3/\rho^{2}\approx {\rm const}$%
~\footnote{Thermodynamics' first principle for an adiabatic
  trasformation of an ideal gas in a volume $V$ is $dU/V = + p d\rho/\rho$ with $U$
  the internal energy and $dU\propto dT$. For a non-relativistic gas
  $P=2/3~U/V$ and for a relativisitc gas $P=1/3~U/V$. Hence, in the
  former case $d\ln T=2/3~d \ln\rho$, and in the latter $d\ln
  T=1/3~ d\ln\rho$.}.
  
In terms of the spatial distribution of $\rho$ and $T$, the merging 
cores have a peculiar structure with two rotating hot spots, 
displaced by an angle of $\sim \pi/2$ with respect to the 
densest regions corresponding to the NS cores (e.g. \cite{Kastaun:2016elu}). 
The expansion of the spiral arms, in combination with their fast orbital 
motion, produces a disk structure around the forming massive NS.
The fast motion of the merging cores and of spiral arms inside this newly 
formed envelope heats it up, mainly via shocks. 
The subsequent increase of entropy determines a spread in the $\rho-T$ diagram.
The EOS is dominated by the non-relativistic baryons,
$T \propto \left( \exp{\left( 2s \right)}/\rho^2 \right)^{1/3}$ 
(see, e.g., \cite{Cox:1968pss..book.....C}) and 
matter experiences a temperature increase up to 
a few tens of MeV (snapshots (3) and (4)).
This behaviour explains the peak structure and evolution observed in the 
$\rho-T$ diagram, where a maximum temperature of a several tens of MeV is 
reached by a substantial amount of matter with density around or immediately 
below nuclear saturation density.
The corresponding electron fraction is significantly larger ($\sim$~0.05-0.10) than the 
low $Y_{e}$ expected from $\nu$-less, cold weak equilibrium ($\lesssim 0.05$).
This is only partially due to an excess of electron antineutrinos
emission (see below), 
while it traces back the origin of this matter to larger densities where
the increase of the symmetry energy produces larger $Y_e$ already in $\nu$-less weak equilibrium conditions.

Shocks inside the disk produce further matter expansion (snaphshots (3)-(5)). 
As soon as the density drops below $10^{10}{\rm g~cm^{-3}}$ and the temperature becomes of the
order of a few MeV, the EOS is dominated by the relativistic electrons and photons.
Thus, adiabatic expansion follows the relation $T^3/\rho \approx {\rm const} \propto s$, 
typical of ultrarelativistic gases.

If the forming massive NS does not collapse (as in the DD2 case, Fig.~\ref{fig:histo_DD2}), 
the cores completely merge on a timescale of several ms and the hot
spots become a spherical annulus at an intermediate
density $\sim \rho_0$ between the denser core and the more diluite envelope, see Fig.~\ref{fig:DD2.M13641364.temp}.
Since this time interval corresponds to several orbital timescales, the envelope around the remnant 
has reached a relatively homogeneous configuration and the spread in the $\rho-T$ plane has
significantly reduced. Moreover, an inspection of the $\rho-Y_{e}$ planes reveals the effect 
of weak reactions on the disk composition: at the location where the last scattering neutrino surface is expected to
be ($\rho \sim \rho_{\rm lim}$), the captures of
positrons and streaming ${\nu}_e$ on free neutrons 
significantly increases the electron fraction.

On the other hand, if a black hole forms (as in the LS220 and SFHo cases, see the fifth row of 
Figs.~\ref{fig:histo_LS135135}-\ref{fig:histo_SFHo}), the evolution of the thermodynamical conditions inside 
the remnant changes  substantially. Within a ms, the formation of an apparent horizon removes all the high density part 
of the system, down to densities of $10^{12}{\rm g~cm^{-3}}$, leaving
a rather cold disk ($T \lesssim 10 {\rm MeV}$)
with a significantly reprocessed electron fraction, i.e. $Y_{e} \sim 0.25$.

In Fig.~\ref{fig:histo_entropy}, we present histograms of the thermodynamical conditions from
the third and fifth snaphshots for all the models presented in Figs.~\ref{fig:histo_DD2}-\ref{fig:histo_SFHo},
binned in density and specific entropy, $s$. For the latter, we consider 50 logarithmically spaced bins 
in the interval
$$
s \in \left[ 0.1,100 \right]~k_{\rm B}{\rm baryon^{-1}}.
$$
For all models, the cold unshocked cores are characterized by the presence of low entropy material 
($s \lesssim 2~k_{\rm B}{\rm baryon^{-1}}$) for matter densities in excess of a few times $\rho_0$.
Matter squeezed by the merging cores and expanding inside the forming disk is subjects to intense 
hydrodynamical shocks that increase matter entropy up to a few $k_{\rm B}~{\rm baryon}^{-1}$.
The combined effect of matter expansion and shocks resulting from the repeated core bounces
produces a transient phase where the envelope
engulfing the cores has a large spread in entropy (see, for example, the left panel
of the DD2\_M136136 cases). 
However, on a timescale of a few dynamical periods, the action of the 
spiral arms on the innermost part of the disk increases the entropy of the colder streams, producing a 
tighter correlation between the matter density and entropy. The bulk of the remnant outside the merging 
cores has $2 < s \left[ k_{\rm B}~{\rm baryon}^{-1} \right] \lesssim 10$
for matter density decreasing from $10^{14}$ down to $10^{10}~{\rm g~cm^{-3}}$.
Looking at the matter distribution inside this density interval, we conclude that
softer EOSs present a remnant characterized by larger values of the entropy 
(i.e. $\Delta s \sim 2~k_{\rm B}{\rm baryon}^{-1}$).
This is a consequence of the stronger shocks that characterize more violent mergers from more compact NSs.
If the EOS is stiffer and the massive NS survives long enough, 
shock expansion inside the low density part of the disk increases its entropy, up to
$20~k_{\rm B}{\rm baryon^{-1}}$. On the other hand, if a black hole forms, the lack of a persistent source of
shocks partially prevents the disk from reaching a tight $\rho-s$ correlation 
(as visible in the LS220 and SFHo equal mass cases).

Apart from the differences due to the different fates of the remnant,
all five fiducial models show a similar qualitative behavior, and the major features
described above do not depend on the specific EOS nor on the binary mass ratio.
Nevertheless, quantitative differences and temporary features can be noticed depending on the 
specific model. Temperature and entropy are, on average, larger for models employing a softer
EOS. This is a results of the more violent collision that characterize more compact NSs.
Models employing turbulent viscosity show immediately after merger a larger spread in the 
density-temperature histogram, as well as in the density-entropy one. However, they require
less time to produce disks with homogeneous properties. In the asymmetric model LS220\_M140120,
the cores merge in a very asymmetric way. In particular, the core of
the lighter NS is tidally deformed by the core of the heavier one and
its matter is subject to a violent temperature increase during its
decompression (see the second raw of 
Fig.~\ref{fig:histo_LS140120_VIS}). However, on a longer timescale,
the outer part of the lighter core contributes to the formation of the
hot envelope around the merging cores and the evolution of all
termodynamical quantities resembles the one of more symmetric
mergers. 

Finally, it is interesting to note that our histograms track the evolution of the dynamical
ejecta in its different components. 
Immediately after merger, a few ejection episodes
develops from matter with initially low temperature (a few MeV), low $Y_{e}$ ($\lesssim 0.1$),
and low entropy ($\lesssim 10 k_B$).
They expands very fast according to $T^3/\rho^{2} \sim {\rm const}$, down to densities 
at the edge of our domain. Positron and electron neutrino captures on free $n$ increase $Y_{e}$
during the expansion, but only marginally ($Y_{e} \lesssim
0.25$). 
We tentatively identify this ejecta as the tidal component. 
After $\sim$~1 millisecond, ejection episodes are also visible from the high-temperature 
part of our histograms (corresponding to $s \gtrsim 10~{\rm k_B}~{\rm baryon^{-1}}$), expanding at the same rate.
In the $T-Y_{e}$ plane, 
this ejecta presents a much broader distribution in $Y_{e}$, up to $Y_{e} \lesssim 0.4$. We identify 
this ejecta as the shock-driven one, whose polar component is more significantly influenced 
by neutrino absorption.

\section{Influence of neutrino trapping on matter}
\label{sec:neutrinos}

We now move to the discussion of the potential effect of trapped neutrinos 
using the approached described in Section~\ref{sec:Method_neutrinos}.
We first re-write the expression of the neutrino fractions, Eqs.~\ref{eq: neutrino fractions}-\ref{eq: neutrino energy fractions}, in terms of the 
Fermi temperature, Eq.~\ref{eq: T Fermi}, neglecting the exponential cut in density:
\begin{eqnarray}
\label{eq: neutrino fractions with T Fermi}
 Y_{\nu_i}(\rho,Y_e,T) & = & \frac{3 \sqrt{2}}{8} \left( \frac{T}{T_{\rm F}} \right)^{3/2} \left( \frac{k_{\rm B}T}{m_b c^2} \right)^{3/2} F_2(\eta_{\nu_i}) \,, \\
 Z_{\nu_i}(\rho,Y_e,T) & = & \left( k_{\rm B} T ~ \frac{F_3(\eta_{\nu_i})}{F_2(\eta_{\nu_i})} \right) Y_{\nu_i}(\rho,Y_e,T) \,.
\end{eqnarray}
Since baryons are always non relativistic, $k_{B}T \ll m_{b}c^2$, significant neutrino fractions require
high temepratures for matter in non-degenerate conditions ($T \gg T_{\rm F}$) and/or highly degenerate
neutrino conditions ($\eta_{\nu} \gg 1$).

We have repeated the analysis of the histrograms of the thermodynamical conditions presented in Section~\ref{sec:histograms}
including the effects of trapped neutrinos for $\rho > \rho_{\rm lim}$. Since the variations in $Y_{e}$ and $T$ are only minors
(see below) and the histograms present already a significant spread in all variables, we conclude that the inclusion 
of trapped neutrinos does not qualitatively changes our previous analysis. 

In Fig.~\ref{fig: DD2_nu_post_processing, plane xy} we present the results of our post-processing
analysis for a snapshot of the DD2\_M136136 simulation taken $\sim 11~{\rm ms}$ 
after merger (i.e. close to the peak temperature) along the equatorial plane. 
In the top panels, we show the rest-mass density (left), matter temperature (middle) and electron fraction (right) 
as obtained by the simulation. In the bottom panels, we present $Y_{{\nu}_e}$ (left), 
$Y_{\bar{\nu}_e}$ (middle), and 
$Y_{\rm e,sim} - Y_{\rm e,eq} = Y_{{\nu}_e} - Y_{\bar{\nu}_e}$ (right) 
as obtained in our post-processing analysis.

Deep inside the remnant, where densities are in excess of a few times
nuclear saturation density, $\rho \gtrsim 3 \rho_0$, the temperature is a few MeV high.
Under such conditions, $T \ll T_{\rm F}$ (i.e. $\mu_i \gg T$ for neutrons, protons, and 
electrons, and thermal effects are negligible). Since matter conditions are still close to the initial 
cold, $\nu$-less, weak equilibrium, 
we find $\left| \eta_{\nu_i} \right| \lesssim 3$,
and $F_2({\eta_{\nu_i}}) \lesssim 20 $
(see e.g. \cite{Takahashi:1978}, appendix). Thus,
the production of electron flavor (anti)neutrinos is suppressed
and $Y_{\rm e,eq} \approx Y_{\rm e,sim}$. 
Since $Y_{e} \lesssim 0.1$, neutrons are more degenerate than protons and any increase
in temperature decreases more significantly $\mu_p$ than $\mu_n$. Thus, $\eta_{\nu_e} \lesssim 0$.
This effect is partially compensated by the increase of $\mu_e$ due to matter compression. 
Overall, we find that
$-3 \lesssim \eta_{\nu_e} \lesssim -1 $, 
$ 10^{-6} \lesssim Y_{\nu_e} \lesssim 10^{-4} $,
$ 10^{-4} \lesssim Y_{\bar{\nu}_e} \lesssim 10^{-3} $, and
$ 10^{-6} \lesssim Y_{\nu_x} \lesssim 10^{-4} $,
such that 
$Y_{\nu_e} < Y_{\nu_x} < Y_{\bar{\nu}_e}$ locally.

At lower densities, $ 10^{14}{\rm g~cm^{-3}} \lesssim \rho \lesssim 4 \times 10^{14}{\rm g~cm^{-3}}$,
the temperature increases up to a few tens of MeV, due to the presence of
the hot annulus outside the merging cores.
While the electrons are still highly degenerate, 
the neutrons and (more significantly) the less
abundant protons becomes mildy degenerate ($T \gtrsim T_{\rm F}$).
The (negative) chemical
potential of electron neutrinos decreases, but not the neutrino degeneracy parameter,
due to the higher temperatures:  in this region
we find that $ -2
\lesssim \eta_{\nu_e} \lesssim -1$. Under these
conditions, a significant electron antineutrino  
gas forms $Y_{\bar{\nu}_e} \sim 0.02$, followed by $Y_{\nu_x} \lesssim 0.01$, 
while electron neutrinos are still suppressed by degeneracy ($Y_{{\nu}_e} \lesssim 10^{-3}$). 
Due to the dominant presence of $\bar{\nu}_e$, the initial $Y_{\rm e,sim}$ ($\sim 0.06$) 
increases by $Y_{\bar{\nu}_e}$, i.e. $Y_{\rm e,eq} 
\approx Y_{\rm e,sim} + Y_{\bar{\nu}_e} $, to guarantee lepton number conservation. 

At densities below $10^{14}{\rm g~cm^{-3}}$
the decrease in temperature related with matter expansion produces less significant deviations 
from the initial cold $\nu$-less weak equilibrium and all neutrino fractions become negligible
inside the cold ($T \lesssim 10~{\rm MeV}$), unshocked streams.
In particular, $\eta_{\nu_e} \gtrsim 1$ and $\nu_e$ dominate over 
$\bar{\nu}_e$. However, due to the low temperatures, $T < T_{\rm F}$ and $Y_{\nu_e} \sim 10^{-3}$.
At the same time, spiral arms moving inside this cold and less dense matter 
produce shockes that heat-up matter at the arm interfaces.
Inside these regions, temperatures in excess of 10~MeV for
matter that decompresses down to a few times $10^{12}{\rm g~cm^{-3}}$
(thus, $T \sim $ several $T_{\rm F}$) 
lead to $\eta_{\nu_e} \sim -1 $, producing trapped neutrino gases with 
$Y_{\bar{\nu}_e} \lesssim 0.03$, 
$Y_{{\nu}_x} \lesssim 0.01$, and
$Y_{{\nu}_e} \sim$ a few $10^{-3}$.
As a results, $Y_{e,{\rm eq}}$ increases with respect to $Y_{e,{\rm sim}}$.

The appearance of a trapped neutrino gas is done at the expences of the fluid internal energy, formed by relativistic electrons, 
positrons, photons, and non-relativistic baryons. Since thermodynamical stability requires $\partial e / \partial T > 0$, $T_{\rm eq} < T_{\rm sim}$. 
In the top panel of Fig.~\ref{fig: DD2_nu_post_processing, P and T} we present the ratio between the fluid pressure
after and before the postprocessing for the same configuration presented in Fig.~\ref{fig: DD2_nu_post_processing, plane xy}. 
The inclusion of trapped neutrinos reduces the temperature down to $\sim$93\% of the simulation value
in the hottest regions of the systems.
Neutrinos are intrinsecally ultra-relativistic and for them
$P_{\nu} \propto Z_{\nu}/3$, while for a non-relativistic ideal gas $P \propto 2e/3$. 
Thus, we expect that their appearance decreases the total pressure and
we quantify the relative variation of the pressure to be $\lesssim 0.04$, in particular where 
the equatorial plane intersects the hot annulus outside the merging cores
(as visible in the bottom panel of Fig.~\ref{fig: DD2_nu_post_processing, P and T}).
Based on these results, the internal energy stored in the neutrino field is at most
$\sim$8\% of the fluid internal energy.

Effects similar to the ones we have discussed for high density region ($\rho \gtrsim 10^{14}{\rm g~cm^{-3}}$) 
in the equatorial plane can be seen along any vertical plane passing through the center.
However, far from the equatorial plane, the annulus structure involves matter at slightly lower density for which   
the decrease in the pressure due to the presence of trapped neutrinos is slightly more significant ($P_{\rm eq}/P_{\rm sim} \sim 0.94$).

We finally move to the analysis of trapped neutrinos in the SFHo\_M135135 model. The former is characterized by larger maximum densities and temperatures,
see Fig.~\ref{fig:rhoT}, and by a quick collapse of the central massive NS. Before black hole formation, results of the neutrino post-processing 
analysis are qualitatively similar to the DD2\_M136136 case. Due to the larger temperatures reached during the NS collisions, 
the proton degeneracy decrease is more pronunced and electron antineutrinos can locally form a trapped gas with $Y_{\bar{\nu}_e} \lesssim 0.05$, 
in particular where the cores collide and at the interface between the spiral arms and the forming disk. The variation of the local pressure
to the appearance of neutrinos can reach 10\% of the simulation pressure.
However, these large temperatures are reached only immediately after merger, while during most of the evolution and for the bulk of the matter, 
neutrino fractions are only marginally larger than the ones observed in the DD2\_M136136 model.
After the formation of an apparent horizon, the maximum density and temperature significantly decrease inside the disk and
trapped neutrinos play no role inside the remnant.

\begin{figure*}
  \noindent\makebox[\textwidth]{%
    \includegraphics[width=.33\textwidth]{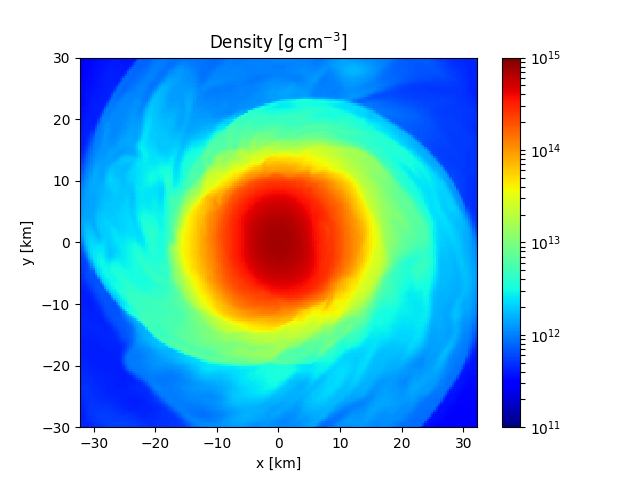}
    \includegraphics[width=.33\textwidth]{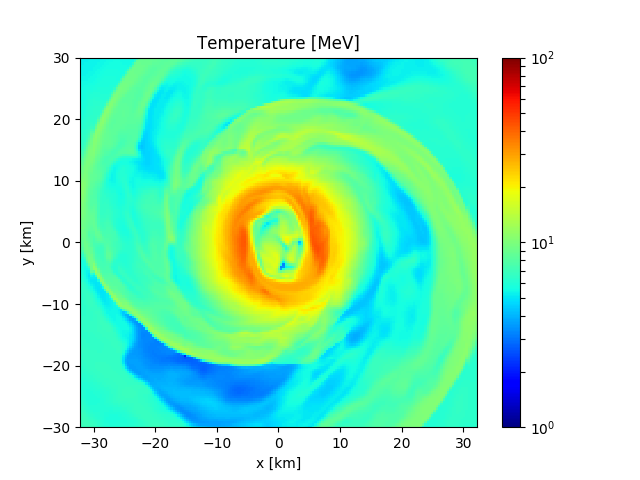}
    \includegraphics[width=.33\textwidth]{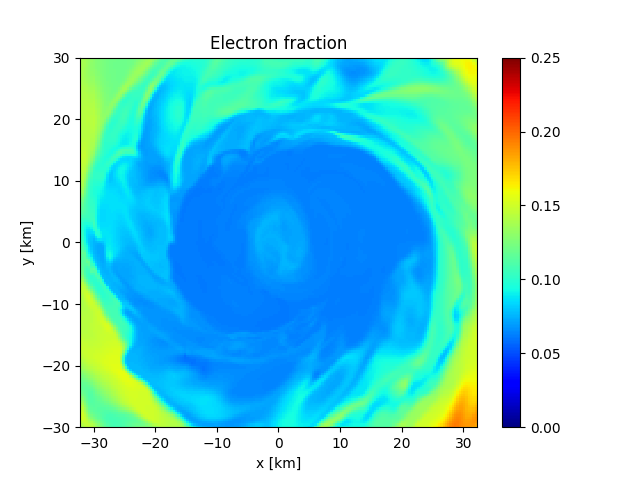}
  }\\
  \noindent\makebox[\textwidth]{%
    \includegraphics[width=.33\textwidth]{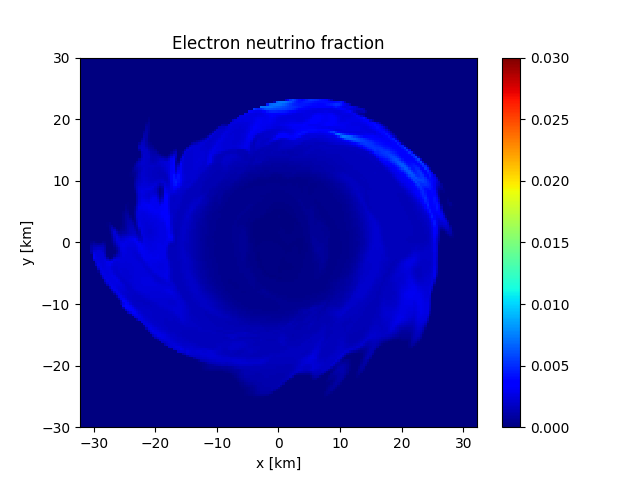}  
    \includegraphics[width=.33\textwidth]{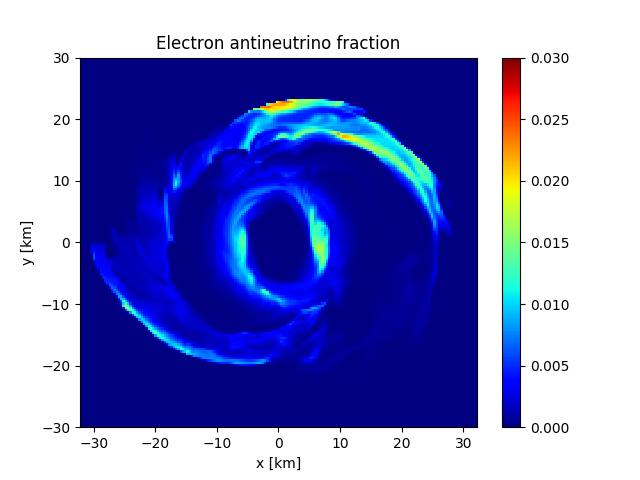}
    \includegraphics[width=.33\textwidth]{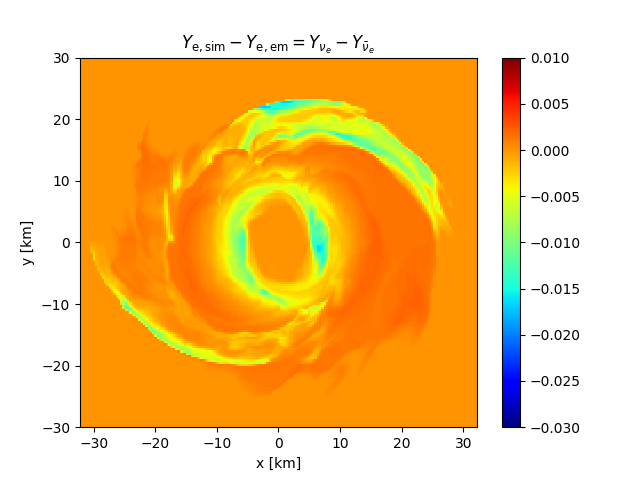}
  }\\
  \caption{Upper row: Rest mass density (left), temperature (middle) and electron fraction (right) on the orbital plane 
  for the DD2\_M136136 model at the temperature peak, as obtained in simulation. Bottom row: electron neutrino fraction ($Y_{{\nu}_e}$,
  left), electron antineutrino fraction ($Y_{\bar{\nu}_e}$, middle), variation of the electron fraction ($Y_{e,{\rm sim}} - Y_{e,{\rm eq}}$
  corresponding to $Y_{{\nu}_e} - Y_{\bar{\nu}_e}$, right) after modelling in postoprocessing the presence of trapped neutrinos 
  at high density. Inside the merging cores, matter degeneracy prevents neutrino formation ($T_F \gg T$). Only for shocked matter
  in semi-degenerate conditions engulfing the cores or developing at spiral arm interfaces, a significant $\bar{\nu}_e$ fractions develops.   
  }
  \label{fig: DD2_nu_post_processing, plane xy}
\end{figure*}

\begin{figure}
    \includegraphics[width=0.49\textwidth]{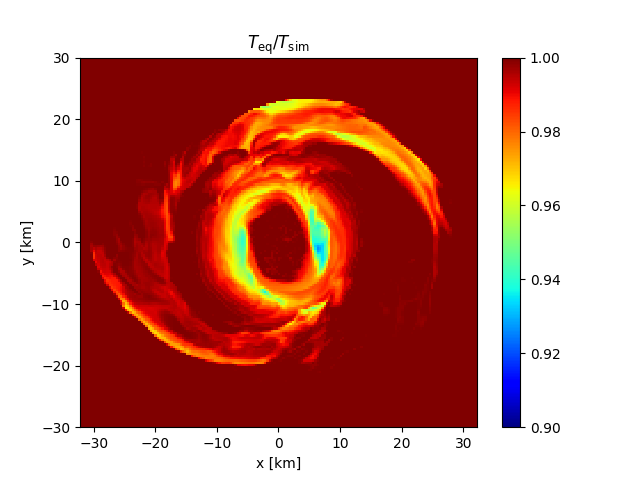}\\
    \includegraphics[width=0.49\textwidth]{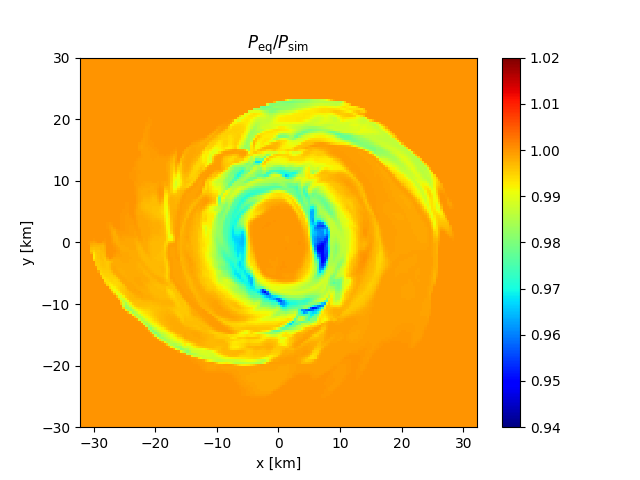}
  \caption{Upper (lower) panel: ratio of equilibrium temperature (pressure) to the one obtained in the 
  simulation. The equilibrium values are calculated in postprocessing including the effect of trapped neutrinos.
  The formation of trapped neutrino gases converts fluid thermal energy into radiation energy, reducing matter temperature 
  down to $\sim$92\% of the simulation value. Since $P_{\nu} \propto Z_{\nu}/3$, the total pressure reduces down to $\sim$94\% of
  its original value.}
  \label{fig: DD2_nu_post_processing, P and T}
\end{figure}

\section{Disks around black holes or NS remnant}
\label{sec:disks}

\begin{figure*}[t]
  \noindent\makebox[\textwidth]{%
  \includegraphics[width=.49\textwidth]{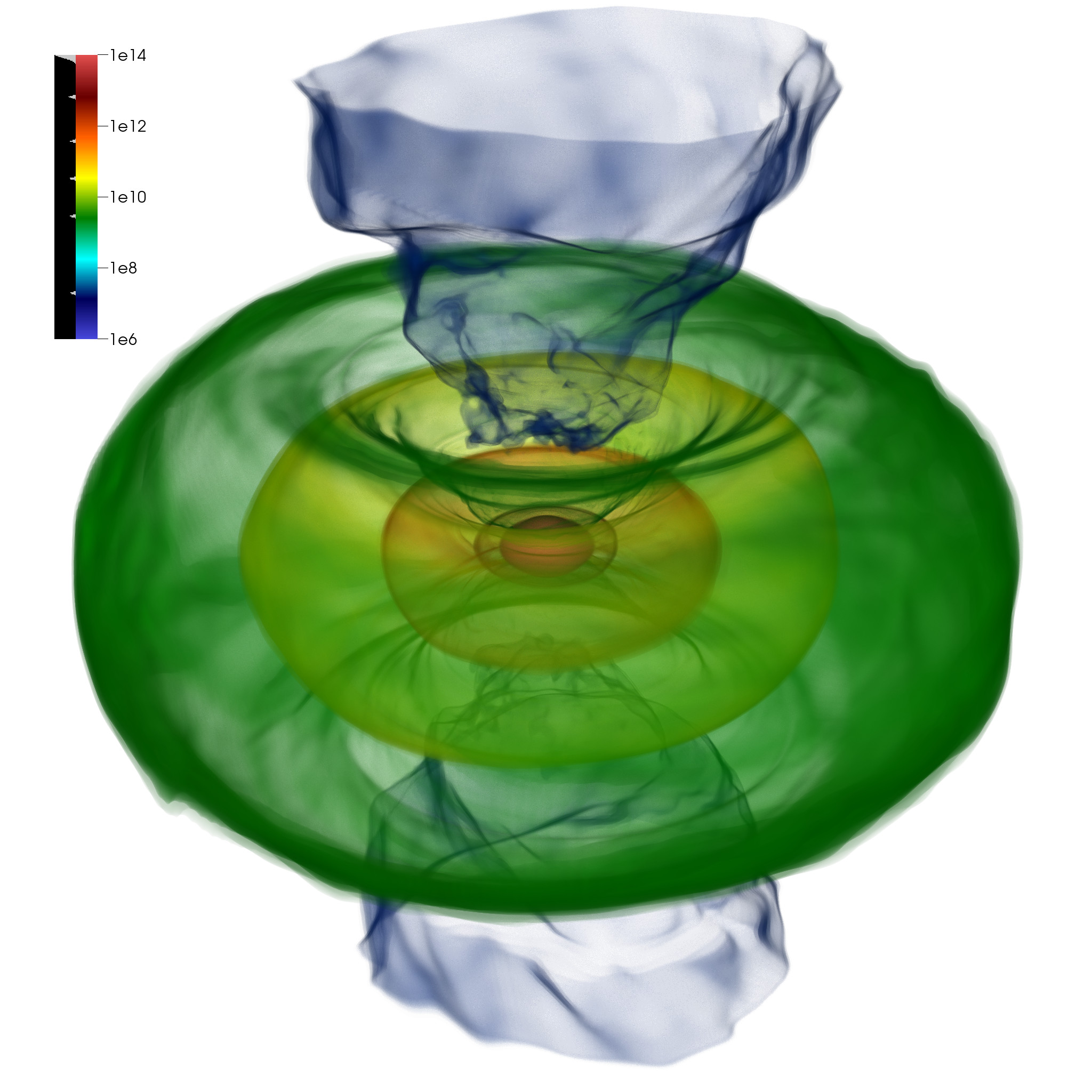}
  \includegraphics[width=.49\textwidth]{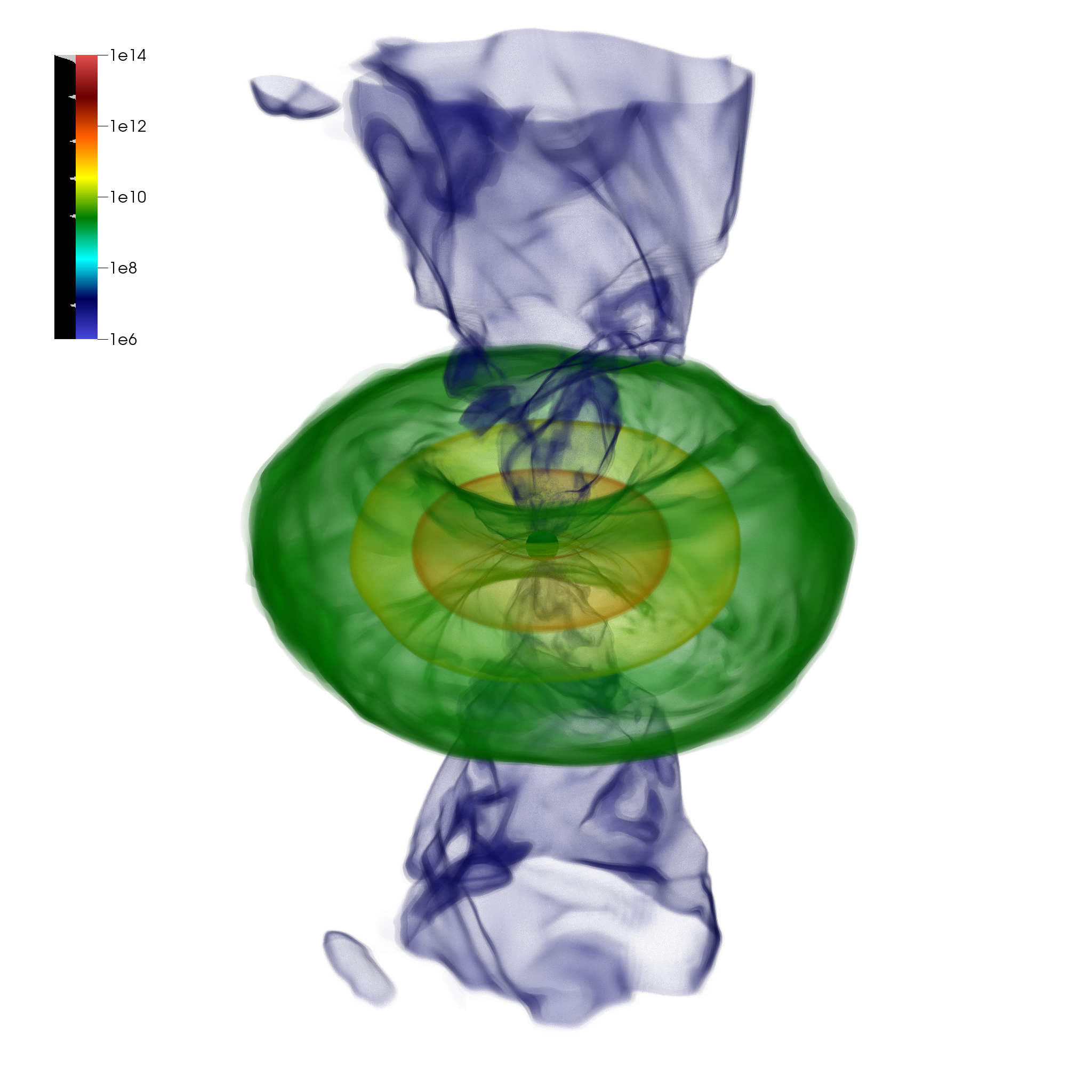}
  }  
  \caption{Volume rendering of the density of the postmerger remnant. 
  Left panel: DD2\_M13641364 at $t = 34.2 {\rm ms}$ after the merger. Right panel: 
  SFHo\_M135135 at $t = 17.2 {\rm ms}$ after the merger. Both image have the same spatial scale and show 
  the data in a box of size ${\sim}750\ {\rm km}$. The black surface denotes the approximate location 
  of the black hole horizon, which we identify as $\alpha = 0.3$, being $\alpha$ the lapse function. Remnants harboring black holes at their centers, 
  such as SFHo\_M135135, produce smaller, compact disks, that are mostly transparent to neutrinos. 
  Conversely, remnants having a massive NS at their center, have more massive, geometrically and optically thick disks. 
  At high latitudes, a low-density funnel is present in both cases.}
  \label{fig:DD2.vs.SFHo.dens}
\end{figure*}

\begin{figure*}
  \begin{minipage}{\textwidth}
  \raisebox{-0.5\height}{
  	\includegraphics[width=.49\textwidth]{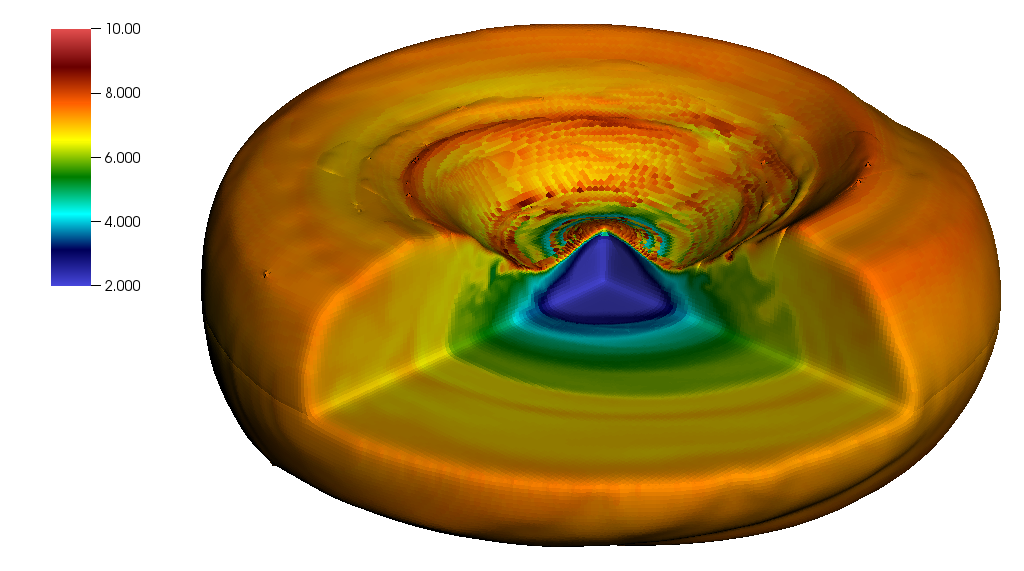}
  }
  \hfill
  \raisebox{-0.5\height}{
  	\includegraphics[width=.49\textwidth]{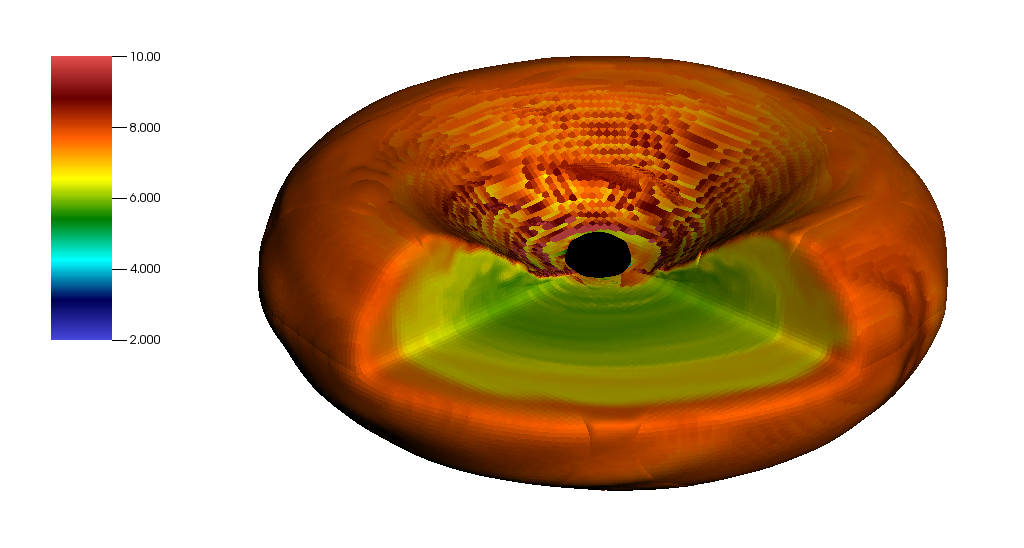}
  }
  \end{minipage}
  \caption{Entropy in $k_{\rm B}$ per baryon of the postmerger
    remnant. Left panel: DD2\_M136136 at $t = 34.2 {\rm ms}$
    after the merger. Right panel: SFHo\_M135135 at $t = 17.2
    {\rm ms}$ after the merger. We only show material with $\rho \geq
    10^{11}\ {\rm g}\, {\rm cm}^{-3}$ corresponding the inner regions
    of Fig.~\ref{fig:DD2.vs.SFHo.dens}. Note that here we are using two
    different spatial scales for the two images (\textit{c.f.,}
    Fig.~\ref{fig:DD2.vs.SFHo.dens}). The black surface denotes the
    approximate location of the black hole horizon, which we identify as
    $\alpha = 0.3$, being $\alpha$ the lapse function. The cores of the merging NSs are not significantly
    shocked during merger, so the central part of the remnant
    maintains a relatively low specific entropy of $s {\lesssim} 2\ k_{\rm B}$. Material squeezed out of 
    the collisional interface between
    the NSs forms the bulk of the disk and has entropy of a few
    $k_{\rm B}$. In the cases in which black hole formation occurs, the disk
    entropy is slightly higher (by a couple of $k_{\rm B}$) than in
    the cases in which a massive NS is still present at the center.}
  \label{fig:DD2.vs.SFHo.entropy}
\end{figure*}

\begin{figure*}[t]
  \noindent\makebox[\textwidth]{%
  \includegraphics[width=.49\textwidth]{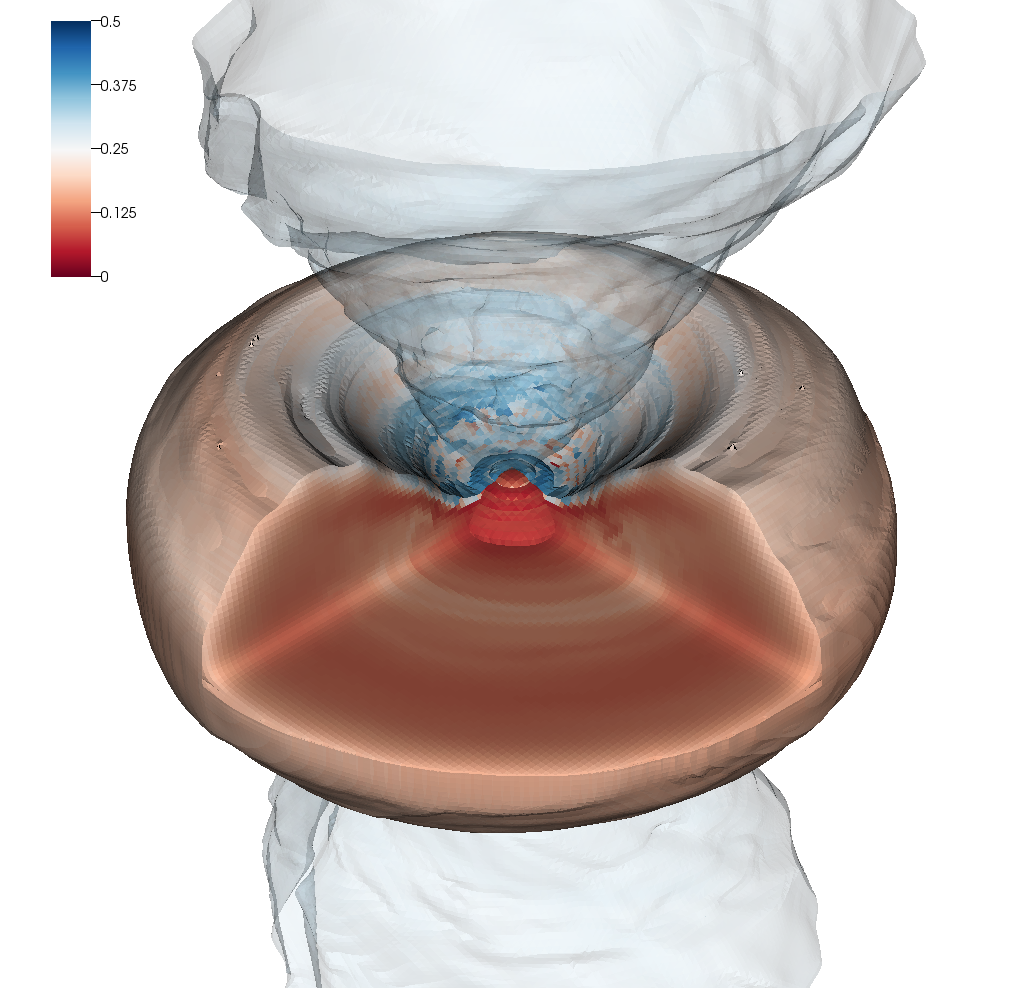}
  \includegraphics[width=.49\textwidth]{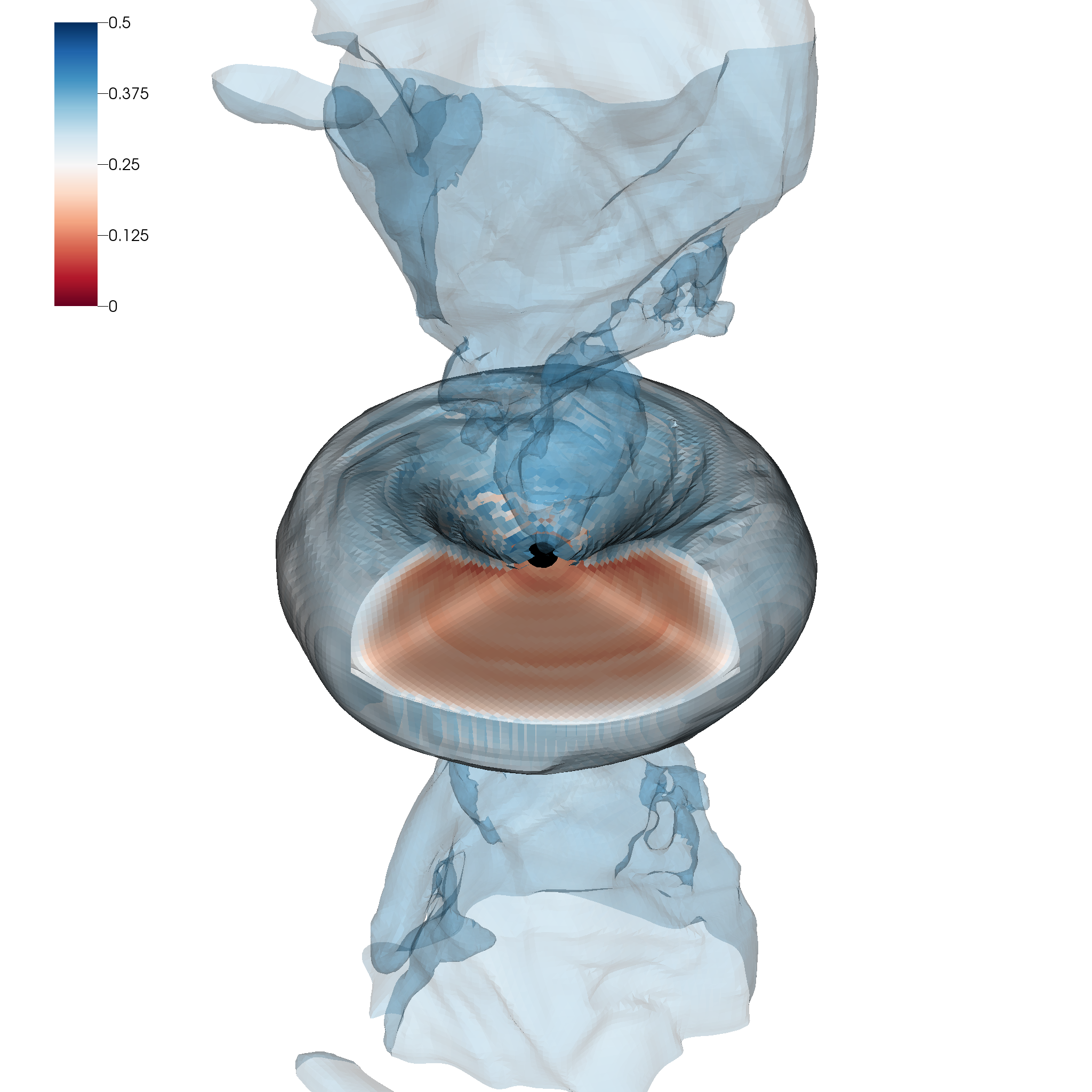}
  }  
  \caption{Visualization of the electron fraction of the merger remnant. Left panel: 
  DD2\_M136136 at $t = 34.2 {\rm ms}$ after the merger. Right panel: SFHo\_M135135 at 
  $t = 17.2 {\rm ms}$ after the merger. Both image have the same spatial scale as in 
  Fig.~\ref{fig:DD2.vs.SFHo.dens} and show the data in a box of size 
  ${\sim}750\ {\rm km}$. The electron fraction is used to color the $10^{7} {\rm g}\ {\rm cm}^{-3}$ 
  (semi-transparent), and the $10^{11} {\rm g}\ {\rm cm}^{-3}$ density iso-surfaces. 
  We also show the electron fraction on the $\rho = 10^{13}  {\rm g}\ {\rm cm}^{-3}$ 
  iso-surface for the DD2\_M13641364 model. The black surface denotes the approximate location of 
  the black hole horizon, which we identify as $\alpha = 0.3$, being $\alpha$ the lapse function. 
  The accretion disks are fairly neutron rich 
  in their bulk, irrespective of the remnant type (massive NS or black hole). The accretion disk coronae are 
  irradiated by neutrinos and are less neutron rich.}
  \label{fig:DD2.vs.SFHo.ye}
\end{figure*}

The analysis of the entropy distributions inside BNS merger remnants 
already revealed important differences between the case in which a black hole
promptly forms and that in which a massive NS survives for several dynamical time
scales (end of Section~\ref{sec:histograms}).  
Further differences in the disk properties are clearly visible in 
Figs.~\ref{fig:DD2.vs.SFHo.dens}-\ref{fig:DD2.vs.SFHo.ye}, where we
present volume rendering of the matter density, and three dimensional
spatial distributions of entropy and electron fraction, for two
simulations employing a stiffer (DD2\_M136136, left panels) and a
softer (SFHo\_M135135, right panels) EOS,  a few tens of milliseconds after
merger.

The disk around the black hole produced at the end of the SFHo\_M135135 simulation 
is less massive and more compact  (i.e., with a significantly smaller spatial extension and a steeper 
density profile) than the disk produced outside the massive NS by the DD2\_M136136 simulation. 
Maximum densities in the former case reach only $\sim 10^{12}~{\rm g~cm^{-3}}$
in a torus around the innermost stable circular orbit. On the other hand, 
in the latter case the spatially more extended disk has a continuously increasing density profile moving towards
the center that joins up with the massive NS density structure. The resulting disk is geometrically and optically thick to
neutrinos (at least during the first tens of ms).
Fig.~\ref{fig:DD2.vs.SFHo.entropy} confirms that a softer EOS produces disks with larger entropies
(possibly, $\Delta s \sim 2~{k_{\rm B}{\rm baryon}^{-1}}$ ) due to the stronger shocks that develop at and after merger.
Neutrino irradiation from the center of the remnant changes the electron fraction above and inside the disk.
On the one hand, in both cases the innermost part of the disk retains its initial neutron richness, while the electron fraction increases 
significantly above 0.3 in the low density funnel above it. On the other hand, significant differences 
characterize $Y_e$ in the outer parts of the disk. Due to the more extended, more diluite and colder structure,
the outer disk obtained in the DD2 run has a lower electron fraction than the more compact
disk resulting from the SFHo run.

\section{Symmary and conclusions}
\label{sec:conclusions}

In this paper, we analyzed the thermodynamical conditions experienced by matter during the merger
of two NSs based on the results of detailed hydrodynamical simulations
in numerical relativity.
Our five fiducial models include the effect of finite-temperature, composition dependent nuclear EOSs in tabular form
and of the emission of neutrinos via a three-species M0 that accounts for the re-absorption of electron flavor 
(anti)neutrinos in optically thin conditions. 

We explored three different hadronic EOSs: DD2, SFHo and LS220. The first two are based
on the RMF approach and on a full distribution of nuclei in NSE, 
while the latter is based on a liquid droplet Skyrme model and, besides free neutrons and protons, includes
$\alpha$ particles and one representative heavy nucleous.
While most of our models refer to equal mass binaries, in one case we have investigate the potential impact 
of an asymmetric mass binary merger. Moreover, in three cases we have included the impact of subgrid-scale 
turbulent angular momentum transport using the general-relativistic large eddy
simulations method (GRLES; \cite{Radice:2017zta}).

Matter stays cold and close to $T=0$, $\nu$-less weak equilibrium during the inspiral (the marginal increase
of temperature at the NS surfaces is a numerical artifact). As soon as the two NS cores touch and merge, the temperature
increases by up to several tens of MeV for densities immediately above nuclear saturation density. This is due to the 
development of hydrodynamical shockes at the interface between the bouncing cores that produce an hot 
mantle around the unshocked, merging bulks. The entropy in the cores stays below $2~k_{\rm B}~{\rm baryon^{-1}}$, while 
it increases to a few $k_{\rm B}~{\rm baryon^{-1}}$ in the hot mantle. The action of rotating spiral arms on the forming 
disk increases temperature and entropy. At first, on a timescale comparable with a few disk dynamical timescale, this produces 
a large spread in temperature and entropy. Later, over several disk orbits, the conditions becomes more homogeneous.
Matter expansion from the merging cores follows $T^3/\rho^2$ profiles, typical of baryon-dominated matter that
expands adiabatically.
The more compact and dense NS cores associated with softer EOS produce the largest densities and temperatures.
Asymmetric NS masses and the inclusion of turbulent viscosity have only a marginal impact on the thermodynamical
properties.

Since our simulations did not include the presence of a trapped neutrino gas in thermal and weak equilibrium with 
the fluid, we analyzed its potential impact in post-processing. We concluded that its inclusion 
can only marginally affect the local thermodynamical properties of dense matter above $10^{12}{\rm g~cm^{-3}}$: the maximum temperature 
is reduced by $\lesssim 10\%$, while the introduction of neutrinos determines a decrease in the total pressure $\lesssim 5\%$ 
and only in the hottest regions of the domain. Due to the initial neutron richness, deviations from cold, $\nu$-less weak equilibrium
due to an increase in temperature favor the production of electron antineutrinos over neutrinos.

If according to our results the inclusion of trapped neutrinos is not expected to significantly change the dynamical 
properties of the remnant, the appearance of additional degrees of freedom such as hyperons and quarks for densities
well above $\rho_0$ could have a more significant impact. BNS merger simulations performed with the BHB$\Lambda\phi$ 
EOS model \cite{Banik:2014qja} (which is identical to DD2 apart for the appearance of hyperons at $\rho \gtrsim 2 \rho_0 $)
present a qualitatively different evolution of the remnant (see, e.g., \cite{Radice:2016rys}).
Hyperons in the BHB$\Lambda\phi$ models soften the EOS and can lead to a fast collapse
of the massive NS into a black hole, whereas identical initial conditions evolved with the DD2 EOS produce a stable
rotating NS. For merging NSs with $M_{\rm NS} \lesssim 1.35 M_{\odot}$ the appearance of hyperons is 
expected only after merger and mainly inside the cold, unshocked cores.
If a black hole does not form, the qualitative behavior of the thermodynamical conditions is expected to be very similar
to the DD2 case, even if the stronger core bounces related with the EOS softening produce a hotter remnant
in the BHB$\Lambda\phi$ than in the DD2 case.
On the other hand, in more massive NSs hyperons could be present already
during the inspiral phase and their fraction increases faster and more significantly during the early post-merger phases.
This can lead to hotter remnant disks than the corresponding DD2 cases, if the massive NS collapse to a black hole.

The possible appearance of muons ($m_{\mu} \approx 105.6~{\rm MeV}$) and mesons (e.g. pions, 
$m_{\pi}\approx 135-140~{\rm MeV}$) has been explored, especially in cold NS matter, e.g. \cite{Glendenning:1996cost.book.....G}. 
In this case, degeneracy drives their production at densities well above
$\rho_0$. During a BNS merger, if the temperature reaches several tens of MeV in the hot envelope,
they could also be thermally produced at densities around or below $\rho_0$. Their inclusion in finite-temperature, composition
depedent EOS is still missing and their potential impact on BNS mergers has not been explored yet. 

Our fiducial models allowed us to study the properties of disk forming in a BNS merger. We found that the formation of a black hole in
the center instead of a massive NS leads to the most significant differences. In the early BH formation case, the disk is usually hotter (due to
the more violent NS collisions), more compact, and less dense ($\rho \lesssim 10^{12}{\rm g~cm^{-3}}$). However,
the higher temperatures and neutrino luminosities change its electron fraction more significantly.
In the case in which a long-lived remnant is formed, the disk is geometrically and optically thick, it is spatially more extended 
and has less steep density gradients, 
due to the prolonged action of the non-singular central object.
Our post-processing analysis of the trapped-neutrino gas could not directly investigate if the electron fraction of the 
matter forming the disk is largely affected by the temporal transition to finite-temperature weak equilibrium conditions before
quasi-adiabatic expansion. The reduced variation of $Y_e$ observed for $\rho \gtrsim 10^{12}{\rm g~cm^{-3}}$ seems 
to imply that this is not the case and more significant changes in $Y_e$ are expected due to the absorption of electrons and positrons, 
as weel as of $\nu_e$ and $\bar{\nu}_e$ on free baryons. This conclusions is qualitatively consistent with results reported by BNS simulations
including the contributions of trapped neutrinos, e.g. \cite{Sekiguchi:2015dma,Foucart:2015vpa}.

Finally, the histograms of the thermodynamical conditions reported in \cite{Bacca:2011qd,Fischer:2013eka} for CCSN simulations
allow a direct comparison between the thermodynamical conditions inside CCSNe and BNS mergers. 
The electron fraction has opposite trends inside these two scenarios. In CCSNe, symmetric nuclear matter deleptonizes to reach neutron rich conditions
inside the forming proto NS. Cold, $\nu$-less weak equilibrium conditions are achieved only on the 
$\nu$-cooling timescale (much larger than the dynamical and explosion timescale). On the other hand, in BNS mergers the decompression,
heating, and neutrino irradiation of cold nuclear matter drive matter {\it leptonization}. This process is relevant to set the thermodynamical
conditions of the ejecta and, ultimately, its nucleosynthesis yields, 
e.g \cite{Perego:2014fma,Metzger:2014ila,Just:2014fka,Sekiguchi:2015dma,Radice:2016dwd,Foucart:2016rxm,Lippuner:2017bfm}.
This fundamental difference reflects also on the evolution of the density-temperature conditions. 
In CCSN simulations the presence of a single shock front between the proto NS and the cold accreting flow channels most of the hot mantle in
a rather narrow band in the $\rho-T$ plane. 
On the other hand, in BNS mergers the combined presence of hydrodynamical compression, tidal decompression, shocks
and expansion produces a much broader band that narrows over several dynamical timescales.
In CCSNe larger densities and temperatures are achieved for softer EOSs, but the mass of the progenitor star is the most crucial parameter
that influences, for example, the maximum temperature reached during the simulation. 
Indeed, densities of the order of a few $\rho_0$ are reached only after the proto NS formation and cooling, and the details 
of the nuclear interactions are of secondary importance with respect to the strength of the gravitational field.
In this work, we have only partially
explored the dependence on the colliding NS masses and, in particular, we did not consider very massive NSs. However, the densities are significantly
larger in NS mergers and the details of the nuclear interactions influence more deeply the dynamics. Thus, the variability
introduced by the different EOSs in NS merger models is compatible with the variability associated with different progenitors for CCSNe.
We might expect more variability in mergers with larger NS masses, however prompt collapse to black hole sets an intrinsic limit
to the densities reached in the remnant.

\paragraph*{Acknowledgments}

The authors thank Domenico Logoteta for useful discussions.
The authors thank the organizers and participants 
of the INT Program INT-18-72R ``First Multi-Messenger Observation of a Neutron Star Merger
and its Implications for Nuclear Physics INT workshop'' held at
(Seattle, March 2018), 
of the ExtreMe Matter Institute's rapid task force meeting at
GSI/FAIR (Darmstadt, June 2018),
of the GWEOS workshop (Pisa, February 2019),  
for stimulating discussions.
SB acknowledges support by the EU H2020 under ERC Starting Grant, 
no.~BinGraSp-714626. 
DR acknowledges support from a Frank and Peggy Taplin Membership at the
Institute for Advanced Study and the
Max-Planck/Princeton Center (MPPC) for Plasma Physics (NSF PHY-1804048).
Computations were performed 
on the supercomputer SuperMUC at the LRZ Munich (Gauss project
pn56zo), 
on supercomputer Marconi at CINECA (ISCRA-B project number HP10B2PL6K
and HP10BMHFQQ); on the supercomputers Bridges, Comet, and Stampede (NSF XSEDE allocation TG-PHY160025);
on NSF/NCSA Blue Waters (NSF AWD-1811236).

\bibliographystyle{epj.bst}
\bibliography{references}

\begin{thebibliography}{99}

\bibitem{Shibata:1999wm}
M.~Shibata, K.~Uryu, Phys. Rev. \textbf{D61}, 064001 (2000),
  \texttt{gr-qc/9911058}

\bibitem{Sekiguchi:2011mc}
Y.~Sekiguchi, K.~Kiuchi, K.~Kyutoku, M.~Shibata, Phys.Rev.Lett. \textbf{107},
  211101 (2011), \texttt{1110.4442}

\bibitem{Bernuzzi:2015opx}
S.~Bernuzzi, D.~Radice, C.D. Ott, L.F. Roberts, P.~Moesta, F.~Galeazzi, Phys.
  Rev. \textbf{D94}, 024023 (2016), \texttt{1512.06397}

\bibitem{Radice:2018pdn}
D.~Radice, A.~Perego, K.~Hotokezaka, S.A. Fromm, S.~Bernuzzi, L.F. Roberts,
  Astrophys. J. \textbf{869}, 130 (2018), \texttt{1809.11161}

\bibitem{Roberts.etal:2012}
L.F. {Roberts}, G.~{Shen}, V.~{Cirigliano}, J.A. {Pons}, S.~{Reddy}, S.E.
  {Woosley}, Physical Review Letters \textbf{108}, 061103 (2012),
  \texttt{1112.0335}

\bibitem{Lattimer:2012nd}
J.M. Lattimer, Ann. Rev. Nucl. Part. Sci. \textbf{62}, 485 (2012),
  \texttt{1305.3510}

\bibitem{Ozel:2016oaf}
F.~Özel, P.~Freire, Ann. Rev. Astron. Astrophys. \textbf{54}, 401 (2016),
  \texttt{1603.02698}

\bibitem{Oertel:2016bki}
M.~Oertel, M.~Hempel, T.~Klähn, S.~Typel, Rev. Mod. Phys. \textbf{89}, 015007
  (2017), \texttt{1610.03361}

\bibitem{Lattimer:1991nc}
J.M. Lattimer, F.D. Swesty, Nucl. Phys. \textbf{A535}, 331 (1991)

\bibitem{Shen:1998gq}
H.~Shen, H.~Toki, K.~Oyamatsu, K.~Sumiyoshi, Nucl. Phys. \textbf{A637}, 435
  (1998), \texttt{nucl-th/9805035}

\bibitem{Bombaci:2018ksa}
I.~Bombaci, D.~Logoteta, Astron. Astrophys. \textbf{609}, A128 (2018),
  \texttt{1805.11846}

\bibitem{Damour:1983a}
T.~{Damour}, \emph{{Gravitational radiation and the motion of compact bodies}},
  in \emph{Gravitational Radiation}, edited by N.~{Deruelle}, T.~{Piran}
  (North-Holland, Amsterdam, 1983), pp. 59--144

\bibitem{Flanagan:2007ix}
E.E. Flanagan, T.~Hinderer, Phys.Rev. \textbf{D77}, 021502 (2008),
  \texttt{0709.1915}

\bibitem{Hinderer:2009ca}
T.~Hinderer, B.D. Lackey, R.N. Lang, J.S. Read, Phys. Rev. \textbf{D81}, 123016
  (2010), \texttt{0911.3535}

\bibitem{Damour:2009wj}
T.~Damour, A.~Nagar, Phys. Rev. \textbf{D81}, 084016 (2010), \texttt{0911.5041}

\bibitem{Damour:2012yf}
T.~Damour, A.~Nagar, L.~Villain, Phys.Rev. \textbf{D85}, 123007 (2012),
  \texttt{1203.4352}

\bibitem{Bernuzzi:2012ci}
S.~Bernuzzi, A.~Nagar, M.~Thierfelder, B.~Br{\"u}gmann, Phys.Rev. \textbf{D86},
  044030 (2012), \texttt{1205.3403}

\bibitem{Bernuzzi:2014kca}
S.~Bernuzzi, A.~Nagar, S.~Balmelli, T.~Dietrich, M.~Ujevic, Phys.Rev.Lett.
  \textbf{112}, 201101 (2014), \texttt{1402.6244}

\bibitem{Zappa:2017xba}
F.~Zappa, S.~Bernuzzi, D.~Radice, A.~Perego, T.~Dietrich, Phys. Rev. Lett.
  \textbf{120}, 111101 (2018), \texttt{1712.04267}

\bibitem{Hotokezaka:2011dh}
K.~Hotokezaka, K.~Kyutoku, H.~Okawa, M.~Shibata, K.~Kiuchi, Phys.Rev.
  \textbf{D83}, 124008 (2011), \texttt{1105.4370}

\bibitem{Bauswein:2013jpa}
A.~Bauswein, T.~Baumgarte, H.T. Janka, Phys.Rev.Lett. \textbf{111}, 131101
  (2013), \texttt{1307.5191}

\bibitem{Radice:2017lry}
D.~Radice, A.~Perego, F.~Zappa, S.~Bernuzzi, Astrophys. J. \textbf{852}, L29
  (2018), \texttt{1711.03647}

\bibitem{Radice:2018xqa}
D.~Radice, A.~Perego, S.~Bernuzzi, B.~Zhang, Mon. Not. Roy. Astron. Soc.
  \textbf{481}, 3670 (2018), \texttt{1803.10865}

\bibitem{Eichler:1989ve}
D.~Eichler, M.~Livio, T.~Piran, D.N. Schramm, Nature \textbf{340}, 126 (1989)

\bibitem{Nakar:2007yr}
E.~Nakar, Phys. Rept. \textbf{442}, 166 (2007), \texttt{astro-ph/0701748}

\bibitem{Rosswog:2015nja}
S.~Rosswog, Int.J.Mod.Phys. \textbf{D24}, 1530012 (2015), \texttt{1501.02081}

\bibitem{Just:2016ApJ...816L..30J}
O.~{Just}, M.~{Obergaulinger}, H.T. {Janka}, A.~{Bauswein}, N.~{Schwarz},
  Astrophys. J. \textbf{816}, L30 (2016), \texttt{1510.04288}

\bibitem{Radice:2016rys}
D.~Radice, S.~Bernuzzi, W.~Del~Pozzo, L.F. Roberts, C.D. Ott, Astrophys. J.
  \textbf{842}, L10 (2017), \texttt{1612.06429}

\bibitem{Most:2018eaw}
E.R. Most, L.J. Papenfort, V.~Dexheimer, M.~Hanauske, S.~Schramm, H.~Stöcker,
  L.~Rezzolla (2018), \texttt{1807.03684}

\bibitem{Bauswein:2018bma}
A.~Bauswein, N.U.F. Bastian, D.B. Blaschke, K.~Chatziioannou, J.A. Clark,
  T.~Fischer, M.~Oertel, Phys. Rev. Lett. \textbf{122}, 061102 (2019),
  \texttt{1809.01116}

\bibitem{TheLIGOScientific:2017qsa}
B.P. Abbott et~al. (Virgo, LIGO Scientific), Phys. Rev. Lett. \textbf{119},
  161101 (2017), \texttt{1710.05832}

\bibitem{Abbott:2018wiz}
B.P. Abbott et~al. (LIGO Scientific, Virgo), Phys. Rev. \textbf{X9}, 011001
  (2019), \texttt{1805.11579}

\bibitem{Abbott:2018exr}
B.P. Abbott et~al. (LIGO Scientific, Virgo), Phys. Rev. Lett. \textbf{121},
  161101 (2018), \texttt{1805.11581}

\bibitem{De:2018uhw}
S.~De, D.~Finstad, J.M. Lattimer, D.A. Brown, E.~Berger, C.M. Biwer (2018),
  \texttt{1804.08583}

\bibitem{Baiotti:2011am}
L.~Baiotti, T.~Damour, B.~Giacomazzo, A.~Nagar, L.~Rezzolla, Phys. Rev.
  \textbf{D84}, 024017 (2011), \texttt{1103.3874}

\bibitem{Radice:2013hxh}
D.~Radice, L.~Rezzolla, F.~Galeazzi, Mon.Not.Roy.Astron.Soc. \textbf{437}, L46
  (2014), \texttt{1306.6052}

\bibitem{Hotokezaka:2015xka}
K.~Hotokezaka, K.~Kyutoku, H.~Okawa, M.~Shibata, Phys. Rev. \textbf{D91},
  064060 (2015), \texttt{1502.03457}

\bibitem{Nagar:2018zoe}
A.~Nagar et~al., Phys. Rev. \textbf{D98}, 104052 (2018), \texttt{1806.01772}

\bibitem{Abbott:2017dke}
B.P. Abbott et~al. (Virgo, LIGO Scientific), Astrophys. J. \textbf{851}, L16
  (2017), \texttt{1710.09320}

\bibitem{Bauswein:2011tp}
A.~Bauswein, H.T. Janka, Phys.Rev.Lett. \textbf{108}, 011101 (2012),
  \texttt{1106.1616}

\bibitem{Takami:2014zpa}
K.~Takami, L.~Rezzolla, L.~Baiotti, Phys.Rev.Lett. \textbf{113}, 091104 (2014),
  \texttt{1403.5672}

\bibitem{Bernuzzi:2015rla}
S.~Bernuzzi, T.~Dietrich, A.~Nagar, Phys. Rev. Lett. \textbf{115}, 091101
  (2015), \texttt{1504.01764}

\bibitem{Yang:2017xlf}
H.~Yang, V.~Paschalidis, K.~Yagi, L.~Lehner, F.~Pretorius, N.~Yunes (2017),
  \texttt{1707.00207}

\bibitem{Chatziioannou:2017ixj}
K.~Chatziioannou, J.A. Clark, A.~Bauswein, M.~Millhouse, T.B. Littenberg,
  N.~Cornish, Phys. Rev. \textbf{D96}, 124035 (2017), \texttt{1711.00040}

\bibitem{Margalit:2017dij}
B.~Margalit, B.D. Metzger, Astrophys. J. \textbf{850}, L19 (2017),
  \texttt{1710.05938}

\bibitem{Shibata:2017xdx}
M.~Shibata, S.~Fujibayashi, K.~Hotokezaka, K.~Kiuchi, K.~Kyutoku, Y.~Sekiguchi,
  M.~Tanaka, Phys. Rev. \textbf{D96}, 123012 (2017), \texttt{1710.07579}

\bibitem{Rezzolla:2017aly}
L.~Rezzolla, E.R. Most, L.R. Weih, Astrophys. J. \textbf{852}, L25 (2018),
  \texttt{1711.00314}

\bibitem{Ruiz:2017due}
M.~Ruiz, S.L. Shapiro, A.~Tsokaros, Phys. Rev. \textbf{D97}, 021501 (2018),
  \texttt{1711.00473}

\bibitem{Bauswein:2017vtn}
A.~Bauswein, O.~Just, H.T. Janka, N.~Stergioulas, Astrophys. J. \textbf{850},
  L34 (2017), \texttt{1710.06843}

\bibitem{Radice:2018ozg}
D.~Radice, L.~Dai (2018), \texttt{1810.12917}

\bibitem{Demorest:2010bx}
P.~Demorest, T.~Pennucci, S.~Ransom, M.~Roberts, J.~Hessels, Nature
  \textbf{467}, 1081 (2010), \texttt{1010.5788}

\bibitem{Antoniadis:2013pzd}
J.~Antoniadis, P.C. Freire, N.~Wex, T.M. Tauris, R.S. Lynch et~al., Science
  \textbf{340}, 6131 (2013), \texttt{1304.6875}

\bibitem{Typel:2009sy}
S.~Typel, G.~Ropke, T.~Klahn, D.~Blaschke, H.H. Wolter, Phys. Rev.
  \textbf{C81}, 015803 (2010), \texttt{0908.2344}

\bibitem{Hempel:2009mc}
M.~Hempel, J.~Schaffner-Bielich, Nucl. Phys. \textbf{A837}, 210 (2010),
  \texttt{0911.4073}

\bibitem{Steiner:2012rk}
A.W. Steiner, M.~Hempel, T.~Fischer, Astrophys. J. \textbf{774}, 17 (2013),
  \texttt{1207.2184}

\bibitem{LoreneCode}
{Eric Gourgoulhon, Philippe Grandcl\'{e}ment, Jean-Alain Marck, J\'{e}r\^{o}me
  Novak and Keisuke Taniguchi}, \url{http://www.lorene.obspm.fr/}, {Paris
  Observatory, Meudon section - LUTH laboratory}

\bibitem{Bernuzzi:2009ex}
S.~Bernuzzi, D.~Hilditch, Phys. Rev. \textbf{D81}, 084003 (2010),
  \texttt{0912.2920}

\bibitem{Hilditch:2012fp}
D.~Hilditch, S.~Bernuzzi, M.~Thierfelder, Z.~Cao, W.~Tichy et~al., Phys. Rev.
  \textbf{D88}, 084057 (2013), \texttt{1212.2901}

\bibitem{Radice:2012cu}
D.~Radice, L.~Rezzolla, Astron. Astrophys. \textbf{547}, A26 (2012),
  \texttt{1206.6502}

\bibitem{Radice:2013xpa}
D.~Radice, L.~Rezzolla, F.~Galeazzi, Class.Quant.Grav. \textbf{31}, 075012
  (2014), \texttt{1312.5004}

\bibitem{Radice:2015nva}
D.~Radice, L.~Rezzolla, F.~Galeazzi, ASP Conf. Ser. \textbf{498}, 121 (2015),
  \texttt{1502.00551}

\bibitem{Berger:1984zza}
M.J. Berger, J.~Oliger, J.Comput.Phys. \textbf{53}, 484 (1984)

\bibitem{Berger:1989a}
M.J. {Berger}, P.~{Colella}, Journal of Computational Physics \textbf{82}, 64
  (1989)

\bibitem{Reisswig:2012nc}
C.~Reisswig, R.~Haas, C.D. Ott, E.~Abdikamalov, P.~Mösta, D.~Pollney,
  E.~Schnetter, Phys. Rev. \textbf{D87}, 064023 (2013), \texttt{1212.1191}

\bibitem{Schnetter:2003rb}
E.~Schnetter, S.H. Hawley, I.~Hawke, Class.Quant.Grav. \textbf{21}, 1465
  (2004), \texttt{gr-qc/0310042}

\bibitem{Radice:2017zta}
D.~Radice, Astrophys. J. \textbf{838}, L2 (2017), \texttt{1703.02046}

\bibitem{Shakura:1972te}
N.I. Shakura, R.A. Sunyaev, Astron. Astrophys. \textbf{24}, 337 (1973)

\bibitem{Kiuchi:2017zzg}
K.~Kiuchi, K.~Kyutoku, Y.~Sekiguchi, M.~Shibata, Phys. Rev. \textbf{D97},
  124039 (2018), \texttt{1710.01311}

\bibitem{Rosswog:2003rv}
S.~Rosswog, M.~Liebendoerfer, Mon.Not.Roy.Astron.Soc. \textbf{342}, 673 (2003),
  \texttt{astro-ph/0302301}

\bibitem{Itoh:1996ApJS..102..411I}
N.~{Itoh}, H.~{Hayashi}, A.~{Nishikawa}, Y.~{Kohyama}, Astrophys. J. Suppl.
  \textbf{102}, 411 (1996)

\bibitem{Galeazzi:2013mia}
F.~Galeazzi, W.~Kastaun, L.~Rezzolla, J.A. Font, Phys.Rev. \textbf{D88}, 064009
  (2013), \texttt{1306.4953}

\bibitem{Radice:2016dwd}
D.~Radice, F.~Galeazzi, J.~Lippuner, L.F. Roberts, C.D. Ott, L.~Rezzolla, Mon.
  Not. Roy. Astron. Soc. \textbf{460}, 3255 (2016), \texttt{1601.02426}

\bibitem{Neilsen:2014hha}
D.~Neilsen, S.L. Liebling, M.~Anderson, L.~Lehner, E.~O’Connor et~al.,
  Phys.Rev. \textbf{D89}, 104029 (2014), \texttt{1403.3680}

\bibitem{Ruffert:1995fs}
M.H. Ruffert, H.T. Janka, G.~Sch{\"a}fer, Astron. Astrophys. \textbf{311}, 532
  (1996), \texttt{astro-ph/9509006}

\bibitem{Sekiguchi:2015dma}
Y.~Sekiguchi, K.~Kiuchi, K.~Kyutoku, M.~Shibata, Phys.Rev. \textbf{D91}, 064059
  (2015), \texttt{1502.06660}

\bibitem{Foucart:2015vpa}
F.~Foucart, E.~O'Connor, L.~Roberts, M.D. Duez, R.~Haas, L.E. Kidder, C.D. Ott,
  H.P. Pfeiffer, M.A. Scheel, B.~Szilagyi, Phys. Rev. \textbf{D91}, 124021
  (2015), \texttt{1502.04146}

\bibitem{Foucart:2018gis}
F.~Foucart, M.D. Duez, L.E. Kidder, R.~Nguyen, H.P. Pfeiffer, M.A. Scheel
  (2018), \texttt{1806.02349}

\bibitem{Bruenn:1985en}
S.W. Bruenn, Astrophys. J. Suppl. \textbf{58}, 771 (1985)

\bibitem{Burrows:2004vq}
A.~Burrows, S.~Reddy, T.A. Thompson, Nucl. Phys. \textbf{A777}, 356 (2006),
  \texttt{astro-ph/0404432}

\bibitem{Shapiro:1983du}
S.L. Shapiro, S.A. Teukolsky, \emph{{Black holes, white dwarfs, and neutron
  stars: The physics of compact objects}} (Wiley, New York, USA, 1983)

\bibitem{Perego:2015agy}
A.~Perego, R.~Cabezon, R.~Kaeppeli, Astrophys. J. Suppl. \textbf{223}, 22
  (2016), \texttt{1511.08519}

\bibitem{Kaplan:2013wra}
J.~Kaplan, C.~Ott, E.~O'Connor, K.~Kiuchi, L.~Roberts et~al., Astrophys.J.
  \textbf{790}, 19 (2014), \texttt{1306.4034}

\bibitem{Hotokezaka:2012ze}
K.~Hotokezaka, K.~Kiuchi, K.~Kyutoku, H.~Okawa, Y.i. Sekiguchi et~al.,
  Phys.Rev. \textbf{D87}, 024001 (2013), \texttt{1212.0905}

\bibitem{Bauswein:2013yna}
A.~Bauswein, S.~Goriely, H.T. Janka, Astrophys.J. \textbf{773}, 78 (2013),
  \texttt{1302.6530}

\bibitem{Stergioulas:2011gd}
N.~Stergioulas, A.~Bauswein, K.~Zagkouris, H.T. Janka, Mon.Not.Roy.Astron.Soc.
  \textbf{418}, 427 (2011), \texttt{1105.0368}

\bibitem{Bernuzzi:2013rza}
S.~Bernuzzi, T.~Dietrich, W.~Tichy, B.~Br{\"u}gmann, Phys.Rev. \textbf{D89},
  104021 (2014), \texttt{1311.4443}

\bibitem{Bacca:2011qd}
S.~Bacca, K.~Hally, M.~Liebendorfer, A.~Perego, C.J. Pethick, A.~Schwenk,
  Astrophys. J. \textbf{758}, 34 (2012), \texttt{1112.5185}

\bibitem{Fischer:2013eka}
T.~Fischer, M.~Hempel, I.~Sagert, Y.~Suwa, J.~Schaffner-Bielich, Eur. Phys. J.
  \textbf{A50}, 46 (2014), \texttt{1307.6190}

\bibitem{Lalit:2018dps}
S.~Lalit, M.A.A. Mamun, C.~Constantinou, M.~Prakash, Eur. Phys. J.
  \textbf{A55}, 10 (2019), \texttt{1809.08126}

\bibitem{Perego:2014fma}
A.~Perego, S.~Rosswog, R.~Cabezon, O.~Korobkin, R.~Kaeppeli et~al.,
  Mon.Not.Roy.Astron.Soc. \textbf{443}, 3134 (2014), \texttt{1405.6730}

\bibitem{Kastaun:2016elu}
W.~Kastaun, R.~Ciolfi, A.~Endrizzi, B.~Giacomazzo, Phys. Rev. \textbf{D96},
  043019 (2017), \texttt{1612.03671}

\bibitem{Cox:1968pss..book.....C}
J.P. {Cox}, R.T. {Giuli}, \emph{{Principles of stellar structure }} (Gordon and
  Breach, New York, 1968)

\bibitem{Takahashi:1978}
K.~{Takahashi}, M.F. {El Eid}, W.~{Hillebrandt}, Astron.Astrophys. \textbf{67},
  185 (1978)

\bibitem{Banik:2014qja}
S.~Banik, M.~Hempel, D.~Bandyopadhyay, Astrophys. J. Suppl. \textbf{214}, 22
  (2014), \texttt{1404.6173}

\bibitem{Glendenning:1996cost.book.....G}
N.K. {Glendenning}, \emph{{Compact Stars}} (Springer-Verlag, New York, 1996)

\bibitem{Metzger:2014ila}
B.D. Metzger, R.~Fernández, Mon.Not.Roy.Astron.Soc. \textbf{441}, 3444 (2014),
  \texttt{1402.4803}

\bibitem{Just:2014fka}
O.~Just, A.~Bauswein, R.A. Pulpillo, S.~Goriely, H.T. Janka, Mon. Not. Roy.
  Astron. Soc. \textbf{448}, 541 (2015), \texttt{1406.2687}

\bibitem{Foucart:2016rxm}
F.~Foucart, E.~O'Connor, L.~Roberts, L.E. Kidder, H.P. Pfeiffer, M.A. Scheel,
  Phys. Rev. \textbf{D94}, 123016 (2016), \texttt{1607.07450}

\bibitem{Lippuner:2017bfm}
J.~Lippuner, R.~Fernández, L.F. Roberts, F.~Foucart, D.~Kasen, B.D. Metzger,
  C.D. Ott, Mon. Not. Roy. Astron. Soc. \textbf{472}, 904 (2017),
  \texttt{1703.06216}

\end{thebibliography}

\end{document}